%% file: OR-Benchmark.tex
\newif\iffulledition
\fulleditiontrue

\newif\ifarXiv
\arXivtrue

\documentclass[journal]{IEEEtran}

\usepackage{amsmath,amssymb}

\usepackage{graphicx}
\usepackage{subfig}

\ifCLASSOPTIONonecolumn
\fi
\newlength\figwidth
\iffulledition
\else
    \ifCLASSOPTIONonecolumn
    \else
    \fi
\fi
\newlength\figheight
\usepackage{pgf}
\usepackage{pgfplots}
\pgfplotsset{
label style={font=\footnotesize},
legend style={font=\footnotesize},
tick label style={font=\footnotesize},
width=\figwidth,
height=\figheight,
xlabel near ticks,
ylabel near ticks,
legend style={legend columns=3,legend cell align=left,align=left,fill=none,draw=none}
}

\usepackage{cite}
\usepackage{url}

\usepackage{color}
\definecolor{dgreen}{rgb}{0,.6,0}
\usepackage{soul}
\usepackage{comment}
\specialcomment{delete}{\begingroup\color{red}}{\endgroup}

\newif\ifcomments
\commentsfalse
\ifcomments
    \newcommand\comments[1]{\textcolor{blue}{[\ul{Shujun's comment(s)}: \textit{#1}]}} 
    \newcommand\deleted[1]{\textcolor{red}{\st{#1}}} 
    \newcommand\modified[2]{\deleted{#1}\textcolor{dgreen}{#2}} 
    \newcommand\reply[1]{\textcolor{cyan}{[\ul{Hui's reply}: \textit{#1}]}} 
    \specialcomment{add}{\begingroup\color{dgreen}}{\endgroup}
\else 
    \newcommand\comments[1]{} 
    \newcommand\deleted[1]{} 
    \excludecomment{delete}
    \newcommand\modified[2]{#2} 
    \newcommand\reply[1]{} 
    \specialcomment{add}{\begingroup}{\endgroup}
\fi

\begin{document}

\title{OR-Benchmark: An Open and Reconfigurable Digital Watermarking Benchmarking Framework%
}

\author{Hui~Wang,~
        Anthony~TS~Ho,~\IEEEmembership{Senior Member,~IEEE}
        and~Shujun~Li,~\IEEEmembership{Senior Member,~IEEE}%
\thanks{All authors are with the Department of Computing and the Surrey Centre for Cyber Security (SCCS), University of Surrey, UK.}
\thanks{Contact information: Hui Wang (\protect\url{h.wang80s@hotmail.com}), Anthony TS Ho (\protect\url{a.ho@surrey.ac.uk}), Shujun Li (\protect\url{http://www.hooklee.com/}).}}


\maketitle

\begin{abstract}
Benchmarking digital watermarking algorithms is not an easy task because different applications of digital watermarking often have very different sets of requirements and trade-offs between conflicting requirements. While there have been some general-purpose digital watermarking benchmarking systems available, they normally do not support complicated benchmarking tasks and cannot be easily reconfigured to work with different watermarking algorithms and testing conditions. In this paper, we propose OR-Benchmark, an open and highly reconfigurable general-purpose digital watermarking benchmarking framework, which has the following two key features: 1) all the interfaces are public and general enough to support all watermarking applications and benchmarking tasks we can think of; 2) end users can easily extend the functionalities and freely configure what watermarking algorithms are tested, what system components are used, how the benchmarking process runs, and what results should be produced. We implemented a prototype of this framework as a MATLAB software package and used it to benchmark a number of digital watermarking algorithms involving two types of watermarks for content authentication and self-restoration purposes. The benchmarking results demonstrated the advantages of the proposed benchmarking framework, and also gave us some useful insights about existing image authentication and self-restoration watermarking algorithms which are an important but less studied topic in digital watermarking.
\end{abstract}

\begin{IEEEkeywords}
Digital Watermarking, Benchmarking, Performance Evaluation, Reconfigurability, Content Authentication, Self-restoration
\end{IEEEkeywords}

\IEEEpeerreviewmaketitle

\section{Introduction}
\label{sec:intro}

\IEEEPARstart{D}{igital} watermarking, a branch of information hiding, involves research on the process of embedding digital information (watermark) within a cover signal to achieve different (often security-related) functionalities related to the cover signal and/or its consumption by end users \cite{Cox2007}. Since the late 1980s a large number of digital watermarking algorithms have been proposed for many applications with different system requirements mostly for protecting different types of multimedia data such as still images, audio, video, 3-D models \cite{VoyatzisPitas1999, WateramrkingSurvey:SPM2001, HalftoningWatermarking:IEEETMM2006, Li:AudioWateramrking:IEEETMM2006, Uhl:H264Wateramrking:IEEETMM2014, 3DWateramrkingSurvey:IEEETMM2008, NetworkWatermarking:IEEEACMTN2014}. For instance, due to the convenience of transmission and storage for digital multimedia data on the Internet, copyright protection of digital multimedia content has become one main application of digital watermarking. In this application, robust watermarking schemes \cite{SSSWateramrking:IEEETIP97, RSTWateramrking:IEEETIP2001, Li:AudioWateramrking:IEEETMM2006, Zhu:ImageRobust4Wateramrking:IEEETMM2014} are desired to embed copyright information as a watermark in the digital media that can be hard to remove. Other applications of digital watermarking include content authentication, transaction tracking, usage control, self-restoration, broadcast monitoring, \emph{etc}. In some multimedia content authentication applications, fragile watermarking schemes are desired because of the need to capture \emph{any} change to the content, which is often achieved via fragility of digital watermarks embedded \cite{Ho:FragileWateramrking:IEEETIFS2008, Zhang:FragileWateramrking:IEEETMM2008, He:FragileWateramrking:IEEETIFS2011}. In some other multimedia content authentication applications, however, semi-fragile watermarking schemes \cite{Ho2004ICA, Maeno:SemiFragileWateramrking:IEEETMM2006, Ni:SemiFragileWateramrking:IEEETCSVT2008} are desired to tolerate benign signal processing operations on watermarked multimedia data while malicious alterations should still be detectable, which is important for today's multimedia systems involve complicated processes between the sender and the receiver of multimedia contents.

There are a number of properties associated with a digital watermarking algorithm depending on different application requirements. It is well accepted that imperceptibility and robustness are the two most important but normally conflicting requirements. Besides, embedding capacity/efficiency, security (\emph{i.e.} the ability to resist malicious attacks) and computational complexity are also important properties for most digital watermarking systems. However, the importance of each property is different in different applications. Some properties also overlap with each other, \emph{e.g.} security is often linked to robustness against malicious signal processing (attacks). \iffulledition For instance, in copyright protection applications, the requirement on robustness is critical as the digital watermark need to survive both benign signal processing and malicious attacks, however, in content authentication applications, fragility (\emph{i.e.} lack of robustness) is required to detect malicious content manipulations.\fi There are also some additional application-oriented properties, \emph{e.g.} reliability (normally measured using decoding error rates or correlation of decoded watermark with the original watermark) in copyright protection applications, accuracy (normally measured using false positive and false negative rates) in content authentication applications, and perceptual quality of the recovered cover in self-restoration applications.

As in many other multimedia systems, a general-purpose, flexible and fair benchmarking environment with appropriate test criteria is of particular importance for performance evaluation and comparison of digital watermarking algorithms. In the literature most researchers compared their digital watermarking algorithms with competitive ones by looking at a number of selected testing criteria for one or more target applications. However, because of different testing conditions and the lack of details of the experimental setups, it is hardly possible to depend on published results to do performance comparison. Thus, it is often needed to repeat the performance evaluation process for previous algorithms under the same testing condition to have a fair comparison with a new algorithm proposed, which calls for the need of a general-purpose benchmarking system that can facilitate the performance evaluation/comparison process and maximize the reuse of previous results. With a properly-designed benchmarking system, end users and researchers can conduct performance evaluation of a given algorithm and compare performance of multiple algorithms more easily and fairly to know more about pros and cons of different algorithms and to draw more insights about how to further improve existing algorithms. Since the 1990s, a number of digital watermarking benchmarking systems have been proposed \cite{PetitcolasAK1998, Petitcolas2000, SolachidisTNTNP2001, PereiraVMMPP2001, VorbruggenF2002, MacqDD2004}.

Generally speaking, benchmarking performance of digital watermarking algorithms is not an easy task because different digital watermarking applications often have very different sets of requirements and trade-offs among conflicting requirements. When multiple digital watermarking algorithms with changeable parameters have to be evaluated against each other, the benchmarking task becomes more complicated. Furthermore, for systems involving more than one type of watermarks, \emph{e.g.} content authentication watermarking with the capability of self-restoration, the complexity of the benchmarking task becomes even higher. While there have been some general-purpose digital watermarking benchmarking systems available, most of them can be applied to only certain digital watermarking systems for a limited range of applications. In addition, existing benchmarking systems normally do not support complicated benchmarking tasks and cannot be easily reconfigured to work with different algorithms and testing conditions. It is thus still a challenge to design an efficient and general-purpose benchmarking system that can be used to benchmark different digital watermarking algorithms.

In this paper, we propose OR-Benchmark, an open and highly reconfigurable general-purpose framework for benchmarking digital watermarking algorithms, which is designed to meet the needs of different digital watermarking algorithms and various benchmarking tasks. Its main features include:
\begin{itemize}
\item The framework has \emph{open} interfaces for (re)configuring different parts of the benchmarking system and addition of new modules.\footnote{Adding new source code is unavoidable for new functionalities, so our focus is how easy one can add own source code to extend its functionalities.} We plan to release the implemented prototype of the framework as an \emph{open}-source tool.


\item The framework defines a unified procedure of benchmarking different digital watermarking algorithms against different attacks and using different performance indicators to make the comparison fairer and more systematic.

\item The framework is designed to be independent of the media type, so it can be applied to digital watermarking algorithms for different media types although in this paper we will only demonstrate it for image watermarking.
\end{itemize}

We implemented a prototype of the proposed OR-Benchmark framework as a MATLAB software package. To demonstrate how the framework can be used to benchmark digital watermarking systems, we used the implemented prototype system to benchmark three recently proposed semi-fragile image watermarking algorithms for content authentication and self-restoration. Those benchmarked digital watermarking systems use two types of watermarks (one for content authentication and the other for self-restoration), so are among the most complicated digital watermarking systems one may need to benchmark. The results on one hand proved the advantages of the proposed framework, and on the other hand led to some insights about how to better compare performance of such complicated digital watermarking systems and further improve their performance.

The rest of the paper is organized as follows. In Section~\ref{sec:related_work}, related work on digital watermarking benchmarking is introduced. Section~\ref{sec:framework} gives a detailed description of our proposed benchmarking framework, including our abstract modelling of digital watermarking systems, important evaluation criteria, the proposed OR-Benchmark framework, and comparison with other existing digital watermarking benchmarking systems. Next, in Section~\ref{sec:experimental_results}, we describe how we implemented a first prototype of OR-Benchmark in MATLAB, and results of using the implemented prototype system to benchmark a number of digital image watermarking systems for content authentication and self-restoration. In Section~\ref{sec:discussion}, we discuss some subtle aspects about digital watermarking benchmarking and how we currently handle them in OR-Benchmark. The paper is concluded by Section~\ref{sec:col_fu} with future work.

\section{Related Work}
\label{sec:related_work}

Benchmarking of digital watermarking algorithms is the process of evaluating and comparing their performance under a fair and normally (semi-)automated environment. While there have been a substantial number of digital watermarking algorithms proposed for different applications and usage scenarios, there are relatively less research on digital watermarking benchmarking especially general-purpose frameworks capable of handling multiple applications with different sets of requirements. Most existing digital watermarking benchmarking systems focus on some well-defined sub-areas among which image watermarking received the most attention. In this section, we briefly overview some representative work.

\subsection{StirMark}
\label{sec:stirmark}

StirMark, one of the earliest and the most well-known digital watermarking benchmarking systems, was firstly proposed by Petitcolas \emph{et al.} in 1998 \cite{PetitcolasAK1998} as a generic tool for benchmarking digital image watermarking algorithms against various attacks, which was later contributed by more researchers in 2001 \cite{PetitcolasSRDFF2001} to become a more general framework for benchmarking digital watermarking algorithms. Subsequently, several enhanced versions of StirMark were developed to include more attacks and cover audio watermarking \cite{SteinebachPRDFSFF2001, SMBA}. The main aim of StirMark is to develop a fully automated evaluation service, which could encapsulate different performance evaluation indicators and allow continuous development of new attacks to be integrated into the whole system. Since StirMark is among the most widely-used benchmarking systems by the digital watermarking community, we discuss it in greater detail below.

\iffulledition
\begin{figure*}[!htb]
\centering
\includegraphics[width=0.8\textwidth]{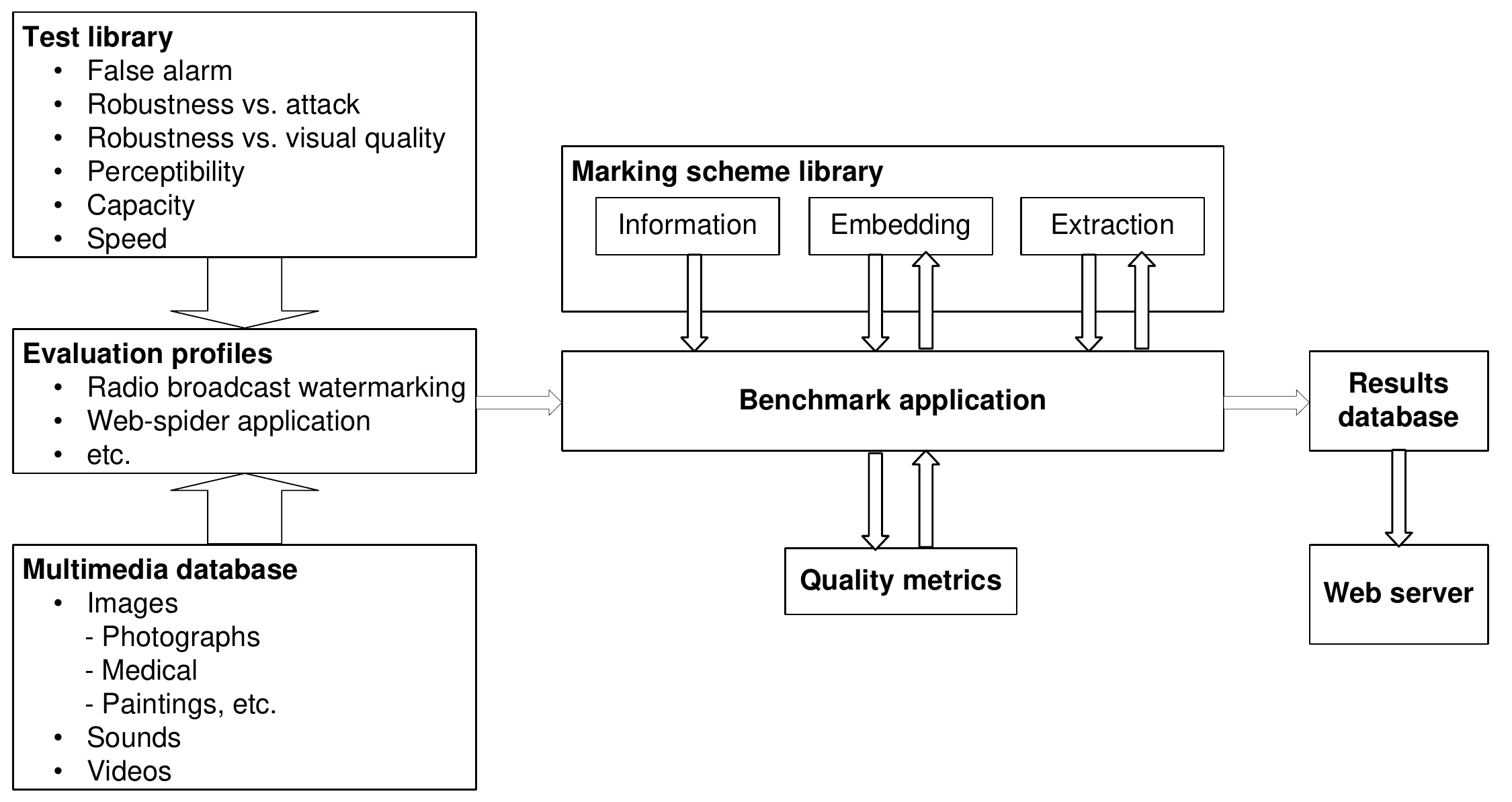}
\caption{The architecture of StirMark for watermarking evaluation \cite{PetitcolasSRDFF2001}.}
\label{fig:stirmark_frame}
\end{figure*}
\fi

\subsubsection{Interfaces}
\label{sec:stir_interface}

To use StirMark for benchmarking a given digital watermarking algorithm, the user is required to supply three functions, \emph{Embed} and \emph{Extract} functions, and one \emph{GetSchemeInfo} function which provides meta-information about the algorithm such as the name, version, author(s), the maximum byte-length of the embedded message, the maximum bite-length of the stego-key, \emph{etc}.

In order to support different use cases and digital watermarking algorithms, several parameters are provided including some mandatory parameters such as the strength for embedding/extraction, the key for embedding/extraction, the watermark to be embedded and extracted, and some optional parameters such as the maximum distortion tolerated and the certainty of extraction (\emph{i.e.} a number between 0 and 100 indicating the probability of an embedded watermark being correctly detected). All the parameters are used to support various types of algorithms, but users cannot easily add new parameters without changing the source code of StirMark.

Since different watermarking algorithms have different evaluation requirements, StirMark divides watermarking algorithms into six categories according to blindness and the output of the \emph{Extract} function. According to the algorithm type, StirMark defines different sets of input and output arguments for watermark embedding and extracting functions, and different sets of tests listed in the evaluation profiles.

\subsubsection{Evaluation Criteria}

The main performance indicators of a digital watermarking algorithm StirMark can evaluate include imperceptibility, capacity, robustness to attacks, false alarm rate and execution speed as discussed below.

\emph{Imperceptibility} is evaluated as the perceptual quality distortion introduced to the cover signal by the watermark embedding process. StirMark uses PSNR as the default perceptual visual quality assessment (PQA) metric and in principle allows the use of other PQA metrics. However, adding other PQA metrics requires modifying the source code of the StirMark implementation related to imperceptibility evaluation\iffulledition\ such as the ``robustness vs. visual quality'' test\fi.

Normally the \emph{embedding capacity} is a fixed constraint, so StirMark does not directly measure this but uses it to inform the robustness testing process where the watermark has a random payload with a given size. StirMark provides tools to analyze relation between capacity and robustness.

Regarding \emph{robustness to attacks}, StirMark implementation models attacks as C++ classes and allows addition of new attacks as new classes to test.

The \emph{false alarm rate} is also known as ``false positive rate'' which contains two cases: 1) the detector reports a mark in a signal without a mark; 2) the detector reports a mark $w'$ in a signal marked with $w\neq w'$. In StirMark, the first case is evaluated by taking some randomly selected signals without any watermark and sending them to the detector to see if the detector reports a watermark (wrongly), and the second case is evaluated by taking some marked signals and run them through the detector to see if a wrong watermark is detected.

In StirMark, the \emph{execution speed} is evaluated by computing the average CPU times spent on the watermark embedding/extraction processes for a given signal of a particular size and on a particular platform.

\subsubsection{Benchmarking Framework}

\iffulledition Figure~\ref{fig:stirmark_frame} shows the architecture of StirMark as a watermarking evaluation framework. There are six main components in the framework\fi\ including the marking scheme library, test library, evaluation profile library, quality metrics library, multimedia database and results database. According to different application requirements, there are different evaluation profiles, each of which is composed by a list of tests or attacks to be applied and a list of multimedia signals required for the test. The end user is required to add the watermarking algorithm under testing (in the form of three C++ functions including \emph{GetSchemeInfo}, \emph{Embed} and \emph{Extract}) to the marking scheme library. In \emph{GetSchemeInfo} function, the end users also selects which evaluation profile will be used. The evaluation profiles are written as INI files with a pre-defined static structure, so although users can define their own profiles they are limited to the static structure. Extending the structure of the evaluation profiles requires changes to the StirMark implementation's source code. According to the information provided by the end user, StirMark runs the defined benchmarking process automatically by using its multimedia database, the tests (attacks) library and the quality metrics library. The results are stored in a database (an SQL server as stated in \cite{PetitcolasSRDFF2001} and simple files as in actual implementations).

StirMark is designed to achieve simplicity (to conduct tests and record results automatically) and customization (to choose different evaluation profiles by the end user). However, the boundaries among watermarking library, evaluation profiles, test library and quality metrics is unclear. For instance, the test library associates with not only the evaluation profile, but also the quality metrics and some information about parameter settings in watermarking scheme library. Although the authors of \cite{PetitcolasSRDFF2001} mentioned that StirMark allows the addition and use of new test and quality metrics, however, it is not easy to do so as the interfaces among different parts of the framework are not all clearly defined and manual changes to core StirMark source code (in C++) are always required. Furthermore, there are only a limited number of evaluation profiles in the current StirMark implementation which can only be used to benchmark some limited types of digital watermarking schemes.

\subsubsection{Implementation}

StirMark was originally developed by\iffulledition\ Markus\fi\ Kuhn in 1997 \cite{stirmark_v1} as a generic software tool for simple robustness testing of image watermarking algorithms. It simulates many common attacks to image watermarking algorithms including random bilinear geometric distortions to de-synchronize watermarking algorithms. In \cite{PetitcolasAK1998} it was suggested that image watermarking algorithms which do not survive StirMark should be considered unacceptably insecure.

Subsequently, further development of StirMark was taken over by\iffulledition\ Fabien\fi\ Petitcolas and it was incrementally improved by Petitcolas and some other researchers for more digital watermarking applications to become a ``fair'' benchmarking system with a longer list of tests and attacks with the release of its 3.1 version in 1999 \cite{KutterPetitcolasEI99}. Later on some more development work took place, including a set of tests for audio watermarking developed by Steinebach \emph{et al.} \cite{SteinebachPRDFSFF2001} and by Lang and Dittmann \cite{SMBA}. There was also efforts of making StirMark a public automated web-based evaluation service made by Petitcolas \emph{et al.} \cite{PetitcolasSRDFF2001} which led to the 4.0 version of Stirmark \cite{stirmark_v4}. The StirMark implementation was written using C++, and it has some level of reconfigurability in terms of an INI file where the end user can define a specified evaluation profile to list all the tests with relevant parameters and all the multimedia objects required for the tests.

\subsubsection{Limitations}

Although StirMark has been widely used as a tool for robustness and security evaluation of digital watermarking algorithms, we feel it has the following limitations.

The modelling and interface for digital watermarking algorithms do not cover all applications. For instance, there are only two types of output for watermark detection (i.e. the \emph{Extract} function): the extracted watermark and a certainty to show the probability whether the watermark is detected correctly. This is obviously not sufficient to support digital watermarking algorithms for tamper localization or image restoration purposes.

StirMark is reconfigurable but the level of reconfigurability is limited. Reconfiguring StirMark for a digital watermarking algorithm can be done by defining the input and output arguments according to one of the six pre-defined types of algorithms, but adding new parameters and extending existing parameter settings will require changing the source code of the StirMark implementation (in C++). For example, the \emph{strength} parameter in StirMark is set to be a single floating-point number with many hard constraints (\emph{e.g.} minimum and maximum values are linked to specific PSNR values), but for digital watermarking algorithms the \emph{strength} could be a more complicated parameter such as a vector comprised of two or more numbers controlling different parts of the watermark embedding process such as the size of single watermark and the number of duplicate watermarks embedded.

Although StirMark allows adding new tests, attacks and PQA metrics, the unclear boundaries among components make it hard to do so without making changes to the source code of the StirMark implementation. Adding some new test, attack and quality metric may require a re-design of the framework, \emph{e.g.} if a non-PSNR PQA metric is introduced the \emph{strength} parameter will need re-defining and many existing components need adapting to the new PQA metric.

The StirMark framework defined in \cite{PetitcolasSRDFF2001}\iffulledition\ and shown in Fig.~\ref{fig:stirmark_frame}\fi\ does not follow a clear data flow, \emph{e.g.} the test library does not really flow into the evaluation profile but read data from the latter and the multimedia database.

In \cite{PetitcolasSRDFF2001} StirMark is described to work with an SQL server to store all the evaluation results which can then be converted into web pages for reporting. However, the SQL-based web service has not been actually implemented. Instead, the latest C++ implementation of StirMark \cite{stirmark_v4} produces a plain data sheet to store the evaluation results which cannot be easily converted into other format or used to do further analysis.

\subsection{Other Benchmarking Systems}

Checkmark was developed by Pereira \emph{et al.} \cite{PereiraVMMPP2001} and downloadable from \url{http://cvml.unige.ch/ResearchProjects/Watermarking/Checkmark/} (now discontinued). Checkmark was based on StirMark with the following main changes. First of all, a number of new attacks, which take into account statistical properties of images and watermarks, are incorporated into Checkmark. The detailed descriptions of most attacks are provided in \cite{VoloshynovskiyPIP2001, VoloshynovskiyPPES2001, PereiraVMMPP2001}. Secondly, weighted PSNR and Watson's metric are used as new metrics for evaluating image quality instead of just PSNR. Thirdly, evaluation results are represented in a flexible XML format and can be automatically converted into HTML web pages. To use Checkmark, users need to supply some original images and their watermarked editions, and then customize two initial functions (one is used to inform Checkmark about the input images, and the other is used to define the watermark detector which should return a binary output indicating the result of the watermark detection process). Despite the changes to StirMark, the reconfigurability of Checkmark remains relatively low so normally users have to make changes to Checkmark's source code.

Optimark \cite{SolachidisTNTNP2001} is a benchmarking software package for image watermarking algorithms downloadable from \cite{optimarkonline}, providing a graphical user interface (GUI) developed using C/C++. To use Optimark for benchmarking a digital watermarking algorithm, users can choose a set of test images, define different watermark embedding keys and watermark messages for multiple trials of the watermarking detector and decoder, and select a set of attacks among 14 types of attacks and attack combinations. It allows evaluation of several statistical characteristics of an image watermarking algorithm, including Receiver Operating Characteristic (ROC) curves\iffulledition\ with false positive and false negative rates\fi\ as watermark detection performance metrics. OptiMark also supports combining multiple ROC curves to measure the overall performance by allowing users to set weights of a number of selected input images and attacks.

Certimark is the outcome of an EU-funded research project (\url{http://www.certimark.org/}, lasting from 2000 to 2002). The objectives of Certimark are to design a benchmarking suite which permits users to assess the appropriateness and to set application scenarios for their needs, and to set up a standard certification process for watermarking technologies \cite{VorbruggenF2002}. The benchmark system is a suite of modules, including image source, watermark embedder/decoder, attack model, comparator model, process-dependent metrics, report writer and result \& certificate module, with the interfaces among different modules to guarantee the consistency along the benchmarking process. Although the reconfigurability level of Certimark is higher than earlier systems, Certimark seems to have been discontinued and there is no source code publicly available.

Watermark Evaluation Testbed (WET) \cite{KimOGD2004, KimLGD2005, GuitartKD2006a, GuitartKD2006b} is a web-based system developed by researchers from the Purdue University to evaluate the performance of image watermarking algorithms. WET consists of three major components: front end, algorithm modules, and image database. To achieve the goal of extensibility, the GNU Image Processing Program (GIMP) is used because it support plug-ins and extensions. Some watermarking algorithms, StirMark 4.0 and some evaluation metrics were implemented as GIMP plug-ins to be part of WET's algorithm modules. The end users can select some images, one or more watermarking algorithms, attacks, and specify needed parameters via a web interface of the front end. The evaluation results can be shown as ROC curves. Similar to other systems, WET has a limited reconfigurability. In addition, its source code is not publicly available.

OpenWatermark \cite{LuganMacq2004, Michiels-Macq:Openwatermark:EUSIPCO2006} is a web-based system for benchmarking of digital watermarking algorithms. It is composed of three parts: 1) a web server and a remote method invocation (RMI) client for users to submit their benchmarking requests with specifications of the benchmarked algorithms, 2) a cluster of RMI benchmark servers automating the benchmarking process, and 3) a SQL database sorting all data used in the benchmarking process and results produced by the benchmark servers. OpenWatermark also contains some reference attacks, evaluation metrics and test images as publicly available resources. OpenWatermark is able to support two typical use cases: watermark extraction test and watermark detection test. OpenWatermark allows benchmarked algorithms to be submitted as Windows/Linux executables or MATLAB/Python scripts and all its components were developed in Java, so it has some reconfigurability. However, to support more features such as benchmarking profiles and other media types its source code has to be modified. To some extent, OpenWatermark is more like an online service for end users to define a sequence of remote calls. OpenWatermark implementation was available to registered members at its website \url{http://www.openwatermark.org/} which is currently unaccessible.

Mesh Benchmark \cite{WangLDBH2010} was proposed for 3D mesh watermarking. It contains three different components: a data set, a software tool and two evaluation protocols. The maximum root mean square error (MRMS) and the mesh structural distortion measure (MSDM) are used as perceptual distortion metrics. The attacks currently included in Mesh benchmark are: file attack, geometry attacks (similarity transformation, noise addition, smoothing, vertex coordinates quantization) and connectivity attacks (simplification, subdivision, cropping). As a benchmarking system focusing on 3D mesh watermarking only, it considers only the payload, distortion and robustness for performance evaluation. Besides, the evaluation protocols are defined with fixed steps and thresholds so the reconfigurability of the mesh benchmark is low.

\section{Proposed OR-Benchmark Framework}
\label{sec:framework}

In this section, our proposed OR-Benchmark framework will be introduced in details. Firstly, we discuss general modelling of digital watermarking systems used in OR-Benchmark in Section~\ref{sec:model_watermark}. Then the evaluation criteria considered in OR-benchmark are discussed in Section~\ref{sec:evlt_creteria}. After that, the architecture of the OR-benchmark framework and its open interfaces for end users are explained in details in Sections~\ref{sec:benchmarking_framework} and \ref{sec:benchmarking_interface}, respectively. This section ends with an comparison between OR-benchmark and all benchmarking systems reviewed in Section~\ref{sec:benchmarking_compare}.

\subsection{Modelling of Watermarking Systems}
\label{sec:model_watermark}

Following the community's common understanding, OR-benchmark models a digital watermarking system as two separate processes: the \emph{Sender} which embeds one or more watermarks into a given cover work to generate a watermarked work; the \emph{Receiver} which extracts and/or detects one or more watermarks that may have been embedded in a received test work. Both the \emph{Sender} and the \emph{Receiver} take at least one input (the cover or test work) but may take more optional inputs (some are parameters), and produce one or more outputs.

\begin{figure}[!htb]
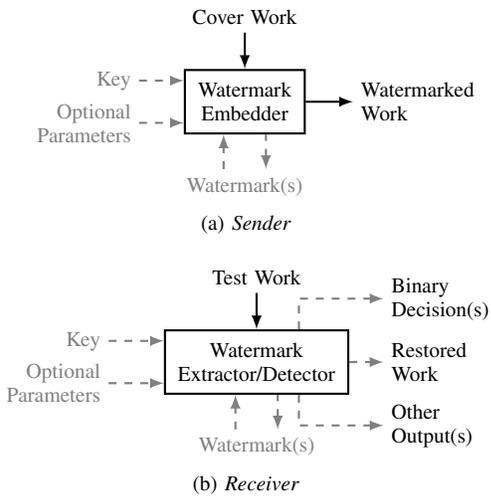

\centering\footnotesize
\subfloat[\emph{Sender}]{
\input figures/proc_embed_sys.tex
}
\\
\subfloat[\emph{Receiver}]{
\input figures/proc_authrec_sys.tex
}
\caption{Modelling of the \emph{Sender} and the \emph{Receiver} in OR-Benchmark. Dashed lines denote optional input/output.}
\label{fig:frame_sys_models}
\end{figure}

The general models of the watermark embedding and extraction/detection processes are shown in Fig.~\ref{fig:frame_sys_models}. As shown in Fig.~\ref{fig:frame_sys_models}(a), the \emph{Sender} will always have the cover work as the input and the watermarked work as the output. There are three groups of optional inputs including the watermark(s) to be embedded, the embedding key, and other optional parameters controlling the embedding process. Note that the watermark(s) in the embedding process can be either an input (supplied by the user) or an output (if generated by the \emph{Sender} automatically), which can be further used for performance evaluation purposes. As shown in Fig.~\ref{fig:frame_sys_models}(b), the \emph{Receiver} takes a minimum input (a test work) and possibly some other inputs and parameters to produce one or more outputs including one or more extracted watermarks, one or more binary decisions (if some given watermark(s) is/are detected), a restored work (if the watermarking algorithm supports self-restoration), and other outputs \emph{e.g.} the confidence level and error rates. We model the inputs and outputs of the \emph{Sender} and the \emph{Receiver} this way to cover different types of digital watermarking algorithms and application scenarios. \iffulledition For example, ``Binary Decision(s)'' as an output could be a single number (to show whether a single given watermark is correctly detected or if the content of the test work is authenticated), or a binary matrix to show the authentication results of individual regions of the test work.\fi

Another important part of performance evaluation of digital watermarking algorithms is the communication channel between the sender and the receiver which can be used to model any intermediate processing on a watermarked work such as channel noises or any other unwanted distortions, benign processing such as re-compression in some applications scenarios, and attacks whose goal is to fail the watermark extraction/detection process. In OR-Benchmark the communication channel is simulated as a black box called ``channel simulator'' taking a single input (a work) and producing a single output (a processed work), which can be used to cover everything that may happen between the sender and the receiver. We will discuss more about this in Sec.~\ref{sec:benchmarking_framework}.

\subsection{Performance Evaluation Criteria}
\label{sec:evlt_creteria}

In OR-benchmark performance evaluation criteria (i.e., indicators) are organized into two categories: 1) built-in performance indicators that can be selected by users directly; 2) user-defined performance indicators that are supported indirectly by generating a comprehensive set of raw results for users to further process. In this section, the commonly required performance indicators for benchmarking digital watermarking algorithms are further discussed.

Similar to StirMark, the properties designers and users of digital watermark algorithms wish to evaluate include imperceptibility, embedding capacity, robustness to benign processing and attacks, false alarm rates and the speed of execution. Since these common criteria have been discussed in Sec.~\ref{sec:stirmark}, here we focus on only two other evaluation properties for content authentication and self-restoration watermarking algorithms that are not (well) supported by StirMark but essential for some application scenarios.

\emph{Authentication Accuracy}: For content authentication watermarking, there are two basic metrics to measure the authentication accuracy of the detection process: the false-positive (FP) rate indicating the level of errors for areas reported as ``tampered'', and the false-negative (FN) rate indicating the level of errors for areas reported as ``untampered''. Many other performance metrics can be derived from the FP and FN rates \emph{e.g.} the average authentication rate\iffulledition\ used in \cite{ZhaoHo2007}\fi\ and the ROC curves\iffulledition\ widely used in the digital watermarking community and the machine learning community more broadly\fi. OR-Benchmark supports the two main metrics and also provide needed raw data in the benchmarking results to allow more complicated user-defined metrics that cannot be derived directly from the FP and FN rates.

\emph{Perceptual Quality of Recovered Work}: For self-restoration watermarking algorithms (which require the use of content authentication watermarks), a key performance indicator is the perceptual quality of the recovered work. The perceptual quality can be measured in the same way as how the perceptual quality of a watermarked work is measured. In OR-Benchmark, some commonly used image quality assessment (IQA) metrics such as PSNR and SSIM are incorporated but users can add their own metrics easily via the open interface discussed in Sec.~\ref{sec:benchmarking_interface}. For self-restoration watermarking algorithms, there is a question on if the perceptual quality should be calculated for the whole work or just the detected regions labelled as ``tampered''. If the latter option is used, the tempered regions falsely reported as ``untampered'' will be missed so the result will be misleading. Therefore, OR-Benchmark measures the quality using the whole work.

\subsection{Our Benchmarking Framework}
\label{sec:benchmarking_framework}

\begin{figure*}[!htb]
\centering
\iffulledition
\includegraphics[width=0.8\textwidth]{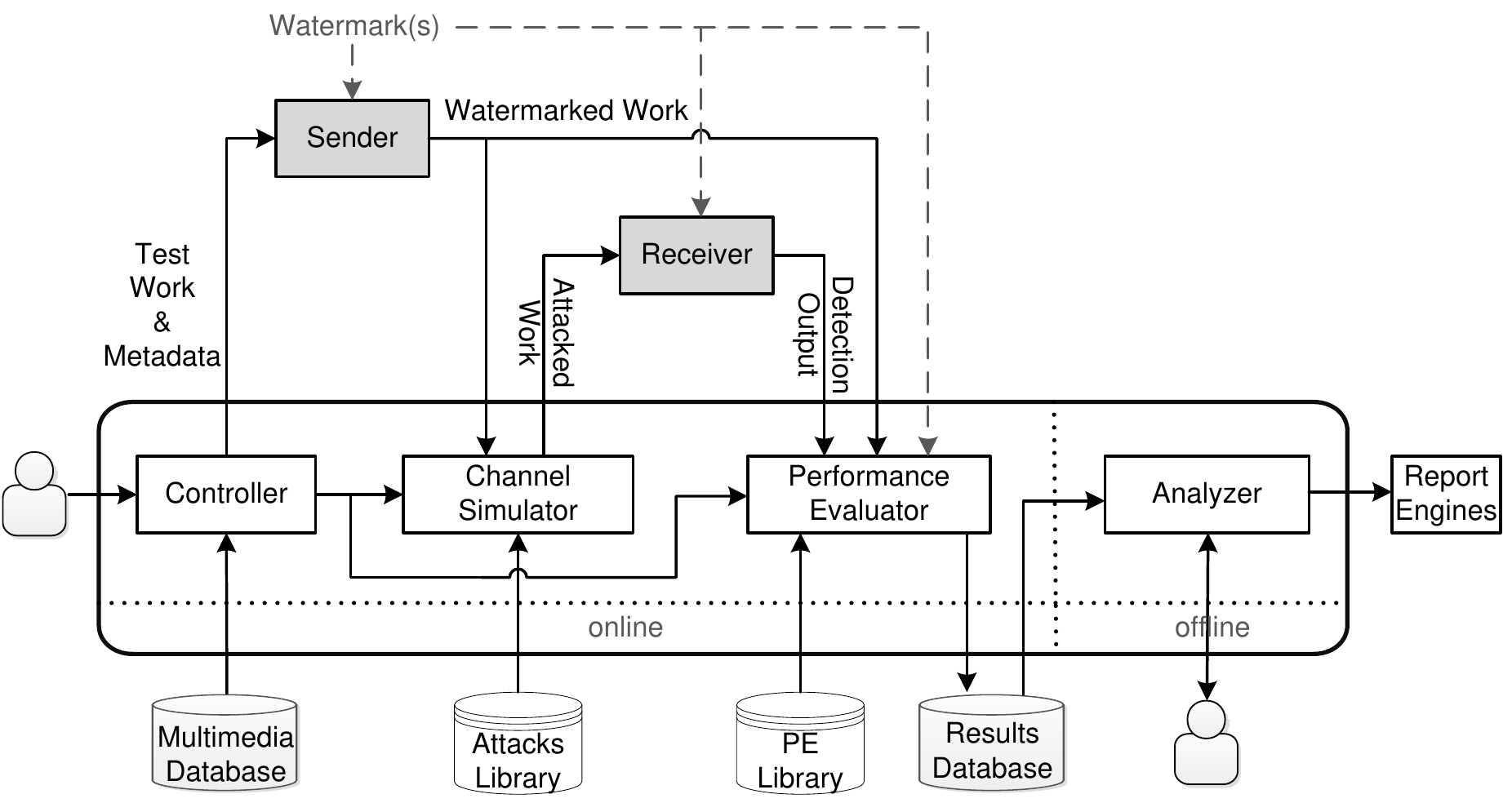}
\fi
\caption{The architecture of the OR-Benchmark framework.}
\label{fig:framework}
\end{figure*}

In this subsection, we introduce the overall architecture of OR-Benchmark in details. Figure~\ref{fig:framework} gives a schematic overview of the framework, which can be split into two parts: an \emph{Online Benchmarker} takes input from the user and automates the benchmarking process to generate results for further analysis, and an \emph{Offline Analyzer} allowing the user to conduct user-specific tasks (\emph{e.g.}\ statistics and visualization) based on the (raw) results produced by the \emph{Online Benchmarker}. The \emph{Offline Analyzer} can be equipped by one or more \emph{Report Engines} to produce more user-friendly reports of benchmarking tasks. The \emph{Report Engines} may also access the results from the \emph{Online Benchmarker} without passing the \emph{Offline Analyzer} (in that case the \emph{Offline Analyzer} can be seen as a simple data forwarder).

The \emph{Online Benchmarker} contains three groups of components: 1) the user-provided components -- the \emph{Sender} and the \emph{Receiver} provided by the user as the subject of benchmarking, 2) a \emph{Multimedia Database} holding the test media, an \emph{Attacks Library} and a \emph{PE library} providing attacks and performance evaluation algorithms, respectively, and 3) the core benchmarker part composed of a central \emph{Controller}, a \emph{Channel Simulator} enabling incorporation of different types of attacks and processing on a watermark work, and a \emph{Performance Evaluator} which produces results to sore in a \emph{Results Database} as the output of the whole benchmarking process. The central \emph{Controller} interacts with the user to define the benchmarking profile, and with other components of the online benchmarker to automatically execute the profile. A benchmarking profile allows automatic testing of multiple parameters of the same digital watermarking algorithm, multiple attacks, multiple PE algorithms and multiple performance indicators. The \emph{Controller} can also automatically determine default settings based on information given by the user to reduce the burden of the user to define the benchmarking profile.

The whole benchmarking process works as follows\iffulledition\ from an end user's point of view\fi. The user first interacts with the \emph{Controller} to define a benchmarking profile, for which (s)he provides own \emph{Sender} and \emph{Receiver} functions for benchmarking and defines what to benchmark. The user may also define the watermark(s) to be embedded if user-specific watermarks are required. It is possible to define how the watermarks are formatted so that the \emph{Controller} can generate them automatically. The user also needs to select test media from the \emph{Multimedia Database}, possibly by extending the database with own test works. Based on the benchmarking profile, the \emph{Controller} feeds selected test multimedia works and any meta-data to the \emph{Sender}, selected attacks to the \emph{Channel Simulator}, attacked works to the \emph{Receiver}, and then selected PE algorithms to the output of the \emph{Receiver} to produce data stored in the \emph{Results Database}.

\subsection{Open Interfaces}
\label{sec:benchmarking_interface}

OR-Benchmark is designed to have open interfaces so that users can easily (re)configure and extend the framework and define different benchmarking tasks easily. Observing Fig.~\ref{fig:framework}, there are mainly the following interfaces.

\emph{The interfaces between the Sender/Receiver and the core benchmarker} allow user-defined digital watermarking algorithms to be benchmarked. Following the general models of the \emph{Sender} and the \emph{Receiver} discussed in Sec.~\ref{sec:model_watermark}, the interfaces are materialized as the input and output interfaces of two functional units: \texttt{\emph{Sender}: (Original Cover Work, [Watermark(s)], [Key], [...])} $\to$ \texttt{(Watermarked Work, [Watermark(s)])}, \texttt{\emph{Receiver}: (Test Work, [Watermark(s)], [Key], [...])} $\to$ \texttt{([Watermark(s)], [Decision], [Restored Work], [...])}, where arguments in the square brackets are optional and ``...'' denotes more optional arguments. A proper mechanism is required to inform the \emph{Controller} about valid values each input argument can take and other meta information\iffulledition (\emph{e.g.} the display name of each argument)\fi, in order to create benchmarking profiles for enumerating all values for any input argument of interest. The implementation of the interfaces differ depending on the programming language used, \emph{e.g.} for object-oriented programming (OOP) languages they can be implemented as a class with methods representing the two functional units and member variables representing inputs, outputs and meta information\iffulledition, and for non-OOP languages two separate functions with optional parameters can be defined achieve the same goal\fi.

\emph{The interface between the Multimedia Database and the core benchmarker} allows users to reconfigure and extend the \emph{Multimedia Database}. This can be achieved by an agreed structure of the \emph{Multimedia Database} such as a hierarchy structure of folders and files or using a human-readable configuration file (such as XML) to allow the system and end users to find test multimedia works. Note that OR-Benchmark can support any media types so the \emph{Multimedia Database} can be a mixture of different types of media files\iffulledition\ including audio tracks, images, video sequences, 3D models and others\fi.

\emph{The interface between the Attacks Library and the core benchmarker} allows users to reconfigure and extend the \emph{Attacks Library} used by the \emph{Channel Simualtor}. As discussed in Sec.~\ref{sec:model_watermark}, an attack in the \emph{Attacks Library} is a simple functional unit as follows: \texttt{\emph{Attack}: (Input Work, [...])} $\to$ \texttt{(Output Work)}. Again, a mechanism is needed to convey meta information about any optional input arguments.

\emph{The interface between the PE Library and the core benchmarker} allows users to reconfigure and extend the \emph{PE Library} used by the \emph{Performance Evaluator}. There are different types of PE algorithms depending on the performance indicators used, so there are different input and output interfaces. An important class of PE algorithms are perceptual quality assessment (PQA) metrics defined as follows: \texttt{\emph{PQA}: (Work1, Work2, [...])} $\to$ \texttt{(Metric)}, where the output metric is a numeric rating of the perceptual quality. Again, a mechanism is needed to convey meta information about optional input arguments. PQA algorithms are generally objective ones based on automated computer programs, but it is possible to define a \emph{virtual} functional unit where human experts (\emph{e.g.}\ from crowdsourcing websites) are involved to rate the quality subjectively.

\emph{System search paths} can be set up for all the above interfaces so that the \emph{Controller} and other components of the core benchmarker can automatically discover candidate algorithms and test multimedia works. Each path can be a combination of local file paths and URLs including web addresses. \iffulledition When web addresses are involved, a local caching mechanism may be created to allow fast retrieval of contents from remote resources.\fi

\emph{The interface between the core benchmarker and the Results Database} allows users to reconfigure and extend the format of the results used by the \emph{Offline Analyzer} and \emph{Report Engines}. This can be achieved by a configuration file (\emph{e.g.} an XML file following a pre-defined schema) indicating the format of the results of a particular benchmarking profile.

\emph{The interface between the user and the Controller} allows creation of benchmarking profiles. Core elements of a benchmarking profile include digital watermarking algorithm(s) tested and candidate values of input parameters, test multimedia works, selected attacks, selected PE algorithms, and format of the results. This can be implemented as a graphical user interface (GUI) and/or a human-readable configurable file.

\emph{The interface between the user and the Offline Analyzer and Report Engines} allows users to investigate the raw results recorded in the \emph{Results Database} in an interactive way and to produce more user-friendly reports. The interface for the \emph{Offline Analyzer} can be implemented as a GUI, but the \emph{Report Engines} could be just command-line tools invoked from the \emph{Offline Analyzer}'s GUI. The format of the produced reports can be defined using a human-readable configurable file.

\subsection{Implementation}

We implemented a prototype of OR-Benchmark as a MATLAB software package which includes all key components shown in Fig.~\ref{fig:framework} and the interfaces listed in Sec.~\ref{sec:benchmarking_interface}. The prototype is built on MATLAB standard functions and toolboxes, and does not depend on any third-party libraries. MATLAB is selected considering its wide use in the digital watermarking community and the ease to dynamically extend the implemented system without compiling the whole source code. The cross-platform nature of MATLAB also makes the OR-Benchmark prototype more accessible to researchers using different operating systems. Although the prototype is fully functional (see a case study in Sec.~\ref{sec:experimental_results}), we are still refining it and plan to release a beta edition under an open source license once this paper is accepted for publication.

The MATLAB prototype by default uses a number of pre-defined folders to store files and data in the \emph{Multimedia Database}, a library of differen digital watermarking algorithms (each including a \emph{Sender} and a \emph{Receiver} functions), the \emph{Attacks Library}, the \emph{PE Library}, and the \emph{Results Database}. The user can freely add new functions following the interfaces discussed in Sec.~\ref{sec:benchmarking_interface} to the corresponding folders to extend the system. The prototype can also be configured to use a search path including multiple folders for each of the above listed components. There is another folder keeping MATLAB scripts implementing the \emph{Controller}, the \emph{Channel Simulator}, the \emph{Performance Evaluator}, and the \emph{Offline Analyzer}. The \emph{Controller} has both a GUI for creating the benchmarking profiles\iffulledition\ (see Fig.~\ref{fig:snapshot_GUI1})\fi\ and a benchmarking scheduler for automatically executing benchmarking profiles. Given the flexible interfaces of and the meta information about digital watermarking algorithms, the \emph{Controller} allows the user to define test multimedia works, candidate values of input arguments in the \emph{Sender} and \emph{Receiver} functions, selected attacks and PE algorithms, in order to schedule and launch a number of repeated runs of the digital watermarking process to produce all raw data and performance indicators recorded in the \emph{Results Database}. Each benchmarking profile created by the \emph{Controller} is stored as a MATLAB structure variable in the workspace, and once the benchmarking task is completed the benchmarking profile and the benchmarking results are saved as another MATLAB variable in a MATLAB data file in a designated folder of the \emph{Results Database}. Here, the benchmarking profile is kept to inform the \emph{Offline Analyzer} about the format of the results.

\iffulledition
\begin{figure}[!htb]
\centering
\includegraphics[width=\figwidth]{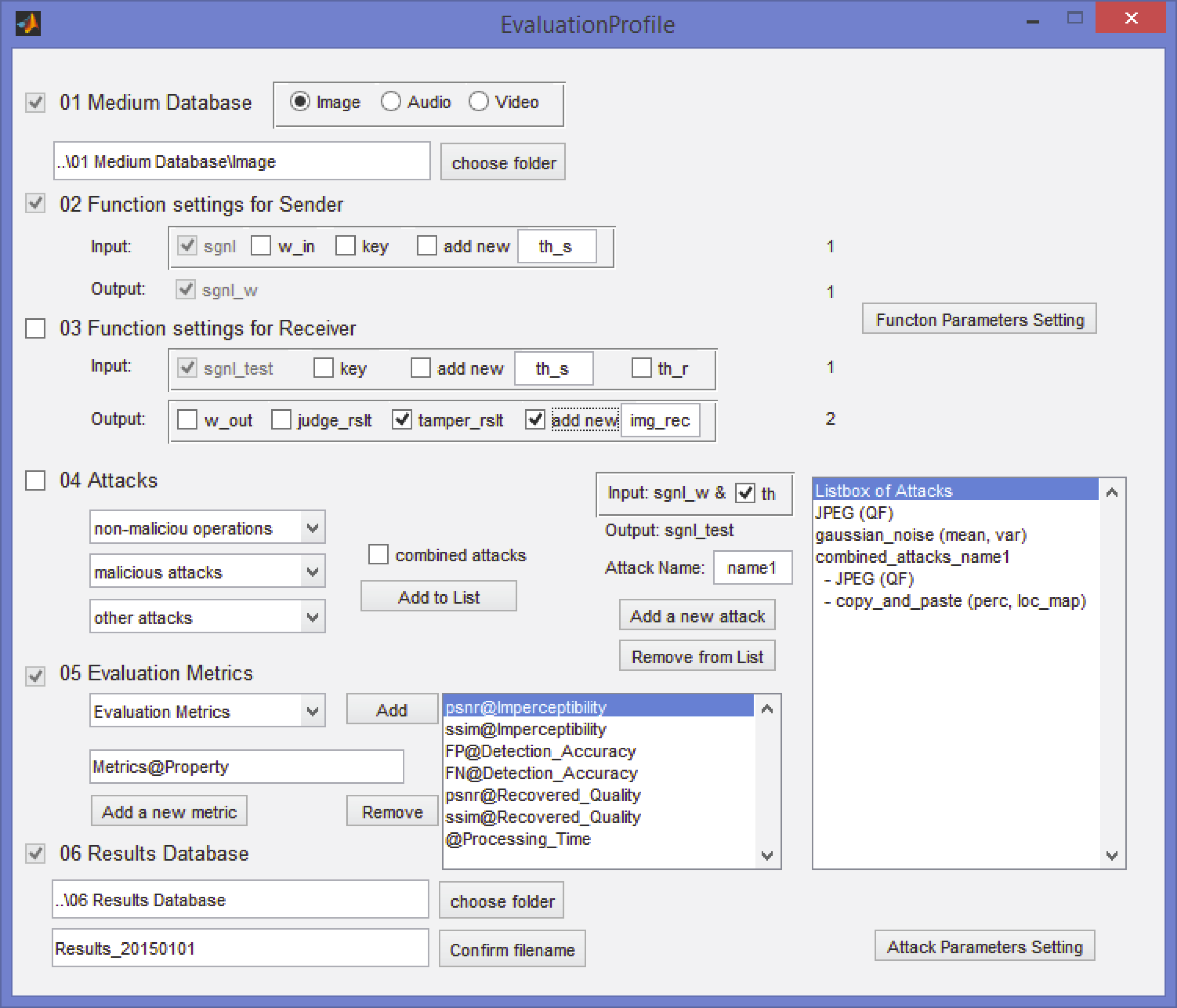}
\caption{The main GUI of the central \emph{Controller} for defining benchmarking profiles as currently implemented in OR-Benchmark.}
\label{fig:snapshot_GUI1}
\end{figure}
\fi

The \emph{Offline Analyzer} has a GUI for producing different kinds of 2-D plots based on raw data in the \emph{Results Database}\iffulledition\ (see Fig.~\ref{fig:snapshot_GUI2})\fi. At the current stage of development, the \emph{Offline Analyzer} is designed to showcase what one can do with the OR-benchmark prototype (see a case study in Sec.~\ref{sec:experimental_results}), so it is not a complete solution for all digital watermarking schemes yet. We plan to design a plug-in interface to allow different analysis and plotting functions to be incorporated into the \emph{Offline Analyzer}. It deserves noting that the user can develop his/her own \emph{Offline Analyzer} easily since the \emph{Results Database} contains all needed data in a structured and directly accessible way.

\iffulledition
\begin{figure}[!htb]
\centering
\includegraphics[width=\figwidth]{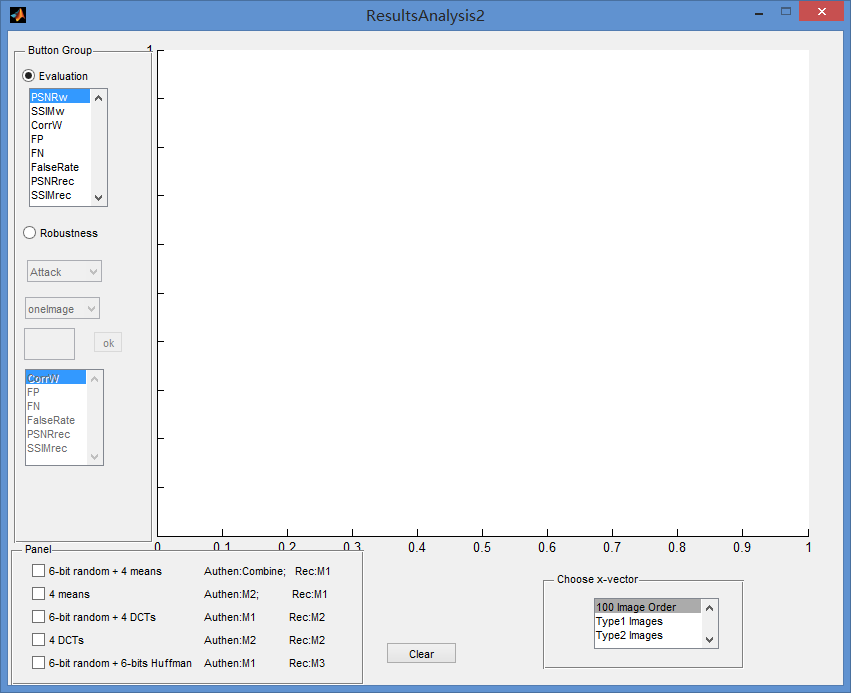}
\caption{The main GUI of the \emph{Offline Analyzer} as currently implemented in OR-Benchmark.}
\label{fig:snapshot_GUI2}
\end{figure}
\fi

\subsection{Comparison with Other Benchmarking Systems}
\label{sec:benchmarking_compare}

Comparing with other digital watermarking benchmarking systems and frameworks, OR-Benchmark has the most generic modelling of digital watermarking systems which allows it to support all types of digital watermarking algorithms at the level of system modelling. While most other benchmarking systems can be extended to cover more types of digital watermarking algorithms, often significant changes to the source code of their implementations are required. Some other benchmarking systems model digital watermarking algorithms in a way such that it is hard to cover multiple watermarks (especially of different types) in the same cover work. This advantage of OR-Benchmark can be seen from the case study we will discuss in Sec.~\ref{sec:experimental_results}, which is about benchmarking three image authentication and self-restoration digital watermarking algorithms involving two different types of watermarks for a single cover (one type for image authentication and the other for self-restoration). Such digital watermarking algorithms are among the most complicated ones and to our best knowledge no any other benchmarking system/framework can properly cover them in their current system models and implementations. This was actually one of the main reasons why we decided to develop our own framework.

Different from many other digital watermarking benchmarking systems, OR-Benchmark is designed to have openness and reconfigurability by design. The framework separates users, data, algorithms, the online benchmarker and the offline analyzer to achieve a more user-friendly data flow in the whole benchmarking process. Comparing \iffulledition Figs.~\ref{fig:framework} with \ref{fig:stirmark_frame}\fi, we can see OR-Benchmark has a clearer separation of different components and a clearer data-flow path from the \emph{Sender} to the \emph{Performance Evaluator}. OR-Benchmark also has more clearly-defined interfaces to support different user-specific operations including creating benchmarking profiles, (re)configuring and extending the benchmarking system. Most other benchmarking systems also allow limited (re)configuration often via definition of a user-specific evaluation profile\iffulledition\ (\emph{e.g.} StirMark using an INI file)\fi, but adding new functional units will normally require updating the source code of their implementation\iffulledition\ (\emph{e.g.} StirMark as a C++ based system changes to key header files cannot be avoided)\fi. As a comparison, OR-Benchmark has open interfaces to allow reconfiguration and extension, and our MATLAB implementation allows new functional units to be added without touching any other parts of the core system\iffulledition\ (not even any configuration file since available functional units can be automatically discovered in the search paths of different components following the defined open interfaces)\fi.

Another unique feature of OR-Benchmark is its support of \emph{all} media types with a \emph{single} model and process. In OR-Benchmark, digital watermarking of any media type can be handled in exactly the same way and the user does exactly the same steps to benchmark digital watermarking algorithm(s). Many functional units can be shared across different media types \emph{e.g.} many PE algorithms can be applied to multiple media types. On the other hand, most other benchmarking systems focus on one or two particular media types (mostly digital images and some extended to cover audio) and the implementations are very much tuned to support the one or two media types. This is another reason why extensibility of other benchmarking systems is lower than OR-Benchmark.

Our selection of using MATLAB to implement the OR-Benchmark prototype also contributes to the reconfigurability of OR-Benchmark. Most other benchmarking systems were developed based on compiled programming languages especially C/++, which makes incorporation of source code written in other programming languages harder or impossible. MATLAB has built-in support for functional units written in most mainstream programming languages such as C/C++, Java, and Python, thus making the extension much more easier.

\section{Case Study}
\label{sec:experimental_results}

In this section, we demonstrate how our implemented OR-Benchmark prototype can be used via a case study on benchmarking three image watermarking algorithms for content authentication and self-restoration. Such algorithms are among the most complicated ones with two types of watermarks per block of the cover work and are not supported by other benchmarking systems. While this section is mainly a case study for showcasing usefulness of OR-Benchmark, the watermarking algorithms benchmarked have never been compared in such a depth like we report here (which was a harder task due to the lack of proper benchmarking tools).

\subsection{Experimental Setup and User Operations}

The three image authentication and self-restoration digital watermarking algorithms benchmarked are the following: Lin and Chang's scheme \cite{LinChang2000} (M1), Li \emph{et al.}'s scheme \cite{LiPeiChen2009} (M2) and Wang \emph{et al.}'s scheme \cite{WangHoZhao2011} (M3), which all use two watermarks separately for image authentication and restoration of each $8\times 8$ block. For each scheme, a \emph{Sender} and a \emph{Receiver} MATLAB functions were written following the interfaces for the two components and then copied to the folder holding all such functions. Those functions were then selected as the target of the benchmarking task via the \emph{Controller}'s GUI. The GUI allows use of multiple candidate values of each parameter of each scheme, but for this case study we tuned the three schemes' parameters so that the average perceptual quality of the watermarked images is roughly aligned to make the comparison fairer (see below for more details).

For attacks, we chose simple ``copy and paste attack'', JPEG compression, additive and multiplicative Gaussian white noises as four separate attacking algorithms each of which is injected into the \emph{Channel Simulator} to create attacked watermarked images sent to the \emph{Receiver}. All the attacks were implemented as separate MATLAB functions with additional input parameters. Those functions were added to the folder holding the \emph{Attacks Library} and then selected (with different values of input parameters) via the \emph{Controller}'s GUI. For the ``copy and paste attack'' 10\% randomly-selected region of the whole image was copied and pasted to other regions of the same image. For JPEG compression, the QF (quality factor) is the only input parameter with values 100, 95, 90, ..., 50. For additive Gaussian white noise, the mean (with the only value 0) and the variance (with the values 1, 3, 5, ..., 39 using 255 as the peak pixel value) are used as input parameters. For multiplicative Gaussian white noise, the same input parameters (the mean and the variance) and the variance's values are different (1, 10, 20, ..., 240). The ``copy and paste attack'' was used as an always-on attack and optionally combined with one of other attacks for benchmarking robustness against attacks.

For performance indicators, we considered imperceptibility (\emph{i.e.}, perceptual quality of watermarked images), authentication accuracy (in terms of FP and FN rates), perceptual quality of recovered images (with and without attacks), and processing times of the \emph{Sender} and the \emph{Receiver} functions. For perceptual quality we chose PSNR and SSIM, which are the two most widely-used IQA metrics. Each performance indicator is represented by one MATLAB function which was added to the folder holding the \emph{PE Library}. The selection of the PE algorithms were also done via the \emph{Controller}'s GUI.

For the test images, we collected 100 8-bit gray-scale images of size $256\times256$, $384\times256$ and $512\times512$, which were added to a sub-folder of the folder holding the \emph{Multimedia Database}. The images cover a broad range of image types \emph{e.g.} outdoor or indoor scenes images, portraits, photos of natural or man-made objects, and texture images. The test images were selected by setting the test multimedia works to be all files from the corresponding sub-folder via the \emph{Controller}'s GUI.

All the above choices allowed the \emph{Controller} to create a benchmarking profile. For the format of the results, we depended on the \emph{Controller} to automatically create a default format to capture all raw data and performance indicators using a MATLAB variable including the benchmarking profile itself. After the benchmarking profile was set up, we instructed the \emph{Controller} to automatically run the benchmarking task to generate the results. The machine running the benchmarking task is a PC with an Intel Core 2 Duo CPU (3.16GHz) and 2GB RAM. The concurrency support of the dual-core CPU was not enabled to get a better estimate of the processing times. The MATLAB version used is MATLAB R2012a.

After the results were produced by the core benchmarker, the \emph{Offline Analyzer} was used to generate some 2-D plots for a better understanding of the performance of the three benchmarked image watermarking schemes. Considering all the results we observed, it is clear that M3 has the best performance, followed by M1 and then M2. In the following, we show some selected benchmarking results we obtained.

\subsection{Imperceptibility}
\label{sec:results_imperceptibility}

\begin{figure}[!htb]
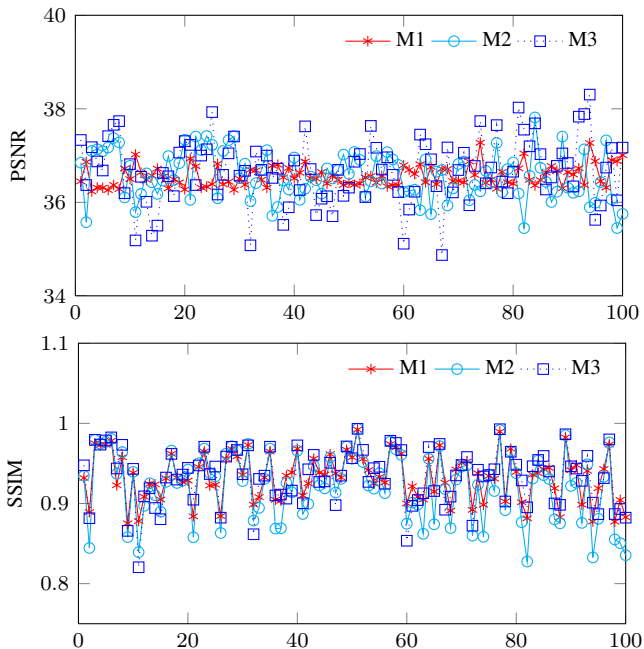

\centering
\input figures/systm_psnr_w.tex
\\
\input figures/systm_ssim_w.tex
\caption{The quality comparison of watermarked images produced by the three different watermarking schemes. The x-axis is the image index and the y-axis is the PSNR/SSIM value.}
\label{fig:sys_imperceptibility}
\end{figure}

Figure~\ref{fig:sys_imperceptibility} shows the PSNR and SSIM values of all the 100 test images after going through each of the three digital watermark embedding processes. Although we tried to align the perceptual quality to make the comparison fairer, there are noticeable fluctuations cross different test images due to the complexity of visual quality assessment. We managed to make the average PSNR values of the three digital watermarking schemes all between 36.5 and 36.7 dB (36.56~dB, 36.57~dB, 36.68~dB, respectively). One interesting observation is that M3's PSNR values are more fluctuated than M1 and M2, while their SSIM values have a similar level of fluctuation. Since SSIM is an IQA metric matching subjective quality better \cite{WangBSS2004}, we thus consider the three digital watermarking schemes are aligned well. Note that the alignment process of the imperceptibility actually involved running the three digital watermarking schemes through all the test images using different parameters and then calculating the average IQA value for each parameter setting of each scheme. This process itself is actually a set of simple benchmarking profiles with only two performance indicators (average PSNR and SSIM values of all watermarked images).

\subsection{Authentication Accuracy}
\label{sec:auth_rec_compr}

To evaluate authentication accuracy of an image authentication watermarking scheme, attacks manipulating contents of watermarked images should be considered. To this end, we applied the 10\% ``copy and paste attack'' to each watermarked image and calculated the FP and FN rates by counting wrongly reported $8\times8$ blocks by the \emph{Receiver}. Other attacks are not considered here so that we focus on the base line FP/FN rates.

For the FP rate, M1 and M3 have an almost zero rate for all images, and M2 has an average FP rate of 1.36\%. For the FN rate, M1 has an almost zero rate, M3 has an average FN rate of 1.59\%, and M2 has the worst rate of 3.02\%. \iffulledition See Fig.~\ref{fig:systm_fp_fn} for the FP and FN rates of all the 100 test images for the three digital watermarking schemes.\fi

\iffulledition
\begin{figure}[!htb]
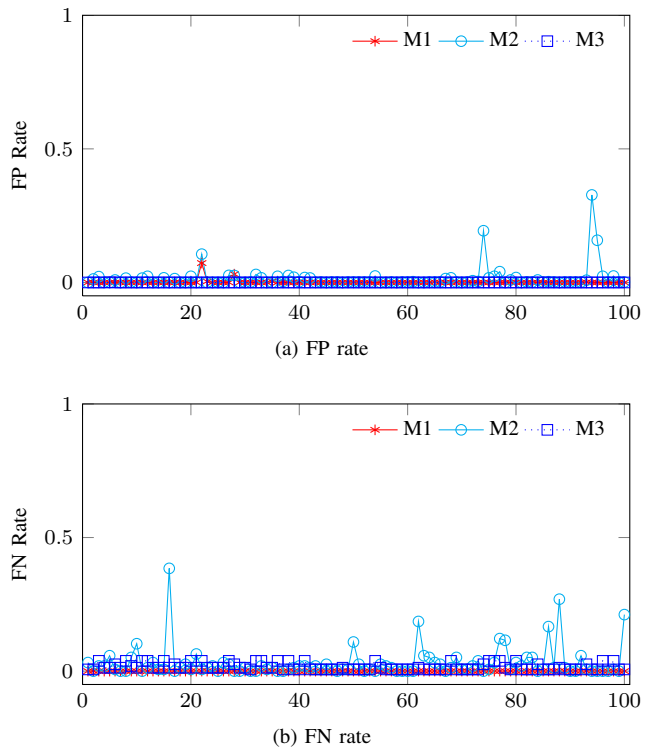

\centering
\subfloat[FP rate]{
\input figures/systm_fp.tex
}\\
\subfloat[FN rate]{
\input figures/systm_fn.tex
}
\caption{The FP and FN ratess of the three different watermarking systems. The x-axis is the image index and the y-axis is the FP/FN rate.}
\label{fig:systm_fp_fn}
\end{figure}
\fi

\subsection{Recovered Image Quality}

Similar to the case of authentication accuracy, for perceptual quality of recovered images we also focused on the condition where the 10\% ``copy and paste attack'' is applied without other attacks. The mean PSNR values of 100 recovered images for M1, M2 and M3 are 27.80, 27.92 and 32.32~dB, respectively. The mean SSIM values of 100 recovered images for M1, M2 and M3 are 0.9249, 0.9270 and 0.9506, respectively. The results showed that M3 is the best scheme with a significantly better capability of recovering manipulated images.

\subsection{Processing Time}

Except the embedding process of M1 which took around 2.6 seconds in average, all other processes of the three digital watermarking schemes consumed less than 1 second. Considering MATLAB is much less effective than other compiled programming languages, the results suggest that all the three schemes are practical for real-world applications.

\subsection{Robustness}
\label{sec:robust_compr}

\newlength\figstarwidth
\setlength\figstarwidth{0.3\textwidth}

\iffulledition
\pgfplotsset{
width=\figstarwidth,
height=0.9\figstarwidth
}
\fi

\begin{figure*}[!htb]
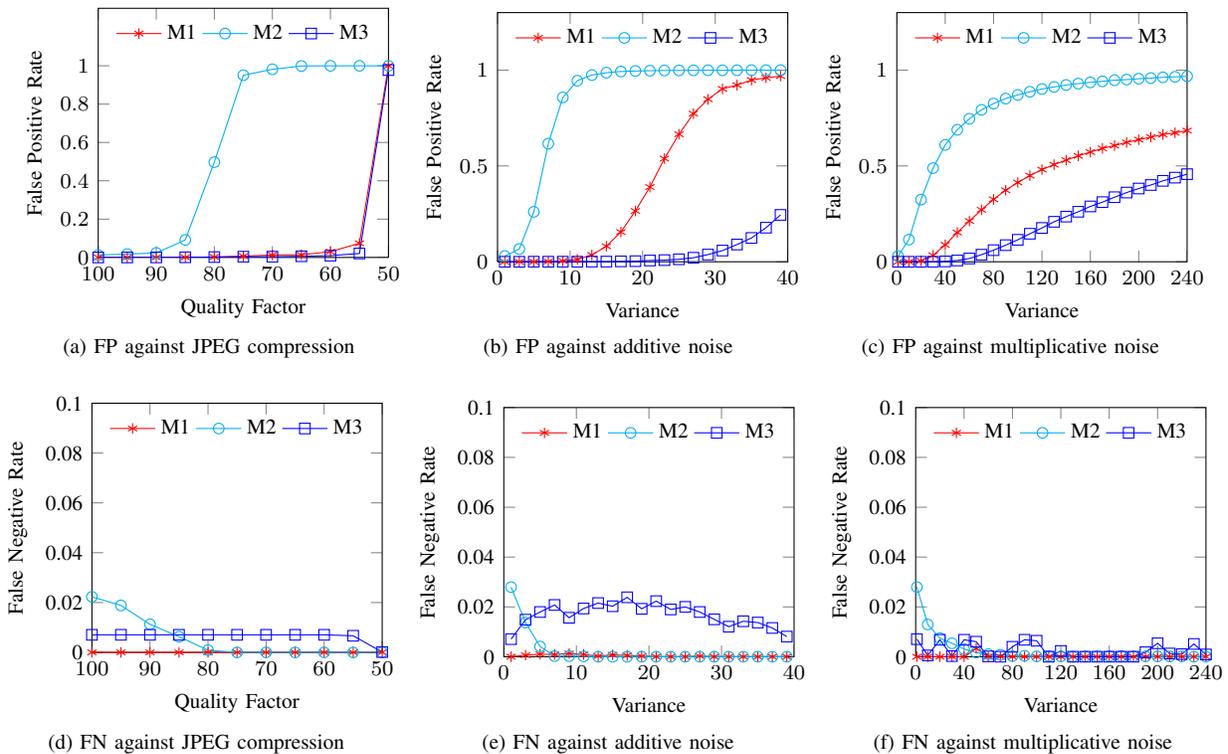

\centering
\subfloat[FP against JPEG compression]{
\input figures/systm_fp_jpeg.tex
\label{fig:sys_jpeg_authrate_fp}}
\subfloat[FP against additive noise]{
\input figures/systm_fp_gaussian.tex
\label{fig:sys_gauss_authrate_fp}}
\subfloat[FP against multiplicative noise]{
\input figures/systm_fp_noise.tex
\label{fig:sys_noise_authrate_fp}}\\
\subfloat[FN against JPEG compression]{
\input figures/systm_fn_jpeg.tex
\label{fig:sys_jpeg_authrate_fn}}
\subfloat[FN against additive noise]{
\input figures/systm_fn_gaussian.tex
\label{fig:sys_gauss_authrate_fn}}
\subfloat[FN against multiplicative noise]{
\input figures/systm_fn_noise.tex
\label{fig:sys_noise_authrate_fn}}
\caption{Average FP and FN rates of M1, M2 and M3 w.r.t.\ different parameter values of attacks.}
\label{fig:sys_authrate}
\end{figure*}

\begin{figure*}[!htb]
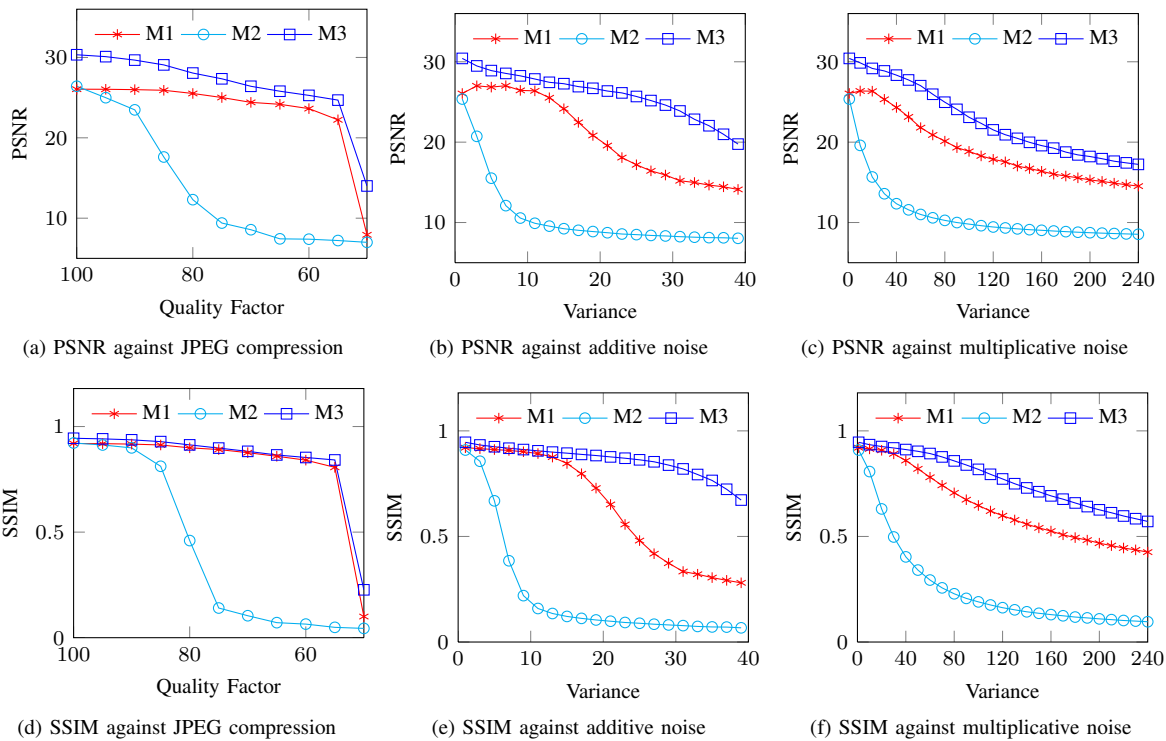

\centering
\subfloat[PSNR against JPEG compression]{
\input figures/systm_psnr_jpeg.tex
\label{sys_psnr_jpeg}}
\subfloat[PSNR against additive noise]{
\input figures/systm_psnr_gaussian.tex
\label{sys_psnr_gauss}}
\subfloat[PSNR against multiplicative noise]{
\input figures/systm_psnr_noise.tex
\label{sys_psnr_noise}}\\
\subfloat[SSIM against JPEG compression]{
\input figures/systm_ssim_jpeg.tex
\label{sys_ssim_jpeg}}
\subfloat[SSIM against additive noise]{
\input figures/systm_ssim_gaussian.tex
\label{sys_ssim_gauss}}
\subfloat[SSIM against multiplicative noise]{
\input figures/systm_ssim_noise.tex
\label{sys_ssim_noise}}
\caption{Average perceptual quality of images recovered by M1, M2 and M3 w.r.t.\ different parameter values of attacks.}
\label{fig:sys_rec}
\end{figure*}

For benchmarking robustness, we combined the 10\% ``copy and paste attack'' with one additional attack (JPEG compression, additive and multiplicative Gaussian white noises) to gauge the robustness of each digital watermarking scheme against each additional attack. \iffulledition Note that each additional attack does not change the contents of watermarked images but tries to fail the authentication process.\fi For each combination, the same performance indicators on authentication accuracy (FP and FN rates) and quality of recovered image were calculated against the parameter of each additional attack (QF for JPEG compression, variance for additive and multiplicative noises). Since now we have more factors to look at, we average the performance indicators cross all 100 images to get the average values which are then shown against the parameter value of each additional attack to see how the strength of the attack influences the performance of each digital watermarking scheme. The results are shown in Figs.~\ref{fig:sys_authrate} and \ref{fig:sys_rec}, and more discussions are given below.

\subsubsection{JPEG Compression}

Figures~\ref{fig:sys_authrate}\subref{fig:sys_jpeg_authrate_fp} and \subref{fig:sys_jpeg_authrate_fn} show average FP and FN rates after JPEG compression is applied to the three digital watermarking schemes. From the results, we can observe that M1 and M3 are very robust to JPEG compression with low FP and FN rates (the FP rate $<10\%$ and the FN rate $\approx0\%$) when QF$>55$. However, when a JPEG compression process with a QF value of 50 is applied, the authentication watermarks in the watermarked images are nearly completely destroyed with a FP rate close to 100\%. We can also observe that, compared with M1 and M3, M2 is less robust against JPEG compression, especially when QF$<85$.

Figures~\ref{fig:sys_rec}\subref{sys_psnr_jpeg} and \subref{sys_ssim_jpeg} show average PSNR and SSIM values corresponding to the three digital watermarking schemes. The results show that M1 and M3 can provide reasonably good image quality if QF$>50$, but M2's performance drops rapidly when QF$<85$. The general trend matches the results on the authentication accuracy since any false detections will influence the quality of the recovered image. Between M1 and M3, we can also observe that M3 performs slightly better in terms of SSIM but significantly better in terms of PSNR.

\subsubsection{Additive \& Multiplicative Gaussian White Noises}

Figures~\ref{fig:sys_authrate}, \subref{fig:sys_gauss_authrate_fp}, \subref{fig:sys_gauss_authrate_fn}, \subref{fig:sys_noise_authrate_fp} and \subref{fig:sys_noise_authrate_fn} show average FP and FN rates when noises are added for the three digital watermarking schemes.

The results on the FP rates show that M3 outperforms M1 significantly and M2 is the worst among the three. The average FN rates of all the three schemes remain close to 0\% so there is no noticeable difference among them.

Figures~\ref{fig:sys_rec}\subref{sys_psnr_gauss}, \subref{sys_ssim_gauss}, \subref{sys_psnr_noise} and \subref{sys_ssim_noise} show average PSNR and SSIM values of the 100 images recovered by the three digital watermarking schemes after different levels of noise are added. As expected the average quality of recovered images largely decreases smoothly as the variance (energy) of noise increases. Among the three schemes, M2 is again the worst performing scheme and M3 outperforms M1 significantly in terms of both PSNR and SSIM (for the latter after the variance of the noise goes beyond a threshold).

\section{More Discussion}
\label{sec:discussion}

The previous section gives evidence about the usefulness of OR-Benchmark. While it is clear that OR-Benchmark has the potential to be a useful framework for the digital watermarking community, we would like to highlight that benchmarking complicated systems like digital watermarking schemes is not a simple matter and a more careful design of the benchmarking task is needed. In other words, the benchmarking task has to be designed on an ad hoc basis by the user of the benchmarking system, which is supported by the high reconfigurability and extensibility of the OR-Benchmark framework. While benchmarking tasks have to be designed individually, there are known common issues that we need to pay special attention to. For instance, it has been well known that using different PQA metrics may lead to different results when comparing different digital watermarking schemes. This has been demonstrated partly from the results shown in Sec.~\ref{sec:experimental_results} where PSNR and SSIM do not always give the same results (\emph{e.g.} for M3).

To further highlight the subtlety of performance evaluation of digital watermarking schemes, in this section we show a concrete example related to the ``visual quality of recovered image \emph{vs.} JPEG compression attack'' issue, which is about the use of QF as the control factor of JPEG compression to compare performance of digital watermarking schemes as shown in Sec.~\ref{sec:robust_compr}. While this is a common practice to use QF as the control factor, it can be reasonably argued that QF is not necessarily a good factor for this purpose because it has different impacts on different images. One alternative is bit per pixel (bpp), which is a more direct measure of compression efficiency than QF. Now we will show what will happen if we switch from QF to bpp for the same benchmarking task described in Sec.~\ref{sec:experimental_results}. To simplify the discussion, we focus on the visual quality of recovered images only.

Switching from QF to bpp immediately raises a problem: we can control QF directly to have a fixed set of values for all digital watermarking schemes, but we cannot control bpp directly as it is not an encoding parameter but a post-compression metric. The fixed set of values is important because we need to calculate average performance indicators which can be done if all performance indicators are aligned. When the values of performance indicators are not aligned, we will need to find a way to average the results cross all test images. One approach is to fit a curve for each image covering a continuous range of the control factor and then to average all those curves produced for all test images. Let us show how this can be done using PSNR as an example. The task here is to get a continuous bpp-PSNR function for each test image based on a finite number of (bpp, PSNR) points, and then average the bpp-PSNR functions of all images to get an average bpp-PSNR function. To this end, denote the bpp-PSNR function for the $i$-th test image by $f_i(\cdot)$. Since we have no knowledge of each individual function $f_i(\cdot)$, we simply connect all the (bpp, PSNR) points to form a piecewise linear function. We also limit the domain of $f_i(\cdot)$ to $[\min_i(\text{bpp}),\max_i(\text{bpp})]$, the range between the minimum and maximum bpp values observed.

\begin{figure}[!htb]
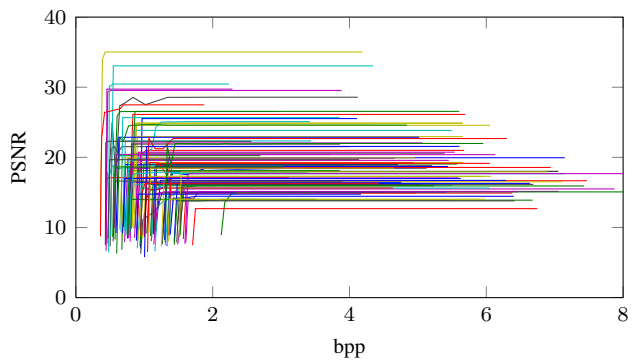

\centering
\input figures/bpp.tex
\caption{The bpp-PSNR functions of 100 images recovered by the digital watermarking scheme M1.}
\label{fig:bpp100}
\end{figure}

Figure~\ref{fig:bpp100} shows 100 bpp-PSNR functions estimated from 100 images for the digital watermarking scheme M1. As expected, those functions do not have aligned domains since the minimum and maximum bpp values vary from image to image. In order to align all the functions, we extend all their domains to $(-\infty,\infty)$ and assign $f_i(x)=0$ when the bpp value $x$ goes out of $[\min_i(\text{bpp}),\max_i(\text{bpp})]$.

\pgfplotsset{
width=\figwidth,
height=\figheight
}

After making the above preparation, the average bpp-PSNR function $\bar{f}(x)$ for all $N$ test images can be defined as follows:
\begin{equation}
\bar{f}(x)=\frac{\sum_{i=1}^{N}f_{i}(x)}{\sum_{i=1}^{N}\text{sign}(f_{i}(x))},
\label{eq:mean_bpp_psnr}
\end{equation}
where $\text{sign}(x)=0$ for $x\leq 0$ and 1 otherwise, which is used to count only bpp-PSNR functions covering $x$.

\begin{figure}[!htb]
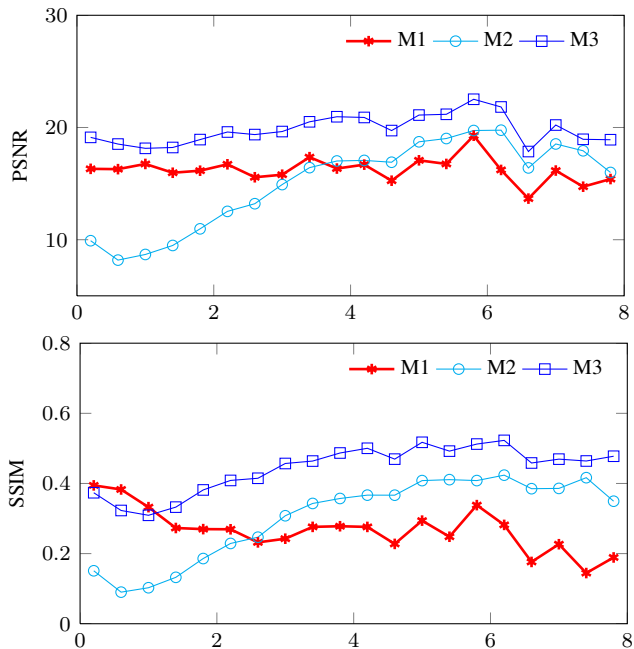

\centering
\input figures/bpp_vs_psnr_sys.tex
\\
\input figures/bpp_vs_ssim_sys.tex
\caption{Reproduction of the results in Figs.~\ref{fig:sys_rec}(a) and (d) by replacing the control factor QF by bpp. The x-axis is the bpp value and the y-axis is the PSNR/SSIM value.}
\label{fig:bpp}
\end{figure}

\modified{Based on the above approach to calculating the average bpp-PSNR function, we added a new PE algorithm}{The above approach can be easily generalized to any IQA metrics. We added two new PE algorithms} to the \emph{PE Library} and produced the performance comparison results for the average recovered image quality w.r.t.\ JPEG compression as shown in Figure~\ref{fig:bpp}. Compared with the results shown in Figures~\ref{fig:sys_rec}\subref{sys_psnr_jpeg} and \subref{sys_ssim_jpeg}, we can see some clear differences in the conclusion of the performance comparison: while M3 remains the best scheme as a whole, M2 now outperforms M1 when the bpp value goes above a threshold. This example demonstrates the big impact of benchmarking details in how the performance indicators are handled on the final results.

\section{Conclusion and Future Work}
\label{sec:col_fu}

In this paper, we present OR-Benchmark, an open and highly reconfigurable general-purpose benchmarking framework, to meet the needs of benchmarking different digital watermarking schemes. To the best of our knowledge, this is the first and the only benchmarking framework supporting all known types of digital watermarking schemes including complicated ones involving multiple types of watermarks. We implemented a prototype as a MATLAB software package, and give a case study on three image authentication and self-restoration watermarking schemes to showcase the usefulness of OR-Benchmark as a convenient and flexible tool.

Although OR-Benchmark as a general framework can easily support any media type, attacks, test multimedia datasets, and PE algorithms, our current implementation has mainly built-in functional units for digital images. The \emph{Offline Analyzer} is also tailored towards our own needs for benchmarking some special types of digital watermarking schemes. In future we plan to add more functional units to the prototype so that users can use it without adding too many user-defined algorithms but focus on the digital watermarking schemes themselves. We also plan to release our MATLAB prototype under an open source license and call for contributions from the whole digital watermarking community. A dedicated website will be set up to host related documents and the source code of our prototype implementation.

\ifCLASSOPTIONcaptionsoff
  \newpage
\fi

\iffulledition
    \ifarXiv
        \input OR-Benchmark-bbl-full.tex

    \else
        \bibliographystyle{IEEEtran}
        \bibliography{OR-Benchmark-full}
    \fi
\fi

\end{document}

%% file: figures/proc_embed_sys.tex
\begin{tikzpicture}[>=latex, thick] 
\node[rectangle, draw, align=center, text width=5em, minimum height=3em] (Embedder) at (0,0) {Watermark Embedder};
\node[align=center, text width=5em] (Cover) at (0,4em) {Cover Work};
\node[gray, align=right, anchor=east] (Key) at (-5em,1em) {Key};
\node[gray, align=right, text width=5em, anchor=east] (Parameters) at (-5em,-1em) {Optional Parameters};
\node[gray, align=center] (Watermark) at (0,-4em) {Watermark(s)};
\node[text width=5em, align=left] (Watermarked) at (8em,0) {Watermarked Work};
\draw[->] (Cover) -- (Embedder);
\draw[->] (Embedder) -- (Watermarked);
\draw[->, gray, dashed] (Key) -- ([yshift=1em]Embedder.west);
\draw[->, gray, dashed] (Parameters) -- ([yshift=-1em]Embedder.west);
\draw[->, gray, dashed] ([xshift=-1em]Watermark.north) -- ([xshift=-1em]Embedder.south);
\draw[<-, gray, dashed] ([xshift=1em]Watermark.north) -- ([xshift=1em]Embedder.south);
\end{tikzpicture} 

%% file: figures/proc_authrec_sys.tex
\begin{tikzpicture}[>=latex, thick] 
\node[rectangle, draw, align=center, text width=8em, minimum height=3em] (Detector) at (0,0) {Watermark Extractor/Detector};
\node[align=center, text width=5em] (Test) at (0,4em) {Test Work};
\node[gray, align=center] (Watermark) at (0,-4em) {Watermark(s)};
\node[gray, align=right, anchor=east] (Key) at (-7em,1em) {Key};
\node[gray, align=right, text width=5em, anchor=east] (Parameters) at (-7em,-1em) {Optional Parameters};
\node[text width=5em, align=left, anchor=west] (Decision) at (6em,3em) {Binary Decision(s)};
\node[text width=5em, align=left, anchor=west] (Restored) at (6em,0) {Restored Work};
\node[text width=5em, align=left, anchor=west] (Other) at (6em,-3em) {Other Output(s)};
\draw[->] (Test) -- (Detector);
\draw[->, gray, dashed] ([xshift=-1em]Watermark.north) -- ([xshift=-1em]Detector.south);
\draw[<-, gray, dashed] ([xshift=1em]Watermark.north) -- ([xshift=1em]Detector.south);
\draw[->, gray, dashed] (Key) -- ([yshift=1em]Detector.west);
\draw[->, gray, dashed] (Parameters) -- ([yshift=-1em]Detector.west);
\draw[->, gray, dashed] ([xshift=2em]Detector.north) |- (Decision);
\draw[->, gray, dashed] (Detector) -- (Restored);
\draw[->, gray, dashed] ([xshift=2em]Detector.south) |- (Other);
\end{tikzpicture} 

%% file: figures/systm_psnr_w.tex
%
%
%
\begin{tikzpicture}

\begin{axis}[%
xmin=0,
xmax=100,
ymin=34,
ymax=40,
ylabel={PSNR},
label style={font=\footnotesize}
]
\addplot [color=red,solid,mark=asterisk,mark options={solid}]
  table[row sep=crcr]{%
1	36.4473305100751\\
2	36.864322940358\\
3	36.2311607148381\\
4	36.3152349879202\\
5	36.31268796409\\
6	36.2609713327073\\
7	36.3711083124214\\
8	36.2821649403255\\
9	36.7223602720456\\
10	36.4966471107644\\
11	37.0224023971089\\
12	36.6377271180469\\
13	36.4805823685296\\
14	36.5726856611077\\
15	36.7207799580079\\
16	36.5480147123932\\
17	36.2975688525542\\
18	36.4936971340048\\
19	36.4265381252328\\
20	36.2736916669716\\
21	36.9356497062373\\
22	36.7709429717417\\
23	36.3116885345351\\
24	36.3329582600652\\
25	36.3892301730654\\
26	36.824871199245\\
27	36.3844900879622\\
28	36.4422809907899\\
29	36.2665439896426\\
30	36.5046294767433\\
31	36.3723863598786\\
32	36.6950688450159\\
33	36.6842692381506\\
34	36.4827933753404\\
35	36.3116885345351\\
36	36.7860477002202\\
37	36.8047911782193\\
38	36.5346781644725\\
39	36.7343221768824\\
40	36.5130397372064\\
41	36.6308017567129\\
42	36.8729752708128\\
43	36.5166696356294\\
44	36.5066187193549\\
45	36.6489437816482\\
46	36.4022469880088\\
47	36.5779598159294\\
48	36.4827933753404\\
49	36.3694728130494\\
50	36.4136503384947\\
51	36.3760981439424\\
52	36.3999894803308\\
53	36.5327176927638\\
54	36.5612807641512\\
55	36.4191927178284\\
56	36.5646909266094\\
57	36.3411640406793\\
58	36.3649016899008\\
59	36.3683744802425\\
60	36.7932301768514\\
61	36.6918053665064\\
62	36.6049249007931\\
63	36.8145685450594\\
64	36.427933329985\\
65	36.7571944886128\\
66	36.3869687011795\\
67	36.7170217246143\\
68	36.7119186087267\\
69	36.4516891334038\\
70	36.4574944212734\\
71	36.4261666435959\\
72	36.8739343980876\\
73	36.5837663185056\\
74	37.2752381475566\\
75	36.416340638628\\
76	36.4888031165707\\
77	36.2906556040107\\
78	36.651430754146\\
79	36.4259369078032\\
80	36.3974130247699\\
81	36.7252381984824\\
82	37.0447142213016\\
83	36.4800563199839\\
84	36.3614136173275\\
85	36.4751229492508\\
86	36.6352937667128\\
87	36.7759979178451\\
88	36.6739062562285\\
89	36.2985889919878\\
90	36.6450039845137\\
91	36.5935324277339\\
92	36.7272715635606\\
93	36.3684842589631\\
94	37.2678781742218\\
95	36.8789716386363\\
96	36.4539811898738\\
97	36.3155102053362\\
98	36.9012740716779\\
99	36.8789782617201\\
100	37.0099000533916\\
};
\addlegendentry{M1};

\addplot [color=cyan,solid,mark=o,mark options={solid}]
  table[row sep=crcr]{%
1	36.8468725880678\\
2	35.5738100464807\\
3	37.097967643186\\
4	37.1349698880163\\
5	37.0924791599774\\
6	37.1716472396008\\
7	37.3738510124483\\
8	37.2819515996303\\
9	36.1031793042503\\
10	36.7802413083447\\
11	35.7906994537428\\
12	36.1785835311588\\
13	36.619299568875\\
14	36.4252626885048\\
15	36.1791796478622\\
16	36.4638262362655\\
17	36.988535804942\\
18	36.4768621229129\\
19	36.8372650110649\\
20	37.3388641362422\\
21	36.053718148979\\
22	37.4064766692161\\
23	37.1151867788289\\
24	37.418071547362\\
25	37.2256232527978\\
26	36.0877227973403\\
27	37.0966983451991\\
28	37.3303239435538\\
29	37.4119150217038\\
30	36.5150697615977\\
31	36.8260131777438\\
32	36.0258233778783\\
33	36.4095942320489\\
34	36.6879362382259\\
35	37.1151867788289\\
36	35.7105455246168\\
37	35.8994975962857\\
38	36.4675375862478\\
39	36.2738501097827\\
40	36.9582639980367\\
41	36.0523433231671\\
42	36.2908109485149\\
43	36.5499886267968\\
44	36.4731137693638\\
45	36.1864043756811\\
46	36.7259094873412\\
47	36.509762645666\\
48	36.6879362382259\\
49	37.0209924600759\\
50	36.6061340940961\\
51	37.0767448505086\\
52	36.9074979978043\\
53	36.1240614604448\\
54	36.4462228031573\\
55	36.9883661222521\\
56	36.4036477665988\\
57	37.0726778710595\\
58	36.9173058168671\\
59	36.8002196689226\\
60	36.2219606785583\\
61	36.2419214022693\\
62	36.2690705614202\\
63	35.8310258813481\\
64	36.9185421254392\\
65	35.7479050048752\\
66	36.6576082017474\\
67	36.6913074138518\\
68	35.9322448411208\\
69	36.2970450119974\\
70	36.819393633537\\
71	36.8712725149813\\
72	36.0451808748751\\
73	36.3713963006638\\
74	36.2355333824005\\
75	36.6720326968921\\
76	36.4169613703212\\
77	37.2680587489599\\
78	36.2203441017912\\
79	36.7114613587837\\
80	36.8532494149543\\
81	36.1881947835318\\
82	35.448198586326\\
83	36.566924763954\\
84	37.8155073760256\\
85	36.7390909463289\\
86	36.4392039039372\\
87	36.0076869432258\\
88	36.2259744816844\\
89	37.4058933908593\\
90	36.3113645231424\\
91	36.2032730583645\\
92	36.2337045992128\\
93	37.1328867208844\\
94	35.891744933736\\
95	35.9893493624935\\
96	36.4293946719465\\
97	37.3245613425855\\
98	36.0637617172174\\
99	35.4496018460981\\
100	35.752008409013\\
};
\addlegendentry{M2};

\addplot [color=blue,dotted,mark=square,mark options={solid}]
  table[row sep=crcr]{%
1	37.332810527538\\
2	36.3683126520697\\
3	37.1789142220474\\
4	36.8736515988884\\
5	36.6804748316609\\
6	37.4145806942985\\
7	37.6575937572685\\
8	37.7365546948012\\
9	36.1958564065666\\
10	36.8158477485649\\
11	35.1851935142254\\
12	36.5454229385873\\
13	36.0111603157031\\
14	35.2858801068743\\
15	35.5015724999158\\
16	36.6279409337604\\
17	36.5569700439003\\
18	36.1328073105324\\
19	37.0636745318015\\
20	37.3089817491695\\
21	37.2276560425077\\
22	36.368870609593\\
23	36.9994955216837\\
24	37.134153283315\\
25	37.9299472200672\\
26	36.1670019314061\\
27	36.5872963710747\\
28	37.0469289047894\\
29	37.408054437072\\
30	36.5697125095244\\
31	36.6739560077575\\
32	35.0801357809919\\
33	37.0853329983712\\
34	36.6940583641874\\
35	36.9994955216837\\
36	36.5215864057876\\
37	36.717152529911\\
38	35.5176907973905\\
39	35.8931592376951\\
40	36.892586827425\\
41	36.2624411735109\\
42	37.6241073475394\\
43	36.711177792413\\
44	35.726714407729\\
45	36.0768548631011\\
46	36.1287396959352\\
47	35.7025189609874\\
48	36.6940583641874\\
49	36.1384240481135\\
50	36.3136839894725\\
51	36.8681762025364\\
52	37.0326690301654\\
53	36.1127764862717\\
54	37.6340828428859\\
55	37.1665689546494\\
56	36.7007973617523\\
57	36.9631096906654\\
58	36.5870471149633\\
59	36.2193112877009\\
60	35.1142987677019\\
61	35.8457000795103\\
62	36.2265354635583\\
63	37.4481060541822\\
64	37.2421566434533\\
65	36.9204708011746\\
66	36.3450102846464\\
67	34.8723841536413\\
68	37.1730208955425\\
69	36.2113273641201\\
70	36.6901880217541\\
71	37.0591464122279\\
72	35.9369740886234\\
73	36.7770070257331\\
74	37.7369031154751\\
75	36.4055033066311\\
76	36.7442838968041\\
77	37.6509622481876\\
78	36.4438887749586\\
79	36.1955646855373\\
80	36.6545394100816\\
81	38.0264709469614\\
82	37.5536229784598\\
83	37.1975046313103\\
84	37.6885891324272\\
85	37.0299972327541\\
86	36.2735217074731\\
87	36.5316017508331\\
88	36.7789875154923\\
89	37.0453851437759\\
90	36.8410582198051\\
91	36.3398622412279\\
92	37.8313987539987\\
93	37.8870937493255\\
94	38.3036361155089\\
95	35.6246090518349\\
96	35.932208891922\\
97	36.742558641571\\
98	37.1620958033457\\
99	36.0413016476701\\
100	37.1698576847112\\
};
\addlegendentry{M3};

\end{axis}
\end{tikzpicture}%

%% file: figures/systm_ssim_w.tex
%
%
%
\begin{tikzpicture}

\begin{axis}[%
xmin=0,
xmax=100,
ymin=0.75,
ymax=1.1,
ylabel={SSIM},
label style={font=\footnotesize}
]
\addplot [color=red,solid,mark=asterisk,mark options={solid}]
  table[row sep=crcr]{%
1	0.931894462380732\\
2	0.889261666268576\\
3	0.975204584207282\\
4	0.972650209222675\\
5	0.973163028524783\\
6	0.978178150436219\\
7	0.922763337080548\\
8	0.957766397888436\\
9	0.875291694304239\\
10	0.938176535957841\\
11	0.877987261813174\\
12	0.908403930066298\\
13	0.924395913252087\\
14	0.923959540610216\\
15	0.904987247787803\\
16	0.931425763170352\\
17	0.962042024198925\\
18	0.932102558082556\\
19	0.926758070068192\\
20	0.928522644553693\\
21	0.883972761547627\\
22	0.945490719965541\\
23	0.964820313678825\\
24	0.922451460104894\\
25	0.923434198182246\\
26	0.884270019664759\\
27	0.955852317141009\\
28	0.96788286731923\\
29	0.958855383533122\\
30	0.93562074796779\\
31	0.972979090114015\\
32	0.898630976142238\\
33	0.908037091807131\\
34	0.93239168874827\\
35	0.964820313678825\\
36	0.904651099618029\\
37	0.903354411457203\\
38	0.93476194237815\\
39	0.938818231674621\\
40	0.968437908990759\\
41	0.91000249660741\\
42	0.925081743934983\\
43	0.955910495966074\\
44	0.938438859265126\\
45	0.93617984911771\\
46	0.961595202000363\\
47	0.937556509751304\\
48	0.93239168874827\\
49	0.966999391418278\\
50	0.957587436770316\\
51	0.992266523565702\\
52	0.955582028707024\\
53	0.941671423161922\\
54	0.924866862195574\\
55	0.933000767687962\\
56	0.926128591031825\\
57	0.974151330776746\\
58	0.970007791409488\\
59	0.962363863191752\\
60	0.900030883330672\\
61	0.921520535090809\\
62	0.908305352134606\\
63	0.90491496165461\\
64	0.95586031137129\\
65	0.914876784486769\\
66	0.972765916459406\\
67	0.925948057955688\\
68	0.89106662089824\\
69	0.942388078351103\\
70	0.951012877902846\\
71	0.952429829725932\\
72	0.892094544246728\\
73	0.937682894926533\\
74	0.89840782266581\\
75	0.934336428353988\\
76	0.930987970646821\\
77	0.989584687690904\\
78	0.902684375030881\\
79	0.968338629776198\\
80	0.939383163351615\\
81	0.901628717836251\\
82	0.881569312385978\\
83	0.936118471530753\\
84	0.937664101774891\\
85	0.943008264157943\\
86	0.944445673110437\\
87	0.918939697066573\\
88	0.882856753387296\\
89	0.982496872811282\\
90	0.941516731763382\\
91	0.947867965883498\\
92	0.898512240844991\\
93	0.940767679387316\\
94	0.877564058116863\\
95	0.919066676401836\\
96	0.943015846000886\\
97	0.973478502609149\\
98	0.877278240410293\\
99	0.904458510228052\\
100	0.882968760258266\\
};
\addlegendentry{M1};

\addplot [color=cyan,solid,mark=o,mark options={solid}]
  table[row sep=crcr]{%
1	0.934544109371021\\
2	0.844533648339168\\
3	0.97943249888608\\
4	0.97468796277994\\
5	0.97693945926409\\
6	0.98072853691293\\
7	0.937026412359435\\
8	0.964065579927994\\
9	0.858053954991308\\
10	0.939389323883897\\
11	0.839227366070774\\
12	0.897029186925693\\
13	0.925215111674902\\
14	0.910275104931923\\
15	0.889800773765813\\
16	0.926961205589171\\
17	0.966099605792259\\
18	0.925605717792636\\
19	0.928287093581278\\
20	0.942418334126733\\
21	0.85784704872736\\
22	0.950205471545061\\
23	0.968491182674127\\
24	0.938157773198269\\
25	0.933635768042091\\
26	0.863218573336553\\
27	0.962150566339618\\
28	0.971500331311655\\
29	0.966210597787441\\
30	0.927955069937435\\
31	0.974374440962303\\
32	0.878725491991689\\
33	0.894506776820041\\
34	0.930702160012819\\
35	0.968491182674127\\
36	0.868690395324093\\
37	0.869149683162564\\
38	0.925566847610655\\
39	0.914034070302646\\
40	0.965839977238961\\
41	0.886914092450935\\
42	0.899444664305526\\
43	0.947032855693075\\
44	0.922554425132873\\
45	0.918798711159449\\
46	0.957656331462263\\
47	0.913593599127199\\
48	0.930702160012819\\
49	0.970971545681255\\
50	0.95578485654898\\
51	0.993220219282936\\
52	0.951640476792246\\
53	0.922309515679379\\
54	0.918484495984092\\
55	0.937684320970936\\
56	0.912901097309872\\
57	0.977266150159018\\
58	0.967699998033083\\
59	0.960615859358005\\
60	0.874749191754495\\
61	0.900981917081434\\
62	0.898563178874137\\
63	0.862163755353682\\
64	0.955813148540425\\
65	0.874049691170446\\
66	0.971577141737762\\
67	0.918008270885381\\
68	0.869276871182927\\
69	0.929782143501592\\
70	0.945974013224576\\
71	0.94759439344123\\
72	0.859863765057954\\
73	0.923371631760394\\
74	0.858440614524036\\
75	0.936004794066746\\
76	0.915556860422183\\
77	0.991606156834411\\
78	0.891617621930098\\
79	0.964908998931757\\
80	0.937736202167392\\
81	0.876630404828304\\
82	0.827562003970595\\
83	0.931101888898635\\
84	0.953655318160835\\
85	0.935303273328197\\
86	0.930627788695606\\
87	0.880226732434339\\
88	0.875208210381784\\
89	0.985189277220909\\
90	0.921910307846433\\
91	0.925805646199932\\
92	0.875798518901808\\
93	0.949032303249959\\
94	0.832671052748905\\
95	0.880865876114045\\
96	0.933302118445776\\
97	0.976178176621201\\
98	0.855156121105879\\
99	0.850516866434652\\
100	0.835475021110488\\
};
\addlegendentry{M2};

\addplot [color=blue,dotted,mark=square,mark options={solid}]
  table[row sep=crcr]{%
1	0.947598152081079\\
2	0.88165385368882\\
3	0.979458063708642\\
4	0.973458870506869\\
5	0.978589605139541\\
6	0.982475516375408\\
7	0.943693306277317\\
8	0.973185487278163\\
9	0.86586509858754\\
10	0.943205257253464\\
11	0.820471059633199\\
12	0.908922891427016\\
13	0.918977533357271\\
14	0.894371780992308\\
15	0.880401203667362\\
16	0.93237590942274\\
17	0.961880572201522\\
18	0.929835346824963\\
19	0.936060052284467\\
20	0.944450287068259\\
21	0.904738629056254\\
22	0.947503032398521\\
23	0.970831426939091\\
24	0.93669930615525\\
25	0.946104399100096\\
26	0.882318673178079\\
27	0.958519656756623\\
28	0.970718376154307\\
29	0.966757502957398\\
30	0.936488716885213\\
31	0.972540295718374\\
32	0.861620942737246\\
33	0.930660559485094\\
34	0.934740704477637\\
35	0.970831426939091\\
36	0.91006622002543\\
37	0.911156763221982\\
38	0.906099654054712\\
39	0.916038090178915\\
40	0.972512161479477\\
41	0.900320348151899\\
42	0.942682836689481\\
43	0.96064153587149\\
44	0.926882067854433\\
45	0.927595477508912\\
46	0.950764725233231\\
47	0.898013976431743\\
48	0.934740704477637\\
49	0.971215899247794\\
50	0.956504852072046\\
51	0.993152779013398\\
52	0.96688755939835\\
53	0.926963269853926\\
54	0.944061833136228\\
55	0.945768119494002\\
56	0.92464265146773\\
57	0.978272666323553\\
58	0.975529657200917\\
59	0.966666879857266\\
60	0.853543227090016\\
61	0.897050404694209\\
62	0.904256490638895\\
63	0.912712607113477\\
64	0.970388860167746\\
65	0.915556872892098\\
66	0.974287659210353\\
67	0.892231750064272\\
68	0.908540340050588\\
69	0.935042799562953\\
70	0.947938660039199\\
71	0.957995023144254\\
72	0.872301948667824\\
73	0.942093929229986\\
74	0.933730704475148\\
75	0.93546613078487\\
76	0.942726182673887\\
77	0.99266698691652\\
78	0.902508949553458\\
79	0.964585549410094\\
80	0.94841690384721\\
81	0.92874601257739\\
82	0.895177578541124\\
83	0.946489262660886\\
84	0.95431680763997\\
85	0.959298514130336\\
86	0.939998427602475\\
87	0.90010357772014\\
88	0.89829365348491\\
89	0.986406320484533\\
90	0.946387328299417\\
91	0.94362492203117\\
92	0.928238943175423\\
93	0.959427916609378\\
94	0.901039611863795\\
95	0.886695672327\\
96	0.931673143098893\\
97	0.979980492715996\\
98	0.886829798065996\\
99	0.893080320015866\\
100	0.882293882043799\\
};
\addlegendentry{M3};

\end{axis}
\end{tikzpicture}%

%% file: figures/systm_fp.tex
%
%
%
\begin{tikzpicture}

\begin{axis}[%
xmin=0,
xmax=101,
ymin=-0.05,
ymax=1,
ylabel={FP Rate},
label style={font=\footnotesize}
]
\addplot [color=red,solid,mark=asterisk,mark options={solid}]
  table[row sep=crcr]{%
1	0\\
2	0\\
3	0\\
4	0\\
5	0\\
6	0\\
7	0.000253807106598985\\
8	0\\
9	0\\
10	0\\
11	0\\
12	0\\
13	0\\
14	0\\
15	0\\
16	0\\
17	0\\
18	0\\
19	0\\
20	0\\
21	0\\
22	0.0725806451612903\\
23	0.00380710659898477\\
24	0\\
25	0\\
26	0\\
27	0\\
28	0.0311447811447811\\
29	0\\
30	0\\
31	0\\
32	0\\
33	0\\
34	0\\
35	0.00380710659898477\\
36	0\\
37	0\\
38	0\\
39	0\\
40	0\\
41	0\\
42	0\\
43	0\\
44	0\\
45	0\\
46	0\\
47	0\\
48	0\\
49	0\\
50	0\\
51	0\\
52	0\\
53	0\\
54	0\\
55	0\\
56	0\\
57	0\\
58	0\\
59	0\\
60	0\\
61	0\\
62	0\\
63	0\\
64	0\\
65	0.00144927536231884\\
66	0\\
67	0\\
68	0\\
69	0\\
70	0\\
71	0\\
72	0\\
73	0\\
74	0\\
75	0\\
76	0\\
77	0\\
78	0\\
79	0\\
80	0\\
81	0\\
82	0\\
83	0\\
84	0\\
85	0\\
86	0\\
87	0\\
88	0\\
89	0\\
90	0\\
91	0\\
92	0\\
93	0.000841750841750842\\
94	0\\
95	0\\
96	0\\
97	0\\
98	0\\
99	0\\
100	0\\
};
\addlegendentry{M1};

\addplot [color=cyan,solid,mark=o,mark options={solid}]
  table[row sep=crcr]{%
1	0\\
2	0.0130434782608696\\
3	0.021889400921659\\
4	0\\
5	0\\
6	0.00925925925925926\\
7	0\\
8	0.0161290322580645\\
9	0.000977756049865559\\
10	0\\
11	0.0161290322580645\\
12	0.0230414746543779\\
13	0\\
14	0\\
15	0.0172811059907834\\
16	0\\
17	0.0143097643097643\\
18	0\\
19	0\\
20	0.0230414746543779\\
21	0.00050761421319797\\
22	0.105990783410138\\
23	0\\
24	0\\
25	0\\
26	0\\
27	0.0264976958525346\\
28	0.0260942760942761\\
29	0\\
30	0\\
31	0.00144927536231884\\
32	0.0299539170506912\\
33	0.0172811059907834\\
34	0\\
35	0\\
36	0.0230414746543779\\
37	0.0036231884057971\\
38	0.0260869565217391\\
39	0.0195652173913043\\
40	0.00289855072463768\\
41	0.0184331797235023\\
42	0.0166666666666667\\
43	0.00072463768115942\\
44	0.00144927536231884\\
45	0\\
46	0.00144927536231884\\
47	0.00144927536231884\\
48	0\\
49	0.00144927536231884\\
50	0.00144927536231884\\
51	0.00144927536231884\\
52	0.00144927536231884\\
53	0.00144927536231884\\
54	0.0241935483870968\\
55	0\\
56	0\\
57	0.00144927536231884\\
58	0\\
59	0.00144927536231884\\
60	0.00144927536231884\\
61	0.00289855072463768\\
62	0\\
63	0\\
64	0.00144927536231884\\
65	0.00144927536231884\\
66	0.00217391304347826\\
67	0.0144927536231884\\
68	0.0172811059907834\\
69	0.00144927536231884\\
70	0\\
71	0.00144927536231884\\
72	0.00652173913043478\\
73	0.0036231884057971\\
74	0.193602693602694\\
75	0.0172811059907834\\
76	0.0230414746543779\\
77	0.0404040404040404\\
78	0\\
79	0.00942028985507246\\
80	0.0184331797235023\\
81	0\\
82	0.00217391304347826\\
83	0.00144927536231884\\
84	0.00925925925925926\\
85	0.00144927536231884\\
86	0.00289855072463768\\
87	0.00144927536231884\\
88	0\\
89	0.00144927536231884\\
90	0.00144927536231884\\
91	0.00144927536231884\\
92	0.00144927536231884\\
93	0.00841750841750842\\
94	0.327536231884058\\
95	0.157971014492754\\
96	0.0230414746543779\\
97	0.00144927536231884\\
98	0.0241935483870968\\
99	0.00144927536231884\\
100	0.00144927536231884\\
};
\addlegendentry{M2};

\addplot [color=blue,dotted,mark=square,mark options={solid}]
  table[row sep=crcr]{%
1	0\\
2	0\\
3	0\\
4	0\\
5	0.000761421319796954\\
6	0\\
7	0\\
8	0\\
9	0\\
10	0\\
11	0\\
12	0\\
13	0\\
14	0\\
15	0\\
16	0\\
17	0\\
18	0\\
19	0\\
20	0\\
21	0\\
22	0\\
23	0\\
24	0\\
25	0\\
26	0\\
27	0\\
28	0\\
29	0\\
30	0\\
31	0\\
32	0\\
33	0\\
34	0\\
35	0\\
36	0\\
37	0\\
38	0\\
39	0\\
40	0\\
41	0\\
42	0\\
43	0\\
44	0\\
45	0\\
46	0\\
47	0\\
48	0\\
49	0\\
50	0\\
51	0\\
52	0\\
53	0\\
54	0\\
55	0\\
56	0\\
57	0\\
58	0\\
59	0\\
60	0\\
61	0\\
62	0\\
63	0\\
64	0\\
65	0\\
66	0\\
67	0\\
68	0\\
69	0\\
70	0\\
71	0\\
72	0\\
73	0\\
74	0\\
75	0\\
76	0\\
77	0\\
78	0\\
79	0\\
80	0\\
81	0\\
82	0\\
83	0\\
84	0\\
85	0\\
86	0\\
87	0\\
88	0\\
89	0\\
90	0\\
91	0\\
92	0\\
93	0\\
94	0\\
95	0\\
96	0\\
97	0\\
98	0\\
99	0\\
100	0\\
};
\addlegendentry{M3};

\end{axis}
\end{tikzpicture}%

%% file: figures/systm_fn.tex
%
%
%
\begin{tikzpicture}

\begin{axis}[%
xmin=0,
xmax=101,
ymin=-0.05,
ymax=1,
ylabel={FN Rate},
label style={font=\footnotesize}
]
\addplot [color=red,solid,mark=asterisk,mark options={solid}]
  table[row sep=crcr]{%
1	0\\
2	0\\
3	0\\
4	0\\
5	0\\
6	0\\
7	0\\
8	0\\
9	0\\
10	0\\
11	0\\
12	0\\
13	0\\
14	0\\
15	0\\
16	0\\
17	0\\
18	0\\
19	0\\
20	0\\
21	0\\
22	0\\
23	0\\
24	0\\
25	0\\
26	0\\
27	0\\
28	0\\
29	0\\
30	0\\
31	0\\
32	0\\
33	0\\
34	0\\
35	0\\
36	0\\
37	0\\
38	0\\
39	0\\
40	0\\
41	0\\
42	0\\
43	0\\
44	0\\
45	0\\
46	0\\
47	0\\
48	0\\
49	0\\
50	0\\
51	0\\
52	0\\
53	0\\
54	0\\
55	0\\
56	0\\
57	0\\
58	0\\
59	0\\
60	0\\
61	0\\
62	0\\
63	0\\
64	0\\
65	0\\
66	0\\
67	0\\
68	0\\
69	0\\
70	0\\
71	0\\
72	0\\
73	0\\
74	0\\
75	0\\
76	0\\
77	0\\
78	0\\
79	0\\
80	0\\
81	0\\
82	0\\
83	0\\
84	0\\
85	0\\
86	0\\
87	0\\
88	0\\
89	0\\
90	0\\
91	0\\
92	0\\
93	0\\
94	0\\
95	0\\
96	0\\
97	0\\
98	0\\
99	0\\
100	0\\
};
\addlegendentry{M1};

\addplot [color=cyan,solid,mark=o,mark options={solid}]
  table[row sep=crcr]{%
1	0.032051282051282\\
2	0\\
3	0.0192307692307692\\
4	0.0192307692307692\\
5	0.0576923076923077\\
6	0.00641025641025641\\
7	0\\
8	0\\
9	0.0512820512820513\\
10	0.102564102564103\\
11	0\\
12	0.0128205128205128\\
13	0.032051282051282\\
14	0.00641025641025641\\
15	0.00641025641025641\\
16	0.384615384615385\\
17	0\\
18	0.0128205128205128\\
19	0.0256410256410256\\
20	0.00641025641025641\\
21	0.0641025641025641\\
22	0.00641025641025641\\
23	0.0128205128205128\\
24	0.0192307692307692\\
25	0\\
26	0.032051282051282\\
27	0.0256410256410256\\
28	0\\
29	0\\
30	0.0128205128205128\\
31	0\\
32	0\\
33	0.0192307692307692\\
34	0.00641025641025641\\
35	0.0128205128205128\\
36	0\\
37	0\\
38	0.00641025641025641\\
39	0.0128205128205128\\
40	0.0192307692307692\\
41	0.0192307692307692\\
42	0.0128205128205128\\
43	0.0192307692307692\\
44	0.00641025641025641\\
45	0.0256410256410256\\
46	0.00641025641025641\\
47	0\\
48	0.00641025641025641\\
49	0.00641025641025641\\
50	0.108974358974359\\
51	0.0256410256410256\\
52	0\\
53	0.00641025641025641\\
54	0\\
55	0.0256410256410256\\
56	0.0192307692307692\\
57	0.0128205128205128\\
58	0\\
59	0\\
60	0\\
61	0\\
62	0.185897435897436\\
63	0.0576923076923077\\
64	0.0512820512820513\\
65	0.032051282051282\\
66	0.0256410256410256\\
67	0\\
68	0.0128205128205128\\
69	0.0512820512820513\\
70	0.00641025641025641\\
71	0.00641025641025641\\
72	0.0192307692307692\\
73	0.0384615384615385\\
74	0\\
75	0.00641025641025641\\
76	0.032051282051282\\
77	0.121794871794872\\
78	0.115384615384615\\
79	0.00641025641025641\\
80	0.032051282051282\\
81	0.0192307692307692\\
82	0.0512820512820513\\
83	0.0512820512820513\\
84	0\\
85	0.00641025641025641\\
86	0.166666666666667\\
87	0.0128205128205128\\
88	0.269230769230769\\
89	0.0128205128205128\\
90	0\\
91	0\\
92	0.0576923076923077\\
93	0.00641025641025641\\
94	0\\
95	0\\
96	0\\
97	0\\
98	0.00641025641025641\\
99	0\\
100	0.211538461538462\\
};
\addlegendentry{M2};

\addplot [color=blue,dotted,mark=square,mark options={solid}]
  table[row sep=crcr]{%
1	0.00641025641025641\\
2	0.00641025641025641\\
3	0.0384615384615385\\
4	0.0128205128205128\\
5	0.0128205128205128\\
6	0.0256410256410256\\
7	0.0128205128205128\\
8	0.0384615384615385\\
9	0.0192307692307692\\
10	0.0128205128205128\\
11	0.0384615384615385\\
12	0.0384615384615385\\
13	0.0128205128205128\\
14	0.0128205128205128\\
15	0.0384615384615385\\
16	0.0128205128205128\\
17	0.0256410256410256\\
18	0.0128205128205128\\
19	0.0128205128205128\\
20	0.0384615384615385\\
21	0.0128205128205128\\
22	0.0384615384615385\\
23	0.0128205128205128\\
24	0.0128205128205128\\
25	0.0128205128205128\\
26	0.0128205128205128\\
27	0.0384615384615385\\
28	0.0256410256410256\\
29	0.0128205128205128\\
30	0.0128205128205128\\
31	0.00641025641025641\\
32	0.0384615384615385\\
33	0.0384615384615385\\
34	0.0128205128205128\\
35	0.0128205128205128\\
36	0.0384615384615385\\
37	0.00641025641025641\\
38	0.0384615384615385\\
39	0.00641025641025641\\
40	0.00641025641025641\\
41	0.0384615384615385\\
42	0.00641025641025641\\
43	0.00641025641025641\\
44	0.00641025641025641\\
45	0.00641025641025641\\
46	0.00641025641025641\\
47	0.00641025641025641\\
48	0.0128205128205128\\
49	0.00641025641025641\\
50	0.00641025641025641\\
51	0.00641025641025641\\
52	0.00641025641025641\\
53	0.00641025641025641\\
54	0.0384615384615385\\
55	0.0128205128205128\\
56	0.00641025641025641\\
57	0.00641025641025641\\
58	0.00641025641025641\\
59	0.00641025641025641\\
60	0.00641025641025641\\
61	0.00641025641025641\\
62	0.0128205128205128\\
63	0.00641025641025641\\
64	0.00641025641025641\\
65	0.00641025641025641\\
66	0.00641025641025641\\
67	0.00641025641025641\\
68	0.0384615384615385\\
69	0.00641025641025641\\
70	0.00641025641025641\\
71	0.00641025641025641\\
72	0.00641025641025641\\
73	0.00641025641025641\\
74	0.0256410256410256\\
75	0.0384615384615385\\
76	0.0384615384615385\\
77	0.0384615384615385\\
78	0.0128205128205128\\
79	0.00641025641025641\\
80	0.0384615384615385\\
81	0.00641025641025641\\
82	0.00641025641025641\\
83	0.00641025641025641\\
84	0.0256410256410256\\
85	0.00641025641025641\\
86	0.00641025641025641\\
87	0.00641025641025641\\
88	0.0128205128205128\\
89	0.00641025641025641\\
90	0.00641025641025641\\
91	0.00641025641025641\\
92	0.00641025641025641\\
93	0.0256410256410256\\
94	0.00641025641025641\\
95	0.00641025641025641\\
96	0.0384615384615385\\
97	0.00641025641025641\\
98	0.0384615384615385\\
99	0.00641025641025641\\
100	0.00641025641025641\\
};
\addlegendentry{M3};

\end{axis}
\end{tikzpicture}%

%% file: figures/systm_fp_jpeg.tex
%
%
%
\begin{tikzpicture}

\begin{axis}[%
x dir=reverse,
xmin=50,
xmax=100,
xlabel=Quality Factor,
ymin=0,
ymax=1.3,
xtick={100, 90, 80, 70, 60, 50},
ytick={0, 0.2, 0.4, 0.6, 0.8, 1},
ylabel=False Positive Rate,
label style={font=\footnotesize}
]
\addplot [color=red,solid,mark=asterisk,mark options={solid}]
  table[row sep=crcr]{%
100	2.89855072463768e-05\\
95	2.89855072463768e-05\\
90	0.000101449275362319\\
85	0.000652173913043478\\
80	0.00244927536231884\\
75	0.00685507246376812\\
70	0.0128405797101449\\
65	0.0137536231884058\\
60	0.0300724637681159\\
55	0.0737101449275362\\
50	0.999927536231884\\
};
\addlegendentry{M1};

\addplot [color=cyan,solid,mark=o,mark options={solid}]
  table[row sep=crcr]{%
100	0.0127971014492754\\
95	0.0175072463768116\\
90	0.0235072463768116\\
85	0.0915217391304348\\
80	0.497072463768116\\
75	0.950811594202899\\
70	0.981637681159421\\
65	0.998782608695652\\
60	0.999913043478261\\
55	0.999797101449275\\
50	0.999826086956522\\
};
\addlegendentry{M2};

\addplot [color=blue,solid,mark=square,mark options={solid}]
  table[row sep=crcr]{%
100	0\\
95	0\\
90	0.000130434782608696\\
85	0.000782608695652174\\
80	0.00189855072463768\\
75	0.00391304347826087\\
70	0.00388405797101449\\
65	0.00610144927536232\\
60	0.00923188405797101\\
55	0.0206376811594203\\
50	0.977507246376812\\
};
\addlegendentry{M3};

\end{axis}
\end{tikzpicture}%

%% file: figures/systm_fp_gaussian.tex
%
%
%
\begin{tikzpicture}

\begin{axis}[%
xmin=0,
xmax=40,
xlabel=Variance,
ymin=0,
ymax=1.3,
ylabel=False Positive Rate,
label style={font=\footnotesize}
]
\addplot [color=red,solid,mark=asterisk,mark options={solid}]
  table[row sep=crcr]{%
1	2.89855072463768e-05\\
3	7.2463768115942e-05\\
5	0.000478260869565217\\
7	0.00136231884057971\\
9	0.00456521739130435\\
11	0.0128840579710145\\
13	0.0357536231884058\\
15	0.0810434782608696\\
17	0.155188405797101\\
19	0.265057971014493\\
21	0.389927536231884\\
23	0.538217391304348\\
25	0.66531884057971\\
27	0.773014492753623\\
29	0.847811594202899\\
31	0.902130434782608\\
33	0.921710144927536\\
35	0.948\\
37	0.959797101449276\\
39	0.968840579710146\\
};
\addlegendentry{M1};

\addplot [color=cyan,solid,mark=o,mark options={solid}]
  table[row sep=crcr]{%
1	0.0298115942028985\\
3	0.0666231884057971\\
5	0.260550724637681\\
7	0.616420289855072\\
9	0.857260869565217\\
11	0.944608695652174\\
13	0.974130434782609\\
15	0.986579710144928\\
17	0.992884057971015\\
19	0.99536231884058\\
21	0.996869565217391\\
23	0.998144927536232\\
25	0.998434782608696\\
27	0.998826086956522\\
29	0.999202898550725\\
31	0.999130434782609\\
33	0.999304347826087\\
35	0.999724637681159\\
37	0.99968115942029\\
39	0.999898550724638\\
};
\addlegendentry{M2};

\addplot [color=blue,solid,mark=square,mark options={solid}]
  table[row sep=crcr]{%
1	0\\
3	0\\
5	0.00036231884057971\\
7	0.000333333333333333\\
9	0.000405797101449275\\
11	0.000304347826086957\\
13	0.000898550724637681\\
15	0.000956521739130435\\
17	0.00210144927536232\\
19	0.0031304347826087\\
21	0.0064927536231884\\
23	0.00860869565217391\\
25	0.0130144927536232\\
27	0.021\\
29	0.0374202898550725\\
31	0.0584492753623188\\
33	0.0899565217391304\\
35	0.124202898550725\\
37	0.177811594202899\\
39	0.244536231884058\\
};
\addlegendentry{M3};

\end{axis}
\end{tikzpicture}%

%% file: figures/systm_fp_noise.tex
%
%
%
\begin{tikzpicture}

\begin{axis}[%
xmin=0,
xmax=240,
xtick={0,  40,  80, 120, 160, 200, 240},
xlabel=Variance,
ymin=0,
ymax=1.3,
ylabel=False Positive Rate,
label style={font=\footnotesize}
]
\addplot [color=red,solid,mark=asterisk,mark options={solid}]
  table[row sep=crcr]{%
1	2.89855072463768e-05\\
10	0.000115942028985507\\
20	0.00505797101449275\\
30	0.0333913043478261\\
40	0.0883188405797101\\
50	0.152594202898551\\
60	0.21268115942029\\
70	0.271057971014493\\
80	0.325942028985507\\
90	0.373072463768116\\
100	0.41363768115942\\
110	0.449695652173913\\
120	0.481014492753623\\
130	0.506768115942029\\
140	0.530304347826087\\
150	0.55263768115942\\
160	0.572347826086956\\
170	0.591579710144928\\
180	0.60636231884058\\
190	0.622724637681159\\
200	0.637231884057971\\
210	0.650478260869565\\
220	0.663463768115942\\
230	0.672927536231884\\
240	0.684115942028985\\
};
\addlegendentry{M1};

\addplot [color=cyan,solid,mark=o,mark options={solid}]
  table[row sep=crcr]{%
1	0.0298115942028985\\
10	0.115710144927536\\
20	0.324376811594203\\
30	0.48868115942029\\
40	0.609869565217391\\
50	0.688521739130435\\
60	0.746043478260869\\
70	0.792579710144928\\
80	0.82531884057971\\
90	0.851434782608696\\
100	0.870376811594203\\
110	0.886942028985507\\
120	0.901507246376812\\
130	0.912115942028986\\
140	0.922347826086957\\
150	0.929492753623189\\
160	0.936028985507246\\
170	0.941391304347826\\
180	0.947333333333334\\
190	0.950521739130435\\
200	0.955086956521739\\
210	0.958376811594203\\
220	0.962492753623189\\
230	0.964869565217391\\
240	0.967739130434783\\
};
\addlegendentry{M2};

\addplot [color=blue,solid,mark=square,mark options={solid}]
  table[row sep=crcr]{%
1	0\\
10	0\\
20	0\\
30	0.000347826086956522\\
40	0.00182608695652174\\
50	0.00666666666666667\\
60	0.0170434782608696\\
70	0.0369565217391304\\
80	0.0605652173913043\\
90	0.086695652173913\\
100	0.114550724637681\\
110	0.146289855072464\\
120	0.176376811594203\\
130	0.208594202898551\\
140	0.235579710144928\\
150	0.260985507246377\\
160	0.288710144927536\\
170	0.311260869565217\\
180	0.336898550724638\\
190	0.361144927536232\\
200	0.382159420289855\\
210	0.400579710144927\\
220	0.42231884057971\\
230	0.438985507246377\\
240	0.457028985507246\\
};
\addlegendentry{M3};

\end{axis}
\end{tikzpicture}%

%% file: figures/systm_fn_jpeg.tex
%
%
%
\begin{tikzpicture}

\begin{axis}[%
x dir=reverse,
xmin=50,
xmax=100,
xlabel={Quality Factor},
xtick={100, 90, 80, 70, 60, 50},
ymin=0,
ymax=0.1,
ytick={0, 0.02, 0.04, 0.06, 0.08, 0.1},
yticklabels={0, 0.02, 0.04, 0.06, 0.08, 0.1},
ylabel={False Negative Rate},
label style={font=\footnotesize}
]
\addplot [color=red,solid,mark=asterisk,mark options={solid}]
  table[row sep=crcr]{%
100	0\\
95	0\\
90	0\\
85	0\\
80	0\\
75	0\\
70	0\\
65	0\\
60	0\\
55	0\\
50	0\\
};
\addlegendentry{M1};

\addplot [color=cyan,solid,mark=o,mark options={solid}]
  table[row sep=crcr]{%
100	0.0223076923076923\\
95	0.0188461538461538\\
90	0.0112820512820513\\
85	0.00628205128205128\\
80	0.000897435897435897\\
75	0\\
70	0\\
65	0\\
60	0\\
55	0\\
50	0\\
};
\addlegendentry{M2};

\addplot [color=blue,solid,mark=square,mark options={solid}]
  table[row sep=crcr]{%
100	0.00705128205128205\\
95	0.00705128205128205\\
90	0.00705128205128205\\
85	0.00705128205128205\\
80	0.00705128205128205\\
75	0.00705128205128205\\
70	0.00705128205128205\\
65	0.00705128205128205\\
60	0.00705128205128205\\
55	0.00666666666666666\\
50	0.000128205128205128\\
};
\addlegendentry{M3};

\end{axis}
\end{tikzpicture}%

%% file: figures/systm_fn_gaussian.tex
%
%
%
\begin{tikzpicture}

\begin{axis}[%
xmin=0,
xmax=40,
xlabel={Variance},
ymin=0,
ymax=0.1,
ytick={0, 0.02, 0.04, 0.06, 0.08, 0.1},
yticklabels={0, 0.02, 0.04, 0.06, 0.08, 0.1},
ylabel={False Negative Rate},
label style={font=\footnotesize}
]
\addplot [color=red,solid,mark=asterisk,mark options={solid}]
  table[row sep=crcr]{%
1	0\\
3	0.000641025641025641\\
5	0.00102564102564103\\
7	0.000897435897435897\\
9	0.00128205128205128\\
11	0.000897435897435897\\
13	0.000256410256410256\\
15	0.000769230769230769\\
17	0.000512820512820513\\
19	0.000256410256410256\\
21	0\\
23	0\\
25	0\\
27	0.000256410256410256\\
29	0\\
31	0\\
33	0\\
35	0\\
37	0\\
39	0\\
};
\addlegendentry{M1};

\addplot [color=cyan,solid,mark=o,mark options={solid}]
  table[row sep=crcr]{%
1	0.0279487179487179\\
3	0.0137179487179487\\
5	0.0041025641025641\\
7	0.000256410256410256\\
9	0.000256410256410256\\
11	0.000128205128205128\\
13	0\\
15	0\\
17	0\\
19	0\\
21	0\\
23	0\\
25	0\\
27	0\\
29	0\\
31	0\\
33	0\\
35	0\\
37	0\\
39	0\\
};
\addlegendentry{M2};

\addplot [color=blue,solid,mark=square,mark options={solid}]
  table[row sep=crcr]{%
1	0.00705128205128205\\
3	0.0148717948717949\\
5	0.0179487179487179\\
7	0.0207692307692308\\
9	0.0156410256410256\\
11	0.0193589743589744\\
13	0.0215384615384615\\
15	0.0202564102564103\\
17	0.0238461538461538\\
19	0.0192307692307692\\
21	0.0223076923076923\\
23	0.018974358974359\\
25	0.02\\
27	0.0179487179487179\\
29	0.015\\
31	0.012051282051282\\
33	0.0142307692307692\\
35	0.0137179487179487\\
37	0.0115384615384615\\
39	0.00807692307692308\\
};
\addlegendentry{M3};

\end{axis}
\end{tikzpicture}%

%% file: figures/systm_fn_noise.tex
%
%
%
\begin{tikzpicture}

\begin{axis}[%
xmin=0,
xmax=240,
xtick={0,  40,  80, 120, 160, 200, 240},
xlabel={Variance},
ymin=0,
ymax=0.1,
ytick={0, 0.02, 0.04, 0.06, 0.08, 0.1},
yticklabels={0, 0.02, 0.04, 0.06, 0.08, 0.1},
ylabel={False Negative Rate},
label style={font=\footnotesize}
]
\addplot [color=red,solid,mark=asterisk,mark options={solid}]
  table[row sep=crcr]{%
1	0\\
10	0\\
20	0\\
30	0\\
40	0\\
50	0.00333333333333333\\
60	0\\
70	0\\
80	0\\
90	0\\
100	0\\
110	0\\
120	0\\
130	0\\
140	0\\
150	0\\
160	0\\
170	0\\
180	0\\
190	0\\
200	0\\
210	0\\
220	0\\
230	0\\
240	0\\
};
\addlegendentry{M1};

\addplot [color=cyan,solid,mark=o,mark options={solid}]
  table[row sep=crcr]{%
1	0.0279487179487179\\
10	0.0129487179487179\\
20	0.00743589743589743\\
30	0.00538461538461538\\
40	0.00307692307692308\\
50	0.00166666666666667\\
60	0.00102564102564103\\
70	0.000769230769230769\\
80	0.000256410256410256\\
90	0.000384615384615385\\
100	0.000128205128205128\\
110	0.000256410256410256\\
120	0.000128205128205128\\
130	0\\
140	0.000128205128205128\\
150	0.000128205128205128\\
160	0.000256410256410256\\
170	0.000256410256410256\\
180	0.000128205128205128\\
190	0.000128205128205128\\
200	0.000128205128205128\\
210	0.000128205128205128\\
220	0.000128205128205128\\
230	0\\
240	0\\
};
\addlegendentry{M2};

\addplot [color=blue,solid,mark=square,mark options={solid}]
  table[row sep=crcr]{%
1	0.00705128205128205\\
10	0.000512820512820513\\
20	0.00679487179487179\\
30	0.000256410256410256\\
40	0.00679487179487179\\
50	0.00602564102564102\\
60	0.000128205128205128\\
70	0\\
80	0.00397435897435898\\
90	0.0067948717948718\\
100	0.00641025641025641\\
110	0\\
120	0.00230769230769231\\
130	0\\
140	0\\
150	0\\
160	0\\
170	0\\
180	0\\
190	0.0017948717948718\\
200	0.00538461538461538\\
210	0.00128205128205128\\
220	0.00102564102564103\\
230	0.00512820512820513\\
240	0.000897435897435897\\
};
\addlegendentry{M3};

\end{axis}
\end{tikzpicture}%

%% file: figures/systm_psnr_jpeg.tex
%
%
%
\begin{tikzpicture}

\begin{axis}[%
x dir=reverse,
xmin=50,
xmax=100,
xlabel={Quality Factor},
ymin=5,
ymax=36,
ylabel={PSNR},
label style={font=\footnotesize}
]
\addplot [color=red,solid,mark=asterisk,mark options={solid}]
  table[row sep=crcr]{%
100	26.049728645016\\
95	26.0385376575127\\
90	25.9935746911215\\
85	25.9148209016749\\
80	25.5169705356235\\
75	25.0148397013579\\
70	24.4101539435898\\
65	24.1681486277627\\
60	23.633437746025\\
55	22.2365906916516\\
50	7.93018062500703\\
};
\addlegendentry{M1};

\addplot [color=cyan,solid,mark=o,mark options={solid}]
  table[row sep=crcr]{%
100	26.4150298653849\\
95	24.9931121822237\\
90	23.478987078539\\
85	17.6135960159333\\
80	12.3040230215392\\
75	9.39553925263626\\
70	8.56583778124553\\
65	7.43075345505404\\
60	7.38763000978002\\
55	7.21931248747362\\
50	6.99661581500114\\
};
\addlegendentry{M2};

\addplot [color=blue,solid,mark=square,mark options={solid}]
  table[row sep=crcr]{%
100	30.3337252207087\\
95	30.0992571758663\\
90	29.6692433398967\\
85	29.0727275362446\\
80	28.0571653873114\\
75	27.3419883658668\\
70	26.4126084435782\\
65	25.7980829585801\\
60	25.2796533776036\\
55	24.6917476773653\\
50	14.0093661139642\\
};
\addlegendentry{M3};

\end{axis}
\end{tikzpicture}%

%% file: figures/systm_psnr_gaussian.tex
%
%
%
\begin{tikzpicture}

\begin{axis}[%
xmin=0,
xmax=40,
xlabel={Variance},
ymin=5,
ymax=36,
ylabel={PSNR},
label style={font=\footnotesize}
]
\addplot [color=red,solid,mark=asterisk,mark options={solid}]
  table[row sep=crcr]{%
1	26.0522980176445\\
3	27.0065268662255\\
5	26.8498057833994\\
7	27.0199095903314\\
9	26.4386714989335\\
11	26.3849242487738\\
13	25.5225990556964\\
15	24.1552586759216\\
17	22.4621588008144\\
19	20.834366739931\\
21	19.5741440812999\\
23	18.0783704665525\\
25	17.1827115876826\\
27	16.4278922268608\\
29	15.9095594475105\\
31	15.1960957793199\\
33	14.9803707658161\\
35	14.6584142832084\\
37	14.4454182266543\\
39	14.1169485057557\\
};
\addlegendentry{M1};

\addplot [color=cyan,solid,mark=o,mark options={solid}]
  table[row sep=crcr]{%
1	25.3640357637055\\
3	20.7226993758883\\
5	15.5163237310613\\
7	12.1003649080895\\
9	10.556213092858\\
11	9.90579219244395\\
13	9.54940559470492\\
15	9.22942461842377\\
17	9.03066399388122\\
19	8.87598813127787\\
21	8.73110373016107\\
23	8.56853444863794\\
25	8.49369043712034\\
27	8.39264644345064\\
29	8.32857060518264\\
31	8.23900647740072\\
33	8.1738243671507\\
35	8.11834232420248\\
37	8.09988296921395\\
39	8.03484323436725\\
};
\addlegendentry{M2};

\addplot [color=blue,solid,mark=square,mark options={solid}]
  table[row sep=crcr]{%
1	30.4181563088265\\
3	29.4844833076382\\
5	28.9013847932399\\
7	28.5524077374447\\
9	28.2492143557437\\
11	27.8573621657883\\
13	27.4438253517934\\
15	27.271032783076\\
17	26.9087505599857\\
19	26.6990832378647\\
21	26.3595748891434\\
23	26.1270225856664\\
25	25.6810488588938\\
27	25.173321307354\\
29	24.6131777794305\\
31	23.8778211369665\\
33	22.8404323676987\\
35	22.0430000319807\\
37	20.977405518753\\
39	19.7588033111561\\
};
\addlegendentry{M3};

\end{axis}
\end{tikzpicture}%

%% file: figures/systm_psnr_noise.tex
%
%
%
\begin{tikzpicture}

\begin{axis}[%
xmin=0,
xmax=240,
xtick={0,  40,  80, 120, 160, 200, 240},
xlabel={Variance},
ymin=5,
ymax=36,
ylabel={PSNR},
label style={font=\footnotesize}
]
\addplot [color=red,solid,mark=asterisk,mark options={solid}]
  table[row sep=crcr]{%
1	26.0522980176445\\
10	26.3662886969929\\
20	26.3407089629727\\
30	25.3293537915112\\
40	24.3200953532528\\
50	23.1415070473653\\
60	21.8025156526589\\
70	20.8955087693426\\
80	20.1229996539634\\
90	19.316333182137\\
100	18.8543898807366\\
110	18.2885580976704\\
120	17.8605384514774\\
130	17.5477238734665\\
140	17.0229287615848\\
150	16.7240130918263\\
160	16.3750251900028\\
170	16.0208189578973\\
180	15.7832575955859\\
190	15.5638195118984\\
200	15.2992302648407\\
210	15.1194712344324\\
220	14.8951860757549\\
230	14.7071425543724\\
240	14.5309415823965\\
};
\addlegendentry{M1};

\addplot [color=cyan,solid,mark=o,mark options={solid}]
  table[row sep=crcr]{%
1	25.3640357637055\\
10	19.5990445060165\\
20	15.6652985773074\\
30	13.5809727706004\\
40	12.3384044414629\\
50	11.5775329595093\\
60	11.0022113503455\\
70	10.5859687353717\\
80	10.2668513553226\\
90	9.99583855672236\\
100	9.80099668436933\\
110	9.60583053088713\\
120	9.44366637473539\\
130	9.32391618127167\\
140	9.21030112024369\\
150	9.10921571595157\\
160	9.03149964565421\\
170	8.94353818978306\\
180	8.86458038338228\\
190	8.79247003775288\\
200	8.73088894610494\\
210	8.67852692475884\\
220	8.62780994506521\\
230	8.58273720966857\\
240	8.53669369666541\\
};
\addlegendentry{M2};

\addplot [color=blue,solid,mark=square,mark options={solid}]
  table[row sep=crcr]{%
1	30.4181563088265\\
10	29.8667619669715\\
20	29.1728052434566\\
30	28.8605856134573\\
40	28.3369121251426\\
50	27.7360209348844\\
60	27.0237430411662\\
70	25.9607494819798\\
80	24.9692308054334\\
90	24.0801209743869\\
100	23.097039528204\\
110	22.34088742417\\
120	21.5245726171254\\
130	20.9229693107991\\
140	20.4716438965633\\
150	19.9792581594839\\
160	19.5702151418917\\
170	19.252160524484\\
180	18.761641342544\\
190	18.4154763104366\\
200	18.2130497592727\\
210	17.93952037733\\
220	17.6270347237241\\
230	17.4222513101494\\
240	17.2350134034689\\
};
\addlegendentry{M3};

\end{axis}
\end{tikzpicture}%

%% file: figures/systm_ssim_jpeg.tex
%
%
%
\begin{tikzpicture}

\begin{axis}[%
x dir=reverse,
xmin=50,
xmax=100,
xlabel={Quality Factor},
ymin=0,
ymax=1.18,
ylabel={SSIM},
label style={font=\footnotesize}
]
\addplot [color=red,solid,mark=asterisk,mark options={solid}]
  table[row sep=crcr]{%
100	0.919860979420684\\
95	0.918921770698152\\
90	0.916823539257297\\
85	0.912714243855008\\
80	0.899234938783777\\
75	0.891026027281227\\
70	0.876247377720361\\
65	0.859292935488886\\
60	0.840914891781586\\
55	0.806765123217117\\
50	0.100169145559882\\
};
\addlegendentry{M1};

\addplot [color=cyan,solid,mark=o,mark options={solid}]
  table[row sep=crcr]{%
100	0.922047410237577\\
95	0.912641662916625\\
90	0.898942010596139\\
85	0.811784688456189\\
80	0.459962108501328\\
75	0.140506773240442\\
70	0.104057304005899\\
65	0.0712548838133277\\
60	0.0643587807439294\\
55	0.0490086475407083\\
50	0.0439878132023704\\
};
\addlegendentry{M2};

\addplot [color=blue,solid,mark=square,mark options={solid}]
  table[row sep=crcr]{%
100	0.944511647852611\\
95	0.941632655391427\\
90	0.937108499743883\\
85	0.928583172860079\\
80	0.913107236853611\\
75	0.897977637119233\\
70	0.882576507888856\\
65	0.865496268420539\\
60	0.853450482879527\\
55	0.841713837757832\\
50	0.226770392612176\\
};
\addlegendentry{M3};

\end{axis}
\end{tikzpicture}%

%% file: figures/systm_ssim_gaussian.tex
%
%
%
\begin{tikzpicture}

\begin{axis}[%
xmin=0,
xmax=40,
xlabel={Variance},
ymin=0,
ymax=1.18,
ylabel={SSIM},
label style={font=\footnotesize}
]
\addplot [color=red,solid,mark=asterisk,mark options={solid}]
  table[row sep=crcr]{%
1	0.920208629335235\\
3	0.915956590530038\\
5	0.91153532087081\\
7	0.907278258737621\\
9	0.901266852008899\\
11	0.89261415078592\\
13	0.875943166559692\\
15	0.845597060995863\\
17	0.79707051824384\\
19	0.729009777567836\\
21	0.651435070335588\\
23	0.556827838501695\\
25	0.48043943618332\\
27	0.418254211110794\\
29	0.373458924360365\\
31	0.333872100530823\\
33	0.321021716986618\\
35	0.305127725741069\\
37	0.292089694711404\\
39	0.280383604831234\\
};
\addlegendentry{M1};

\addplot [color=cyan,solid,mark=o,mark options={solid}]
  table[row sep=crcr]{%
1	0.90929754359964\\
3	0.856167416385648\\
5	0.668664250172454\\
7	0.384852723243871\\
9	0.219715928617528\\
11	0.158778496307701\\
13	0.135053095800848\\
15	0.121205479138389\\
17	0.111972534278997\\
19	0.10445152931438\\
21	0.0985196643400427\\
23	0.092515666154357\\
25	0.088866929430442\\
27	0.0843574366758839\\
29	0.0809769094805253\\
31	0.0776693756008427\\
33	0.073825759750758\\
35	0.0714321400589832\\
37	0.0707301628483092\\
39	0.067265716405973\\
};
\addlegendentry{M2};

\addplot [color=blue,solid,mark=square,mark options={solid}]
  table[row sep=crcr]{%
1	0.94582712658732\\
3	0.933218687145931\\
5	0.924798103776428\\
7	0.91781676428078\\
9	0.91145508570451\\
11	0.905492585980061\\
13	0.899430687560586\\
15	0.894431474806902\\
17	0.888577108180244\\
19	0.88324675399925\\
21	0.876718489380931\\
23	0.87047143765098\\
25	0.862926481198907\\
27	0.853597512118402\\
29	0.837926025249799\\
31	0.819833891927906\\
33	0.793276441096474\\
35	0.765159988268897\\
37	0.72342117939751\\
39	0.672419281050598\\
};
\addlegendentry{M3};

\end{axis}
\end{tikzpicture}%

%% file: figures/systm_ssim_noise.tex
%
%
%
\begin{tikzpicture}

\begin{axis}[%
xmin=0,
xmax=240,
xtick={0,  40,  80, 120, 160, 200, 240},
xlabel={Variance},
ymin=0,
ymax=1.18,
ylabel={SSIM},
label style={font=\footnotesize}
]
\addplot [color=red,solid,mark=asterisk,mark options={solid}]
  table[row sep=crcr]{%
1	0.920208629335235\\
10	0.914778925488238\\
20	0.908091510296376\\
30	0.891259005548726\\
40	0.859356682711712\\
50	0.820146100051858\\
60	0.780356755769678\\
70	0.742092570309121\\
80	0.706438251827823\\
90	0.673931372636865\\
100	0.646678348276033\\
110	0.620258994554421\\
120	0.598273073760371\\
130	0.578021268209399\\
140	0.557825078222447\\
150	0.540678753090938\\
160	0.525458435458994\\
170	0.507908475699441\\
180	0.494539969045281\\
190	0.482595963933772\\
200	0.468127167345797\\
210	0.456605642443293\\
220	0.445287455126395\\
230	0.436402347647747\\
240	0.425770953420442\\
};
\addlegendentry{M1};

\addplot [color=cyan,solid,mark=o,mark options={solid}]
  table[row sep=crcr]{%
1	0.90929754359964\\
10	0.807310036259784\\
20	0.630357786032837\\
30	0.497496354260403\\
40	0.403250383319152\\
50	0.340768936301119\\
60	0.293562621846506\\
70	0.256479149806523\\
80	0.228435865407204\\
90	0.206655920760277\\
100	0.190161046393873\\
110	0.17545528898674\\
120	0.162597105242726\\
130	0.152485710117397\\
140	0.143143958424232\\
150	0.136106845123546\\
160	0.129786905980354\\
170	0.124278495117079\\
180	0.118263545537142\\
190	0.114225931261064\\
200	0.109681421511566\\
210	0.106034235036629\\
220	0.101984181097205\\
230	0.0991621644050617\\
240	0.0959726691526729\\
};
\addlegendentry{M2};

\addplot [color=blue,solid,mark=square,mark options={solid}]
  table[row sep=crcr]{%
1	0.94582712658732\\
10	0.934606693663721\\
20	0.925767059962886\\
30	0.919062490664204\\
40	0.91149740731427\\
50	0.903173145700287\\
60	0.892647071148417\\
70	0.876629119390233\\
80	0.857806209156814\\
90	0.838462647873345\\
100	0.817660798815548\\
110	0.794284204657178\\
120	0.772291625166807\\
130	0.749700308089432\\
140	0.729828063273227\\
150	0.711774912336366\\
160	0.692242525640315\\
170	0.676495538562403\\
180	0.658605434085092\\
190	0.641463183733253\\
200	0.626095603010117\\
210	0.612154774346347\\
220	0.596869569283339\\
230	0.584031068196017\\
240	0.571136353068381\\
};
\addlegendentry{M3};

\end{axis}
\end{tikzpicture}%

%% file: figures/bpp.tex
%
%
\definecolor{mycolor1}{rgb}{0.00000,0.75000,0.75000}%
\definecolor{mycolor2}{rgb}{0.75000,0.00000,0.75000}%
\definecolor{mycolor3}{rgb}{0.75000,0.75000,0.00000}%
\begin{tikzpicture}

\begin{axis}[%
xmin=0,
xmax=8,
xlabel=bpp,
ymin=0,
ymax=40,
ylabel=PSNR,
label style={font=\footnotesize}
]
\addplot [color=blue,solid,forget plot]
  table[row sep=crcr]{%
4.10994466145833	25.5302665966855\\
2.781982421875	25.5302665966855\\
2.20263671875	25.5302665966855\\
1.7744140625	25.5302665966855\\
1.66162109375	25.5302665966855\\
1.550048828125	25.5302665966855\\
1.47599283854167	25.5302665966855\\
1.06062825520833	25.5302665966855\\
1.00301106770833	25.5302665966855\\
0.966796875	25.5302665966855\\
0.944498697916667	7.00405957719516\\
};
\addplot [color=black!50!green,solid,forget plot]
  table[row sep=crcr]{%
2.69775390625	20.2395839628032\\
1.62459309895833	20.2395839628032\\
1.259765625	20.2395839628032\\
1.02156575520833	20.2395839628032\\
0.962809244791667	20.2395839628032\\
0.90087890625	20.2395839628032\\
0.850423177083333	20.2395839628032\\
0.630045572916667	20.2395839628032\\
0.583251953125	18.8617835349837\\
0.561848958333333	17.3156648294107\\
0.550211588541667	8.96567942990101\\
};
\addplot [color=red,solid,forget plot]
  table[row sep=crcr]{%
7.4727783203125	16.6676112683185\\
4.8145751953125	16.6676112683185\\
3.5267333984375	16.6676112683185\\
2.8497314453125	16.6676112683185\\
2.4522705078125	16.6676112683185\\
2.15869140625	16.6676112683185\\
1.9732666015625	16.6676112683185\\
1.8087158203125	16.6676948126449\\
1.67724609375	16.6676112683185\\
1.5693359375	16.6563317140119\\
1.482421875	10.4018965245451\\
};
\addplot [color=mycolor1,solid,forget plot]
  table[row sep=crcr]{%
7.14279174804688	17.6709693414059\\
4.54971313476563	17.6709693414059\\
3.29440307617188	17.6709693414059\\
2.62933349609375	17.6709693414059\\
2.23056030273438	17.6709693414059\\
1.94570922851563	17.6709693414059\\
1.76449584960938	17.6709693414059\\
1.6104736328125	17.6709693414059\\
1.48361206054688	17.6709693414059\\
1.38601684570313	17.6637869355917\\
1.3077392578125	7.48570494626518\\
};
\addplot [color=mycolor2,solid,forget plot]
  table[row sep=crcr]{%
7.8758544921875	15.4979892463049\\
5.05145263671875	15.4979892463049\\
3.75637817382813	15.4979892463049\\
3.03271484375	15.4979892463049\\
2.58480834960938	15.4979892463049\\
2.25430297851563	15.4979892463049\\
2.04141235351563	15.4979892463049\\
1.85488891601563	15.4979892463049\\
1.70684814453125	15.4943329320511\\
1.58462524414063	15.4833237835625\\
1.4886474609375	7.59905267868709\\
};
\addplot [color=mycolor3,solid,forget plot]
  table[row sep=crcr]{%
4.19009399414063	35.0204429646208\\
2.05618286132813	35.0204429646208\\
1.27157592773438	35.0204429646208\\
0.946533203125	35.0204429646208\\
0.760284423828125	35.0204429646208\\
0.63336181640625	35.0204429646208\\
0.553924560546875	35.0204429646208\\
0.483154296875	35.0204429646208\\
0.428863525390625	35.0204429646208\\
0.387908935546875	34.0180676194333\\
0.3587646484375	9.4081701192266\\
};
\addplot [color=darkgray,solid,forget plot]
  table[row sep=crcr]{%
5.39071655273438	20.6012730575689\\
3.17010498046875	20.6012730575689\\
2.08731079101563	20.6012730575689\\
1.6060791015625	20.6012730575689\\
1.34609985351563	20.6012730575689\\
1.17041015625	20.6012730575689\\
1.06304931640625	20.6012730575689\\
0.966217041015625	20.6012730575689\\
0.883697509765625	20.6012730575689\\
0.81610107421875	20.6012730575689\\
0.7630615234375	10.3379674187264\\
};
\addplot [color=blue,solid,forget plot]
  table[row sep=crcr]{%
6.2894287109375	16.7161293694943\\
3.9052734375	16.7161293694943\\
2.7982177734375	16.7161293694943\\
2.2672119140625	16.7161293694943\\
1.9459228515625	16.7161293694943\\
1.7200927734375	16.7161293694943\\
1.572509765625	16.7161293694943\\
1.4381103515625	16.7161293694943\\
1.326904296875	16.7161293694943\\
1.23779296875	16.7161293694943\\
1.1700439453125	9.44806104629885\\
};
\addplot [color=black!50!green,solid,forget plot]
  table[row sep=crcr]{%
4.50439453125	18.5134879936184\\
2.43701171875	18.5134879936184\\
1.6590576171875	18.5134879936184\\
1.3096923828125	18.5134879936184\\
1.12451171875	18.5134879936184\\
0.9859619140625	18.5134879936184\\
0.893798828125	18.5134879936184\\
0.818359375	18.5134879936184\\
0.7586669921875	18.5134879936184\\
0.709228515625	18.5134879936184\\
0.67138671875	6.82595464317185\\
};
\addplot [color=red,solid,forget plot]
  table[row sep=crcr]{%
5.67608642578125	20.9976039782178\\
3.36434936523438	20.9976039782178\\
2.28604125976563	20.9976039782178\\
1.80160522460938	20.9976039782178\\
1.5302734375	20.9976039782178\\
1.33743286132813	20.9976039782178\\
1.21380615234375	20.9976039782178\\
1.10687255859375	20.9976039782178\\
1.01971435546875	20.9976039782178\\
0.946929931640625	20.0877413455175\\
0.890899658203125	8.56932602555379\\
};
\addplot [color=mycolor1,solid,forget plot]
  table[row sep=crcr]{%
3.9085693359375	18.8712870526655\\
1.8057861328125	18.8712870526655\\
1.1431884765625	18.8712870526655\\
0.8543701171875	18.8712870526655\\
0.703125	18.4959897892139\\
0.617919921875	18.3743741466825\\
0.5648193359375	18.9443978523952\\
0.5213623046875	18.5878169295719\\
0.48193359375	17.0643444382563\\
0.454833984375	17.856551162674\\
0.43310546875	9.67680449195126\\
};
\addplot [color=mycolor2,solid,forget plot]
  table[row sep=crcr]{%
4.6619873046875	19.8590791381055\\
2.5419921875	19.8590791381055\\
1.7431640625	19.8590791381055\\
1.3814697265625	19.8590791381055\\
1.1861572265625	19.8590791381055\\
1.04296875	19.8590791381055\\
0.94970703125	19.8590791381055\\
0.8724365234375	19.8590791381055\\
0.8106689453125	19.8590791381055\\
0.7598876953125	19.8590791381055\\
0.7117919921875	7.92698852464617\\
};
\addplot [color=mycolor3,solid,forget plot]
  table[row sep=crcr]{%
5.66241455078125	19.2774202375024\\
3.38555908203125	19.2774202375024\\
2.3150634765625	19.2774202375024\\
1.71917724609375	19.2774202375024\\
1.352783203125	19.2774202375024\\
1.1248779296875	19.2774202375024\\
0.994354248046875	19.2774202375024\\
0.885498046875	19.2774202375024\\
0.794403076171875	19.2774202375024\\
0.72314453125	19.2774202375024\\
0.668243408203125	8.28052812708242\\
};
\addplot [color=darkgray,solid,forget plot]
  table[row sep=crcr]{%
5.12689208984375	18.4542796199268\\
2.9600830078125	18.4542796199268\\
2.08154296875	18.4542796199268\\
1.66998291015625	18.4542796199268\\
1.43182373046875	18.4542796199268\\
1.26174926757813	18.4542796199268\\
1.15328979492188	18.4542796199268\\
1.0584716796875	18.4542796199268\\
0.979461669921875	18.4542796199268\\
0.915618896484375	18.4542796199268\\
0.8636474609375	9.9357723725828\\
};
\addplot [color=blue,solid,forget plot]
  table[row sep=crcr]{%
4.707275390625	18.9748928824062\\
2.5750732421875	18.9748928824062\\
1.7598876953125	18.9748928824062\\
1.3958740234375	18.9748928824062\\
1.1878662109375	18.9748928824062\\
1.03662109375	18.9748928824062\\
0.94384765625	18.9748928824062\\
0.8590087890625	18.9748928824062\\
0.791748046875	18.9748928824062\\
0.7275390625	18.9748928824062\\
0.6827392578125	9.18015334839911\\
};
\addplot [color=black!50!green,solid,forget plot]
  table[row sep=crcr]{%
5.60391235351563	26.5541864767879\\
3.3133544921875	26.5541864767879\\
2.22393798828125	26.5541864767879\\
1.70468139648438	26.5541864767879\\
1.42523193359375	26.5541864767879\\
1.23812866210938	26.5541864767879\\
1.119384765625	26.5541864767879\\
1.0142822265625	26.5516638213386\\
0.9305419921875	26.5541864767879\\
0.8631591796875	26.5541864767879\\
0.810546875	9.85136814189297\\
};
\addplot [color=red,solid,forget plot]
  table[row sep=crcr]{%
3.11865234375	17.0954398443437\\
1.4486083984375	17.0954398443437\\
0.9678955078125	17.0954398443437\\
0.7669677734375	17.0954398443437\\
0.6651611328125	17.0954398443437\\
0.5928955078125	17.0954398443437\\
0.55078125	17.0954398443437\\
0.5135498046875	17.0954398443437\\
0.480712890625	17.0954398443437\\
0.4566650390625	17.0954398443437\\
0.43701171875	8.28232390887578\\
};
\addplot [color=mycolor1,solid,forget plot]
  table[row sep=crcr]{%
6.044677734375	15.6191562419374\\
3.69635009765625	15.6191562419374\\
2.55661010742188	15.6191562419374\\
1.9364013671875	15.6191562419374\\
1.59286499023438	15.6191562419374\\
1.37075805664063	15.6191562419374\\
1.23629760742188	15.6191562419374\\
1.12265014648438	15.6191562419374\\
1.0308837890625	15.6191562419374\\
0.955810546875	15.6191562419374\\
0.8970947265625	7.0386658172622\\
};
\addplot [color=mycolor2,solid,forget plot]
  table[row sep=crcr]{%
5.53988647460938	20.7365747601608\\
3.23684692382813	20.7365747601608\\
2.24221801757813	20.7365747601608\\
1.76992797851563	20.7365747601608\\
1.50103759765625	20.7365747601608\\
1.3094482421875	20.7365747601608\\
1.18560791015625	20.7365747601608\\
1.07809448242188	20.7365747601608\\
0.986724853515625	20.7365747601608\\
0.914337158203125	20.7365747601608\\
0.857696533203125	8.53650963123199\\
};
\addplot [color=mycolor3,solid,forget plot]
  table[row sep=crcr]{%
6.053466796875	24.573328334415\\
3.6839599609375	24.573328334415\\
2.530517578125	24.573328334415\\
1.93408203125	24.573328334415\\
1.5970458984375	24.573328334415\\
1.3536376953125	24.573328334415\\
1.200927734375	24.573328334415\\
1.072021484375	24.573328334415\\
0.9676513671875	24.573328334415\\
0.8878173828125	24.573328334415\\
0.8199462890625	10.6898779391465\\
};
\addplot [color=darkgray,solid,forget plot]
  table[row sep=crcr]{%
4.12124633789063	28.5776426970146\\
2.05706787109375	28.5776426970146\\
1.34033203125	28.55423582674\\
1.01666259765625	27.5085555461317\\
0.834808349609375	28.55423582674\\
0.7139892578125	27.7344701416764\\
0.641387939453125	27.3604856569602\\
0.5810546875	17.336009225026\\
0.53167724609375	24.9779729751745\\
0.49542236328125	19.2728119944796\\
0.46868896484375	7.69473899438354\\
};
\addplot [color=blue,solid,forget plot]
  table[row sep=crcr]{%
5.4154052734375	18.5782535979742\\
3.311767578125	18.4588666197503\\
2.3560791015625	18.3041337625011\\
1.8895263671875	18.4070991192223\\
1.61572265625	17.5391688595798\\
1.4271240234375	17.8163058460296\\
1.29833984375	15.7762766454578\\
1.1878662109375	15.4191216220271\\
1.09423828125	15.3661517344866\\
1.0257568359375	14.0271337784257\\
0.9688720703125	8.86512241695288\\
};
\addplot [color=black!50!green,solid,forget plot]
  table[row sep=crcr]{%
7.050048828125	17.9976693496074\\
4.39288330078125	17.9976693496074\\
3.19952392578125	17.9976693496074\\
2.57891845703125	17.9976693496074\\
2.19686889648438	17.9976693496074\\
1.92025756835938	17.9976693496074\\
1.736572265625	17.9976693496074\\
1.57891845703125	17.9976693496074\\
1.44760131835938	17.9976693496074\\
1.343994140625	17.9976693496074\\
1.26089477539063	7.62993607333508\\
};
\addplot [color=red,solid,forget plot]
  table[row sep=crcr]{%
5.69192504882813	26.1230443114606\\
3.39627075195313	26.1230443114606\\
2.2713623046875	26.1230443114606\\
1.72720336914063	26.1230443114606\\
1.430419921875	26.1230443114606\\
1.22970581054688	26.1230443114606\\
1.107421875	26.1230443114606\\
0.998138427734375	26.1230443114606\\
0.906494140625	26.1230443114606\\
0.833587646484375	26.1230443114606\\
0.77642822265625	11.3207419976416\\
};
\addplot [color=mycolor1,solid,forget plot]
  table[row sep=crcr]{%
5.49386596679688	23.8400939934652\\
3.23101806640625	23.8400939934652\\
2.15841674804688	23.8400939934652\\
1.66900634765625	23.8400939934652\\
1.39633178710938	23.8400939934652\\
1.21353149414063	23.8400939934652\\
1.09893798828125	23.8400939934652\\
0.995452880859375	23.8400939934652\\
0.910400390625	23.8400939934652\\
0.84161376953125	23.8400939934652\\
0.789306640625	8.73904056107739\\
};
\addplot [color=mycolor2,solid,forget plot]
  table[row sep=crcr]{%
3.88473510742188	29.5459787191624\\
1.9310302734375	29.5459787191624\\
1.24423217773438	29.5459787191624\\
0.94732666015625	29.5459787191624\\
0.791046142578125	29.5459787191624\\
0.678558349609375	29.5459787191624\\
0.611419677734375	29.5460630777896\\
0.55340576171875	29.5460630777896\\
0.504425048828125	29.5460630777896\\
0.46649169921875	29.3120805700138\\
0.4375	7.8739220059736\\
};
\addplot [color=mycolor3,solid,forget plot]
  table[row sep=crcr]{%
7.114501953125	16.5134676057602\\
4.49072265625	16.5134676057602\\
3.2666015625	16.5134676057602\\
2.6448974609375	16.5134676057602\\
2.27001953125	16.5044511710819\\
2.002685546875	16.5134972802197\\
1.82861328125	16.3977946278205\\
1.67138671875	16.4693757808955\\
1.5445556640625	15.9753984053591\\
1.4390869140625	15.0906839198617\\
1.354736328125	8.05795581667725\\
};
\addplot [color=darkgray,solid,forget plot]
  table[row sep=crcr]{%
4.001708984375	21.5298991401982\\
1.998046875	21.5298991401982\\
1.3033447265625	21.5298991401982\\
1.0181884765625	21.5298991401982\\
0.8525390625	21.5298991401982\\
0.74462890625	21.5193365477649\\
0.6734619140625	21.5298991401982\\
0.6141357421875	20.681342860776\\
0.5662841796875	21.5033037799829\\
0.526611328125	21.3165764351859\\
0.5	7.32164867660775\\
};
\addplot [color=blue,solid,forget plot]
  table[row sep=crcr]{%
7.14642333984375	19.9462962613952\\
4.45529174804688	19.9462962613952\\
3.23089599609375	19.9462962613952\\
2.58731079101563	19.9462962613952\\
2.20367431640625	19.9462962613952\\
1.92654418945313	19.9462962613952\\
1.74447631835938	19.9462962613952\\
1.58743286132813	19.9462962613952\\
1.46002197265625	19.9462962613952\\
1.3526611328125	19.9462962613952\\
1.26824951171875	9.26354668835658\\
};
\addplot [color=black!50!green,solid,forget plot]
  table[row sep=crcr]{%
5.57232666015625	19.0523239875503\\
3.24664306640625	19.0523239875503\\
2.25607299804688	19.0523239875503\\
1.805908203125	19.0523239875503\\
1.55233764648438	19.0523239875503\\
1.36898803710938	19.0523239875503\\
1.25128173828125	19.0523239875503\\
1.14898681640625	19.0523239875503\\
1.0621337890625	19.0523239875503\\
0.99371337890625	19.0523239875503\\
0.937744140625	7.35518194801541\\
};
\addplot [color=red,solid,forget plot]
  table[row sep=crcr]{%
6.37809244791667	14.9112644995836\\
4.384521484375	14.9112644995836\\
3.52921549479167	14.9112644995836\\
2.868896484375	14.9112644995836\\
2.67789713541667	14.9112644995836\\
2.48958333333333	14.9112644995836\\
2.38460286458333	14.9112644995836\\
1.76399739583333	14.914520193557\\
1.67862955729167	14.9080064808959\\
1.60896809895833	14.841491601068\\
1.5703125	8.35981778168099\\
};
\addplot [color=mycolor1,solid,forget plot]
  table[row sep=crcr]{%
4.194580078125	16.1426205452711\\
2.3673095703125	16.1426205452711\\
1.6795654296875	16.1426205452711\\
1.3529052734375	16.1426205452711\\
1.174072265625	16.1426205452711\\
1.0450439453125	16.1426205452711\\
0.965087890625	16.1426205452711\\
0.8880615234375	16.1433438527545\\
0.822509765625	15.8911344004115\\
0.7698974609375	15.6133444184955\\
0.7283935546875	9.58722360586609\\
};
\addplot [color=mycolor2,solid,forget plot]
  table[row sep=crcr]{%
5.068603515625	22.0606323393603\\
2.8839111328125	22.0606323393603\\
2.0115966796875	22.0606323393603\\
1.5938720703125	22.0606323393603\\
1.3583984375	22.0606323393603\\
1.1864013671875	22.0606323393603\\
1.0740966796875	22.0606323393603\\
0.9805908203125	22.0601743564086\\
0.899658203125	22.0109613697294\\
0.842041015625	22.0314514692398\\
0.7928466796875	8.00529631741537\\
};
\addplot [color=mycolor3,solid,forget plot]
  table[row sep=crcr]{%
5.65567016601563	24.8754237295348\\
3.41058349609375	24.8754237295348\\
2.35073852539063	24.8754237295348\\
1.81417846679688	24.8754237295348\\
1.49725341796875	24.8754237295348\\
1.27920532226563	24.8754237295348\\
1.144775390625	24.8754237295348\\
1.03518676757813	24.8754237295348\\
0.947967529296875	24.8754237295348\\
0.87890625	24.8754237295348\\
0.82470703125	10.139071354531\\
};
\addplot [color=darkgray,solid,forget plot]
  table[row sep=crcr]{%
7.050048828125	17.9976693496074\\
4.39288330078125	17.9976693496074\\
3.19952392578125	17.9976693496074\\
2.57891845703125	17.9976693496074\\
2.19686889648438	17.9976693496074\\
1.92025756835938	17.9976693496074\\
1.736572265625	17.9976693496074\\
1.57891845703125	17.9976693496074\\
1.44760131835938	17.9976693496074\\
1.343994140625	17.9976693496074\\
1.26089477539063	7.62993607333508\\
};
\addplot [color=blue,solid,forget plot]
  table[row sep=crcr]{%
4.7584228515625	16.4377364644156\\
2.6219482421875	16.4377364644156\\
1.8372802734375	16.4377364644156\\
1.4661865234375	16.4377364644156\\
1.2498779296875	16.4377364644156\\
1.0933837890625	16.4377364644156\\
0.998291015625	16.4377364644156\\
0.918701171875	16.4377364644156\\
0.85302734375	16.4377364644156\\
0.798583984375	16.4369812795365\\
0.755859375	8.42839697706227\\
};
\addplot [color=black!50!green,solid,forget plot]
  table[row sep=crcr]{%
3.85693359375	18.1064395337908\\
2.47176106770833	18.1064395337908\\
1.81217447916667	18.1064395337908\\
1.60196940104167	18.1064395337908\\
1.35595703125	18.1064395337908\\
1.06388346354167	18.1064395337908\\
1.00724283854167	18.1064395337908\\
0.963297526041667	18.1064395337908\\
0.870279947916667	18.1064395337908\\
0.776204427083333	18.099200673069\\
0.74853515625	9.05606108686618\\
};
\addplot [color=red,solid,forget plot]
  table[row sep=crcr]{%
4.97965494791667	14.9321830601688\\
3.24357096354167	14.9321830601688\\
2.59806315104167	14.9321830601688\\
2.09090169270833	14.9321830601688\\
1.96590169270833	14.9321830601688\\
1.83040364583333	14.9321830601688\\
1.751953125	14.9321830601688\\
1.23551432291667	14.9251922129214\\
1.17529296875	14.8341667876098\\
1.12809244791667	13.975562460917\\
1.10319010416667	8.05135783594777\\
};
\addplot [color=mycolor1,solid,forget plot]
  table[row sep=crcr]{%
5.49845377604167	16.032812454986\\
3.62337239583333	16.032812454986\\
2.685302734375	16.032812454986\\
2.36507161458333	16.032812454986\\
2.03190104166667	16.0320759322704\\
1.63020833333333	16.0110379144893\\
1.54288736979167	15.9098838050831\\
1.47314453125	14.9267910783601\\
1.33072916666667	14.2681998210226\\
1.19189453125	12.9286036926699\\
1.15673828125	6.61482867978417\\
};
\addplot [color=mycolor2,solid,forget plot]
  table[row sep=crcr]{%
7.05802408854167	15.1459543364446\\
4.51114908854167	15.1420887972392\\
3.65234375	15.1420887972392\\
2.95865885416667	15.1406136401501\\
2.777099609375	15.1406136401501\\
2.56917317708333	15.1493600066258\\
2.46280924479167	15.0256950632062\\
1.79451497395833	15.139064936646\\
1.71687825520833	14.058013765838\\
1.64290364583333	14.1764211000889\\
1.60725911458333	7.88965469030958\\
};
\addplot [color=mycolor3,solid,forget plot]
  table[row sep=crcr]{%
5.5189208984375	13.9347310840832\\
3.1605224609375	13.9347310840832\\
2.228271484375	13.9347310840832\\
1.7880859375	13.9347310840832\\
1.5252685546875	13.9347310840832\\
1.3397216796875	13.9347310840832\\
1.2283935546875	13.9347310840832\\
1.1201171875	13.9347310840832\\
1.0333251953125	13.9347310840832\\
0.9619140625	13.9343540267833\\
0.907470703125	8.19362144970878\\
};
\addplot [color=darkgray,solid,forget plot]
  table[row sep=crcr]{%
4.53458658854167	15.0525392019821\\
2.99153645833333	15.0003124921781\\
2.20939127604167	15.0003124921781\\
1.9501953125	14.9936771631603\\
1.66072591145833	14.9496244639654\\
1.32234700520833	14.996003067266\\
1.25651041666667	14.8902015803419\\
1.200927734375	12.8279813780157\\
1.093505859375	11.8982513599776\\
0.978515625	11.1431327478376\\
0.948160807291667	6.27227450842164\\
};
\addplot [color=blue,solid,forget plot]
  table[row sep=crcr]{%
6.40022786458333	14.4494156563233\\
4.30135091145833	14.4494156563233\\
3.20263671875	14.4494156563233\\
2.82373046875	14.4494156563233\\
2.431884765625	14.4486175713877\\
1.94474283854167	14.4505892920394\\
1.84749348958333	14.4462381625453\\
1.76920572916667	14.4494648543099\\
1.60782877604167	14.248267613931\\
1.4228515625	14.1557629216917\\
1.37947591145833	9.00708367656631\\
};
\addplot [color=black!50!green,solid,forget plot]
  table[row sep=crcr]{%
5.23453776041667	15.9546788940072\\
3.51692708333333	15.9546788940072\\
2.59781901041667	15.9546788940072\\
2.294677734375	15.9546788940072\\
1.96435546875	15.9546788940072\\
1.53971354166667	15.9546788940072\\
1.46468098958333	15.9546788940072\\
1.400634765625	15.9546788940072\\
1.26717122395833	15.9546788940072\\
1.12540690104167	15.9546788940072\\
1.09025065104167	8.77811729146347\\
};
\addplot [color=red,solid,forget plot]
  table[row sep=crcr]{%
4.632080078125	15.6274976074504\\
3.106689453125	15.6274976074504\\
2.30777994791667	15.6274976074504\\
2.04549153645833	15.6274976074504\\
1.76847330729167	15.6274976074504\\
1.40511067708333	15.6274976074504\\
1.33203125	15.6274976074504\\
1.27840169270833	15.6274976074504\\
1.15543619791667	15.290517443192\\
1.03751627604167	15.5035267677442\\
1.00382486979167	8.653234081994\\
};
\addplot [color=mycolor1,solid,forget plot]
  table[row sep=crcr]{%
6.04915364583333	17.7285985505885\\
4.02848307291667	17.7285985505885\\
3.26310221354167	17.7285985505885\\
2.59733072916667	17.7285985505885\\
2.44466145833333	17.7285985505885\\
2.26969401041667	17.7285985505885\\
2.18961588541667	17.7285985505885\\
1.52791341145833	17.7335705170905\\
1.46223958333333	17.7285985505885\\
1.39949544270833	17.733438328475\\
1.36832682291667	8.84902905133318\\
};
\addplot [color=mycolor2,solid,forget plot]
  table[row sep=crcr]{%
5.5927734375	17.1877493115885\\
3.57552083333333	17.1877493115885\\
2.62947591145833	17.1877493115885\\
2.321044921875	17.1877493115885\\
1.99259440104167	17.1877493115885\\
1.52994791666667	17.1877493115885\\
1.45654296875	17.1877493115885\\
1.3984375	17.1877493115885\\
1.25138346354167	17.1877493115885\\
1.08976236979167	17.1877493115885\\
1.05647786458333	10.8066376464585\\
};
\addplot [color=mycolor3,solid,forget plot]
  table[row sep=crcr]{%
5.65567016601563	24.8754237295348\\
3.41058349609375	24.8754237295348\\
2.35073852539063	24.8754237295348\\
1.81417846679688	24.8754237295348\\
1.49725341796875	24.8754237295348\\
1.27920532226563	24.8754237295348\\
1.144775390625	24.8754237295348\\
1.03518676757813	24.8754237295348\\
0.947967529296875	24.8754237295348\\
0.87890625	24.8754237295348\\
0.82470703125	10.139071354531\\
};
\addplot [color=darkgray,solid,forget plot]
  table[row sep=crcr]{%
6.68758138020833	16.1503439726505\\
4.46126302083333	16.1503439726505\\
3.61669921875	16.1503439726505\\
2.90478515625	16.1503439726505\\
2.72574869791667	16.1503439726505\\
2.537109375	16.1503798135515\\
2.43611653645833	16.1484134123058\\
1.68196614583333	16.171647221606\\
1.611328125	16.1482177580561\\
1.545654296875	15.9331995886655\\
1.51334635416667	8.66452821213352\\
};
\addplot [color=blue,solid,forget plot]
  table[row sep=crcr]{%
5.42203776041667	15.0546363522796\\
3.70784505208333	15.0546363522796\\
2.96671549479167	15.0546363522796\\
2.41129557291667	15.0546363522796\\
2.2509765625	15.0546363522796\\
2.09261067708333	15.0546363522796\\
1.99300130208333	15.0546363522796\\
1.50382486979167	15.0546363522796\\
1.42561848958333	15.0546363522796\\
1.366943359375	15.0546363522796\\
1.333984375	7.34887714642556\\
};
\addplot [color=black!50!green,solid,forget plot]
  table[row sep=crcr]{%
8.96500651041667	15.0741414663445\\
5.82706705729167	15.0741414663445\\
4.73600260416667	15.0741414663445\\
3.86564127604167	15.0741414663445\\
3.61564127604167	15.0741414663445\\
3.33968098958333	15.0207617885243\\
3.20979817708333	15.0661293187206\\
2.36865234375	14.9895304764293\\
2.27392578125	14.8363472686533\\
2.17521158854167	13.5334656994676\\
2.12548828125	8.90504893944778\\
};
\addplot [color=red,solid,forget plot]
  table[row sep=crcr]{%
6.05305989583333	19.1448734421782\\
4.07625325520833	19.1448734421782\\
2.98347981770833	19.1448734421782\\
2.64737955729167	19.1448734421782\\
2.2109375	19.1446042457195\\
1.65568033854167	19.1407927866513\\
1.57552083333333	19.1297052730387\\
1.51310221354167	19.149177356947\\
1.35148111979167	19.124052368819\\
1.18001302083333	19.0753268510479\\
1.14493815104167	9.25172446672152\\
};
\addplot [color=mycolor1,solid,forget plot]
  table[row sep=crcr]{%
5.25008138020833	16.282637390754\\
3.525634765625	16.282637390754\\
2.60042317708333	16.282637390754\\
2.29939778645833	16.282637390754\\
1.97086588541667	16.282637390754\\
1.56632486979167	16.282637390754\\
1.484619140625	16.282637390754\\
1.42244466145833	16.282637390754\\
1.28466796875	16.2634822496942\\
1.14737955729167	15.3982348244176\\
1.10904947916667	8.48823828985546\\
};
\addplot [color=mycolor2,solid,forget plot]
  table[row sep=crcr]{%
5.4521484375	19.5712582635809\\
3.2203369140625	19.5712582635809\\
2.1351318359375	19.5712582635809\\
1.63134765625	19.5712582635809\\
1.3616943359375	19.5712582635809\\
1.17822265625	19.5712582635809\\
1.07080078125	19.5712582635809\\
0.973388671875	19.5648041378134\\
0.890380859375	19.5712582635809\\
0.823486328125	19.6247106156589\\
0.7755126953125	9.2295575175901\\
};
\addplot [color=mycolor3,solid,forget plot]
  table[row sep=crcr]{%
5.65652465820313	22.9572992997215\\
3.36016845703125	22.9572992997215\\
2.26486206054688	22.9572992997215\\
1.75592041015625	22.9572992997215\\
1.47842407226563	22.9572992997215\\
1.28775024414063	22.9572992997215\\
1.16494750976563	22.9572992997215\\
1.05889892578125	22.9572992997215\\
0.97296142578125	22.9572992997215\\
0.903045654296875	22.9572992997215\\
0.845672607421875	9.87125716557888\\
};
\addplot [color=darkgray,solid,forget plot]
  table[row sep=crcr]{%
5.048583984375	16.1550537949572\\
3.32120768229167	16.1550537949572\\
2.46061197916667	16.1550537949572\\
2.17635091145833	16.1550537949572\\
1.86539713541667	16.1550537949572\\
1.47127278645833	16.1550537949572\\
1.39567057291667	16.1550537949572\\
1.33675130208333	16.1550537949572\\
1.20865885416667	16.1550537949572\\
1.07462565104167	16.1550537949572\\
1.036865234375	9.15356128900543\\
};
\addplot [color=blue,solid,forget plot]
  table[row sep=crcr]{%
6.636962890625	16.2860431794855\\
4.589111328125	16.2860431794855\\
3.40852864583333	16.2860431794855\\
3.010986328125	16.2860431794855\\
2.58536783854167	16.2860431794855\\
2.01603190104167	16.2860431794855\\
1.91715494791667	16.2860431794855\\
1.83675130208333	16.2860431794855\\
1.66552734375	16.2860431794855\\
1.48282877604167	16.2860431794855\\
1.43701171875	7.9712366817234\\
};
\addplot [color=black!50!green,solid,forget plot]
  table[row sep=crcr]{%
7.43123372395833	15.9218465894164\\
4.966552734375	15.9218465894164\\
3.68058268229167	15.9218465894164\\
3.24886067708333	15.9218465894164\\
2.78483072916667	15.9218465894164\\
2.18912760416667	15.9218465894164\\
2.07853190104167	15.9258760769001\\
1.98990885416667	15.9078600950239\\
1.787109375	15.9223381028505\\
1.57023111979167	15.7760280071001\\
1.523681640625	8.87569278411058\\
};
\addplot [color=red,solid,forget plot]
  table[row sep=crcr]{%
6.942626953125	18.5622289228742\\
4.59456380208333	18.5622289228742\\
3.380859375	18.5622289228742\\
2.98811848958333	18.5622289228742\\
2.557373046875	18.5622289228742\\
1.97184244791667	18.5595970117078\\
1.87467447916667	18.5622289228742\\
1.80045572916667	18.5622289228742\\
1.61726888020833	18.5622289228742\\
1.39705403645833	18.5639866597154\\
1.35709635416667	8.90551732584949\\
};
\addplot [color=mycolor1,solid,forget plot]
  table[row sep=crcr]{%
3.44197591145833	22.3432180316161\\
2.2109375	22.3432180316161\\
1.73958333333333	22.3432180316161\\
1.3779296875	22.3432180316161\\
1.30086263020833	22.3432180316161\\
1.21695963541667	22.3432180316161\\
1.16739908854167	22.3432180316161\\
0.795084635416667	22.3432180316161\\
0.7490234375	22.3432180316161\\
0.721761067708333	22.3432180316161\\
0.707926432291667	9.57277668218402\\
};
\addplot [color=mycolor2,solid,forget plot]
  table[row sep=crcr]{%
3.33040364583333	18.4647942699777\\
2.203125	18.4647942699777\\
1.72037760416667	18.4647942699777\\
1.4013671875	18.4647942699777\\
1.31062825520833	18.4647942699777\\
1.22542317708333	18.4647942699777\\
1.16560872395833	18.4647942699777\\
0.867268880208333	18.4647942699777\\
0.818277994791667	18.4647942699777\\
0.789957682291667	18.4647942699777\\
0.771321614583333	8.48496715189268\\
};
\addplot [color=mycolor3,solid,forget plot]
  table[row sep=crcr]{%
4.95953369140625	19.928826524509\\
2.78402709960938	19.928826524509\\
1.81146240234375	19.928826524509\\
1.38601684570313	19.928826524509\\
1.1593017578125	19.928826524509\\
0.99609375	19.928826524509\\
0.89569091796875	19.928826524509\\
0.808258056640625	19.928826524509\\
0.735870361328125	19.928826524509\\
0.682861328125	19.928826524509\\
0.640106201171875	9.72067454058619\\
};
\addplot [color=darkgray,solid,forget plot]
  table[row sep=crcr]{%
3.86311848958333	21.891730764698\\
2.35302734375	21.891730764698\\
1.854248046875	21.891730764698\\
1.49641927083333	21.891730764698\\
1.40348307291667	21.891730764698\\
1.30550130208333	21.891730764698\\
1.24007161458333	21.891730764698\\
0.900472005208333	21.891730764698\\
0.8505859375	21.891730764698\\
0.81494140625	21.891730764698\\
0.796712239583333	9.38594448818198\\
};
\addplot [color=blue,solid,forget plot]
  table[row sep=crcr]{%
5.62516276041667	16.9951400773241\\
3.74088541666667	16.9951400773241\\
3.017822265625	16.9951400773241\\
2.4267578125	16.9951400773241\\
2.27408854166667	16.9951400773241\\
2.112548828125	16.9951400773241\\
2.02376302083333	16.9951039084215\\
1.44278971354167	16.9951400773241\\
1.382568359375	16.9601847560677\\
1.325927734375	16.9877658452562\\
1.29671223958333	8.22565819704845\\
};
\addplot [color=black!50!green,solid,forget plot]
  table[row sep=crcr]{%
3.508056640625	14.0146527459545\\
2.25040690104167	14.0146527459545\\
1.785888671875	14.0146527459545\\
1.42903645833333	14.0146527459545\\
1.34342447916667	14.0146527459545\\
1.25309244791667	14.0146527459545\\
1.20426432291667	14.0107122646631\\
0.858235677083333	14.0107122646631\\
0.808024088541667	14.0129641805177\\
0.778564453125	14.0090567183194\\
0.760904947916667	8.93018685258015\\
};
\addplot [color=red,solid,forget plot]
  table[row sep=crcr]{%
6.74527994791667	12.690420211682\\
4.646484375	12.690420211682\\
3.74739583333333	12.690420211682\\
3.04728190104167	12.690420211682\\
2.84562174479167	12.690420211682\\
2.63525390625	12.6902539128504\\
2.54191080729167	12.6902873461101\\
1.91031901041667	12.6902873461101\\
1.82950846354167	12.689517949185\\
1.74755859375	12.6777797851217\\
1.7060546875	7.46537065622464\\
};
\addplot [color=mycolor1,solid,forget plot]
  table[row sep=crcr]{%
3.42130533854167	25.0942510114775\\
2.11824544270833	25.0942510114775\\
1.65315755208333	25.0942510114775\\
1.33992513020833	25.0461578214579\\
1.253173828125	25.0299582352782\\
1.17041015625	24.313285449057\\
1.10701497395833	21.1606577130969\\
0.837809244791667	19.4368267108727\\
0.776204427083333	19.7493501154837\\
0.747884114583333	14.2982663940042\\
0.730061848958333	8.7665918455801\\
};
\addplot [color=mycolor2,solid,forget plot]
  table[row sep=crcr]{%
3.7095947265625	20.4177280668799\\
1.9774169921875	20.4177280668799\\
1.38037109375	20.4177280668799\\
1.097412109375	20.4177280668799\\
0.9420166015625	20.4177280668799\\
0.82470703125	20.4177280668799\\
0.75146484375	20.4177280668799\\
0.685791015625	20.4177280668799\\
0.63330078125	20.4177280668799\\
0.5936279296875	20.4177280668799\\
0.5616455078125	9.93918462214044\\
};
\addplot [color=mycolor3,solid,forget plot]
  table[row sep=crcr]{%
6.07185872395833	17.2998043742145\\
4.04646809895833	17.2998043742145\\
3.0009765625	17.2998043742145\\
2.64925130208333	17.2998043742145\\
2.27880859375	17.2998043742145\\
1.78084309895833	17.2998043742145\\
1.692626953125	17.2998043742145\\
1.622802734375	17.2998043742145\\
1.46809895833333	17.2998043742145\\
1.28995768229167	17.2998043742145\\
1.25211588541667	7.31366924545169\\
};
\addplot [color=darkgray,solid,forget plot]
  table[row sep=crcr]{%
5.20857747395833	18.8031759648526\\
3.53889973958333	18.8031759648526\\
2.85734049479167	18.8031759648526\\
2.257080078125	18.8031759648526\\
2.12337239583333	18.8031759648526\\
1.97607421875	18.8031759648526\\
1.91251627604167	18.8031759648526\\
1.281982421875	18.8031759648526\\
1.22884114583333	18.8031759648526\\
1.17928059895833	18.8029164722989\\
1.15397135416667	8.67937010940019\\
};
\addplot [color=blue,solid,forget plot]
  table[row sep=crcr]{%
5.439208984375	18.7005512514929\\
3.60286458333333	18.7005512514929\\
2.92138671875	18.7005512514929\\
2.29850260416667	18.7005512514929\\
2.16666666666667	18.7005512514929\\
2.02490234375	18.7005512514929\\
1.956298828125	18.7005512514929\\
1.26009114583333	18.7005512514929\\
1.202880859375	18.7005512514929\\
1.155517578125	18.7005512514929\\
1.1318359375	9.03624680305898\\
};
\addplot [color=black!50!green,solid,forget plot]
  table[row sep=crcr]{%
2.95955403645833	26.5350702238309\\
1.79069010416667	26.5350702238309\\
1.393310546875	26.5350702238309\\
1.10701497395833	26.5350702238309\\
1.04484049479167	26.5350702238309\\
0.982259114583333	26.5350702238309\\
0.938557942708333	26.5350702238309\\
0.648111979166667	26.5271429155896\\
0.603352864583333	25.9097382542552\\
0.583577473958333	20.5800769646497\\
0.572835286458333	8.28900466609054\\
};
\addplot [color=red,solid,forget plot]
  table[row sep=crcr]{%
4.93375651041667	19.2182721979351\\
3.35611979166667	19.2182721979351\\
2.487548828125	19.2182721979351\\
2.19921875	19.2182721979351\\
1.866943359375	19.2182721979351\\
1.46923828125	19.2182721979351\\
1.394287109375	19.2182721979351\\
1.335693359375	17.376194556949\\
1.21085611979167	19.2182721979351\\
1.08194986979167	19.2817113168319\\
1.04703776041667	6.50723866123642\\
};
\addplot [color=mycolor1,solid,forget plot]
  table[row sep=crcr]{%
2.23624674479167	30.4490328289014\\
1.43310546875	30.4490328289014\\
1.11124674479167	30.4490328289014\\
0.901041666666667	30.4490328289014\\
0.851888020833333	30.4490328289014\\
0.801676432291667	30.4520011512367\\
0.755696614583333	30.4520011512367\\
0.541097005208333	30.4520308447079\\
0.505452473958333	30.1659039342749\\
0.489013671875	30.1586332904434\\
0.479410807291667	6.45488154761444\\
};
\addplot [color=mycolor2,solid,forget plot]
  table[row sep=crcr]{%
6.1317138671875	20.3869376563002\\
3.7696533203125	20.3869376563002\\
2.61083984375	20.3869376563002\\
1.9964599609375	20.3869376563002\\
1.6380615234375	20.3869376563002\\
1.4005126953125	20.3869376563002\\
1.2552490234375	20.3869376563002\\
1.13037109375	20.3869376563002\\
1.02734375	20.3869376563002\\
0.94677734375	20.3869376563002\\
0.8839111328125	8.1263970860743\\
};
\addplot [color=mycolor3,solid,forget plot]
  table[row sep=crcr]{%
5.94091796875	17.2845722105694\\
3.5498046875	17.2845722105694\\
2.5313720703125	17.2845722105694\\
2.0015869140625	17.2845722105694\\
1.6981201171875	17.2845722105694\\
1.473876953125	17.2845722105694\\
1.326416015625	17.2845722105694\\
1.2080078125	17.2845722105694\\
1.1142578125	17.2845722105694\\
1.035888671875	17.2845722105694\\
0.97412109375	8.46183910406325\\
};
\addplot [color=darkgray,solid,forget plot]
  table[row sep=crcr]{%
3.9005126953125	16.618319381303\\
2.084716796875	16.618319381303\\
1.441162109375	16.618319381303\\
1.126220703125	16.618319381303\\
0.9442138671875	16.618319381303\\
0.818115234375	16.618319381303\\
0.7371826171875	16.618319381303\\
0.6741943359375	16.618319381303\\
0.6185302734375	16.618319381303\\
0.5770263671875	16.618319381303\\
0.5443115234375	8.01942848964548\\
};
\addplot [color=blue,solid,forget plot]
  table[row sep=crcr]{%
5.01873779296875	22.8645209223638\\
2.90982055664063	22.8645209223638\\
1.81689453125	22.8645209223638\\
1.32403564453125	22.8645209223638\\
1.08770751953125	22.8645209223638\\
0.932159423828125	22.8645209223638\\
0.83636474609375	22.8645209223638\\
0.752044677734375	22.8645209223638\\
0.68292236328125	22.8645209223638\\
0.62847900390625	22.8645209223638\\
0.58709716796875	9.22113931205707\\
};
\addplot [color=black!50!green,solid,forget plot]
  table[row sep=crcr]{%
6.67732747395833	13.8973989911685\\
4.5341796875	13.8973989911685\\
3.36930338541667	13.8973989911685\\
2.97477213541667	13.8973989911685\\
2.563720703125	13.8973989911685\\
2.035400390625	13.8973989911685\\
1.93277994791667	13.8716567408642\\
1.85205078125	13.8140987565893\\
1.67578125	13.7914820890446\\
1.48486328125	13.8459442157142\\
1.43831380208333	7.54206360050326\\
};
\addplot [color=red,solid,forget plot]
  table[row sep=crcr]{%
6.3011474609375	22.6819914897076\\
3.7982177734375	22.6819914897076\\
2.7467041015625	22.6819914897076\\
2.1844482421875	22.6819914897076\\
1.835693359375	22.6819914897076\\
1.58544921875	22.6819914897076\\
1.42138671875	22.6819914897076\\
1.275390625	21.2299490678935\\
1.160400390625	21.2299490678935\\
1.068115234375	22.6816714512041\\
0.9976806640625	8.89377270319402\\
};
\addplot [color=mycolor1,solid,forget plot]
  table[row sep=crcr]{%
3.85294596354167	25.6668871606313\\
2.47957356770833	25.6668871606313\\
1.79874674479167	25.6668871606313\\
1.60286458333333	25.6668871606313\\
1.33870442708333	25.6668871606313\\
0.96923828125	25.6668871606313\\
0.921630859375	25.6668871606313\\
0.88623046875	25.6668871606313\\
0.788818359375	25.6668871606313\\
0.684814453125	25.6668871606313\\
0.659342447916667	9.40051539409192\\
};
\addplot [color=mycolor2,solid,forget plot]
  table[row sep=crcr]{%
2.28873697916667	29.7284906755022\\
1.31209309895833	29.7284906755022\\
0.9951171875	29.7284906755022\\
0.801595052083333	29.7284906755022\\
0.7548828125	29.7284906755022\\
0.707845052083333	29.7284906755022\\
0.66796875	29.7284906755022\\
0.509847005208333	29.7284906755022\\
0.470621744791667	29.7284906755022\\
0.453206380208333	29.7284906755022\\
0.442626953125	6.69622784600271\\
};
\addplot [color=mycolor3,solid,forget plot]
  table[row sep=crcr]{%
4.64957682291667	18.4614963769204\\
3.13151041666667	18.4614963769204\\
2.51318359375	18.4614963769204\\
2.00935872395833	18.4614963769204\\
1.89241536458333	18.4614963769204\\
1.75528971354167	18.4614963769204\\
1.68798828125	18.4614963769204\\
1.19148763020833	18.4614963769204\\
1.136962890625	18.4614963769204\\
1.09375	18.4614963769204\\
1.06689453125	9.35200235829326\\
};
\addplot [color=darkgray,solid,forget plot]
  table[row sep=crcr]{%
2.56461588541667	22.2892532071592\\
1.553466796875	22.2892532071592\\
1.18196614583333	22.2892532071592\\
0.937825520833333	22.2892532071592\\
0.898274739583333	22.2892532071592\\
0.852457682291667	22.2892532071592\\
0.799560546875	22.2892532071592\\
0.498291015625	22.2892532071592\\
0.44873046875	22.2892532071592\\
0.436767578125	22.0101419452738\\
0.430989583333333	7.50111847823481\\
};
\addplot [color=blue,solid,forget plot]
  table[row sep=crcr]{%
5.60791015625	21.5602125038517\\
3.68318684895833	21.5602125038517\\
2.69417317708333	21.5602125038517\\
2.39469401041667	21.5602125038517\\
2.00162760416667	21.5602125038517\\
1.47509765625	21.5602125038517\\
1.40641276041667	21.5602125038517\\
1.3525390625	21.5602125038517\\
1.20564778645833	21.5602125038517\\
1.035400390625	21.5602125038517\\
1.00390625	5.78216189355818\\
};
\addplot [color=black!50!green,solid,forget plot]
  table[row sep=crcr]{%
5.95670572916667	21.9809710548665\\
3.83984375	21.9809710548665\\
3.10286458333333	21.9809710548665\\
2.486083984375	21.9076886529704\\
2.34073893229167	21.9076886529704\\
2.17342122395833	21.9700067401778\\
2.09440104166667	21.8981449972075\\
1.44954427083333	21.982523053791\\
1.39070638020833	18.9055846582729\\
1.33463541666667	21.74145542322\\
1.30550130208333	7.91321010300124\\
};
\addplot [color=red,solid,forget plot]
  table[row sep=crcr]{%
4.42879231770833	16.3005495076775\\
2.83748372395833	16.3005495076775\\
2.27091471354167	16.3005495076775\\
1.80021158854167	16.3005495076775\\
1.69669596354167	16.3005495076775\\
1.5751953125	16.3005495076775\\
1.52571614583333	16.2928399884321\\
1.057373046875	16.3005495076775\\
1.01066080729167	16.254379444912\\
0.968831380208333	15.0278865910363\\
0.948160807291667	8.87649402019074\\
};
\addplot [color=mycolor1,solid,forget plot]
  table[row sep=crcr]{%
4.34457397460938	33.0431331660688\\
2.27239990234375	33.0431331660688\\
1.462646484375	33.0431331660688\\
1.12841796875	33.0431331660688\\
0.942901611328125	33.0431331660688\\
0.812225341796875	33.0431331660688\\
0.72930908203125	33.0431331660688\\
0.657958984375	33.0431331660688\\
0.59735107421875	33.0431331660688\\
0.547332763671875	33.0431331660688\\
0.51055908203125	8.72634168500632\\
};
\addplot [color=mycolor2,solid,forget plot]
  table[row sep=crcr]{%
8.30826822916667	17.6793968526283\\
5.41609700520833	17.6793968526283\\
4.003173828125	17.6793968526283\\
3.52872721354167	17.6793968526283\\
3.03971354166667	17.6793968526283\\
2.34334309895833	17.6793968526283\\
2.228759765625	17.6793968526283\\
2.1396484375	17.6793968526283\\
1.90690104166667	17.6793968526283\\
1.63932291666667	17.6502827815054\\
1.59423828125	7.66416868697323\\
};
\addplot [color=mycolor3,solid,forget plot]
  table[row sep=crcr]{%
5.97574869791667	14.0359494986228\\
3.82975260416667	14.0359494986228\\
3.06819661458333	14.0359494986228\\
2.483154296875	14.0359494986228\\
2.328125	14.0359494986228\\
2.16178385416667	14.0359494986228\\
2.07454427083333	13.9920286378459\\
1.49226888020833	14.0359494986228\\
1.42781575520833	14.0359494986228\\
1.36726888020833	14.0357856613877\\
1.3388671875	7.43095877349761\\
};
\addplot [color=darkgray,solid,forget plot]
  table[row sep=crcr]{%
6.41300455729167	13.7967333071458\\
4.15242513020833	13.7967333071458\\
3.35652669270833	13.7967333071458\\
2.72102864583333	13.7967333071458\\
2.55509440104167	13.7967333071458\\
2.3681640625	13.7967333071458\\
2.27897135416667	13.7967333071458\\
1.64794921875	13.7967333071458\\
1.57438151041667	13.7915709774294\\
1.50911458333333	13.7954158996694\\
1.474365234375	7.88249670665103\\
};
\addplot [color=blue,solid,forget plot]
  table[row sep=crcr]{%
3.57845052083333	16.9804902252875\\
2.33610026041667	16.9804902252875\\
1.70247395833333	16.9804902252875\\
1.51123046875	16.9804902252875\\
1.27718098958333	16.9804902252875\\
0.977132161458333	16.9804902252875\\
0.9326171875	16.9804902252875\\
0.894694010416667	16.9804902252875\\
0.8076171875	16.9804902252875\\
0.716796875	16.9804902252875\\
0.695068359375	10.1067069361274\\
};
\addplot [color=black!50!green,solid,forget plot]
  table[row sep=crcr]{%
4.84066772460938	24.5957805911272\\
2.67083740234375	24.6107844968113\\
1.76467895507813	24.5922972044829\\
1.35427856445313	24.5916886692136\\
1.12542724609375	24.6812060855154\\
0.96343994140625	24.6501724813653\\
0.85833740234375	24.5838604232313\\
0.769683837890625	24.4947591334354\\
0.696990966796875	22.3381986902094\\
0.640106201171875	22.7556485653459\\
0.596038818359375	6.29038728756166\\
};
\addplot [color=red,solid,forget plot]
  table[row sep=crcr]{%
1.87565104166667	27.4857025091104\\
1.15755208333333	27.4857025091104\\
0.901529947916667	27.4857025091104\\
0.725667317708333	27.4857025091104\\
0.691162109375	27.4857025091104\\
0.65234375	26.9668457679518\\
0.608154296875	26.8376695664778\\
0.419677734375	26.4145614933145\\
0.378336588541667	22.8893581282924\\
0.367594401041667	17.7666505515965\\
0.35986328125	8.72356965547795\\
};
\addplot [color=mycolor1,solid,forget plot]
  table[row sep=crcr]{%
3.396484375	17.866694674398\\
2.28133138020833	17.866694674398\\
1.833984375	17.866694674398\\
1.47981770833333	17.866694674398\\
1.38916015625	17.866749524367\\
1.298095703125	17.866749524367\\
1.23860677083333	17.8785912180451\\
0.882893880208333	17.6710519039174\\
0.833089192708333	17.5515996484099\\
0.802164713541667	14.6883108677489\\
0.786702473958333	8.88781391933448\\
};
\addplot [color=mycolor2,solid,forget plot]
  table[row sep=crcr]{%
6.45458984375	15.0933632326316\\
3.982177734375	15.0933632326316\\
2.8663330078125	15.0933632326316\\
2.2867431640625	15.0933632326316\\
1.939208984375	15.0933632326316\\
1.693603515625	15.0933632326316\\
1.538818359375	15.0933632326316\\
1.403076171875	15.0934501218917\\
1.294677734375	15.0811899977598\\
1.2064208984375	14.9118237363746\\
1.1380615234375	7.58298306658389\\
};
\addplot [color=mycolor3,solid,forget plot]
  table[row sep=crcr]{%
6.89225260416667	17.9327683001463\\
4.54630533854167	17.9327683001463\\
3.70792643229167	17.9327683001463\\
2.94930013020833	17.9327683001463\\
2.770263671875	17.9327683001463\\
2.57283528645833	17.9327683001463\\
2.49235026041667	17.9327683001463\\
1.67081705729167	17.9327683001463\\
1.60929361979167	17.9327683001463\\
1.54069010416667	17.9327683001463\\
1.5078125	8.5159769743347\\
};
\addplot [color=darkgray,solid,forget plot]
  table[row sep=crcr]{%
4.140380859375	19.6683738294479\\
2.07568359375	19.6683738294479\\
1.350341796875	19.6683738294479\\
1.02490234375	19.6683738294479\\
0.8480224609375	19.6683738294479\\
0.7213134765625	19.6683738294479\\
0.6466064453125	19.6683738294479\\
0.5849609375	19.6683738294479\\
0.5355224609375	19.6683738294479\\
0.495849609375	19.6683738294479\\
0.4681396484375	8.46919404327023\\
};
\addplot [color=blue,solid,forget plot]
  table[row sep=crcr]{%
4.16853841145833	14.7541779162913\\
2.51985677083333	14.7541779162913\\
1.97591145833333	14.7541779162913\\
1.603271484375	14.7541779162913\\
1.493408203125	14.7541779162913\\
1.38427734375	14.7541779162913\\
1.324951171875	14.7541779162913\\
0.998128255208333	14.7541779162913\\
0.949055989583333	14.7541779162913\\
0.907063802083333	14.7541779162913\\
0.8857421875	8.16198127101281\\
};
\addplot [color=black!50!green,solid,forget plot]
  table[row sep=crcr]{%
3.02555338541667	18.6401911636986\\
1.86173502604167	18.6401911636986\\
1.332763671875	18.6401911636986\\
1.1650390625	18.6401911636986\\
0.994140625	18.6401911636986\\
0.776774088541667	18.6401911636986\\
0.732991536458333	18.6401911636986\\
0.695963541666667	18.6401911636986\\
0.628662109375	18.6401911636986\\
0.55712890625	18.6401911636986\\
0.538899739583333	8.65882235972738\\
};
\end{axis}
\end{tikzpicture}%

%% file: figures/bpp_vs_psnr_sys.tex
%
%
%
\begin{tikzpicture}

\begin{axis}[%
xmin=0,
xmax=8,
ymin=5,
ymax=30,
ylabel=PSNR,
label style={font=\footnotesize}
]
\addplot [color=red,solid,line width=1.0pt,mark=asterisk,mark options={solid}]
  table[row sep=crcr]{%
0.2	16.3157095473123\\
0.6	16.2880191230485\\
1	16.7516664606297\\
1.4	15.9679192652699\\
1.8	16.150064798131\\
2.2	16.7211112862233\\
2.6	15.5661563267648\\
3	15.7928485596187\\
3.4	17.35563488164\\
3.8	16.3416738939183\\
4.2	16.6896586368478\\
4.6	15.2588224480126\\
5	17.0755287358183\\
5.4	16.7522125180427\\
5.8	19.2665986505177\\
6.2	16.2184216710469\\
6.6	13.6636214730216\\
7	16.1639683000853\\
7.4	14.7247113564348\\
7.8	15.406378134155\\
};
\addlegendentry{M1};

\addplot [color=cyan,solid,mark=o,mark options={solid}]
  table[row sep=crcr]{%
0.2	9.91198213474862\\
0.6	8.18248726758061\\
1	8.68488530366975\\
1.4	9.48735954530841\\
1.8	10.9691136240083\\
2.2	12.5246676559427\\
2.6	13.2085635915427\\
3	14.9137099718005\\
3.4	16.4212946675511\\
3.8	17.0151047142076\\
4.2	17.0668366689331\\
4.6	16.9036435675483\\
5	18.7249743814139\\
5.4	19.0173992696455\\
5.8	19.7311020548595\\
6.2	19.7602432211376\\
6.6	16.3980105675353\\
7	18.5329965699843\\
7.4	17.9248140490937\\
7.8	15.9918282508911\\
};
\addlegendentry{M2};

\addplot [color=blue,solid,mark=square,mark options={solid}]
  table[row sep=crcr]{%
0.2	19.1206777276922\\
0.6	18.5248642211696\\
1	18.143590540599\\
1.4	18.2187575333064\\
1.8	18.9223020514168\\
2.2	19.5981910451134\\
2.6	19.3681702412971\\
3	19.6391681394351\\
3.4	20.5065064641138\\
3.8	20.958561663552\\
4.2	20.8947956491533\\
4.6	19.7256685253126\\
5	21.1082337390475\\
5.4	21.1765535614648\\
5.8	22.516371672867\\
6.2	21.8285651393519\\
6.6	17.8580787680265\\
7	20.2198045226641\\
7.4	18.9473913496381\\
7.8	18.9098617558046\\
};
\addlegendentry{M3};

\end{axis}
\end{tikzpicture}%

%% file: figures/bpp_vs_ssim_sys.tex
%
%
%
\begin{tikzpicture}

\begin{axis}[%
xmin=0,
xmax=8,
ymin=0,
ymax=0.8,
ylabel=SSIM,
label style={font=\footnotesize}
]
\addplot [color=red,solid,line width=1.0pt,mark=asterisk,mark options={solid}]
  table[row sep=crcr]{%
0.2	0.394549823997846\\
0.6	0.382862006897106\\
1	0.332462597104338\\
1.4	0.272993596120629\\
1.8	0.269788928611414\\
2.2	0.269344453008571\\
2.6	0.231982672720072\\
3	0.242474362857786\\
3.4	0.276220064081526\\
3.8	0.278048381834043\\
4.2	0.276119839914672\\
4.6	0.227644332103919\\
5	0.294011677279528\\
5.4	0.248099593590905\\
5.8	0.338229292658003\\
6.2	0.281846521019636\\
6.6	0.176256969824195\\
7	0.22676887410775\\
7.4	0.144526932495626\\
7.8	0.189011173333323\\
};
\addlegendentry{M1};

\addplot [color=cyan,solid,mark=o,mark options={solid}]
  table[row sep=crcr]{%
0.2	0.151000831223033\\
0.6	0.0900949384143473\\
1	0.102643826823238\\
1.4	0.132321024111926\\
1.8	0.185872358335385\\
2.2	0.228922142657883\\
2.6	0.246795965856003\\
3	0.307933149817338\\
3.4	0.342999189983936\\
3.8	0.357157666403959\\
4.2	0.366723465468447\\
4.6	0.366589017231428\\
5	0.408168943932337\\
5.4	0.410883051511027\\
5.8	0.408230449545855\\
6.2	0.423698339608773\\
6.6	0.385098665368691\\
7	0.385897393408797\\
7.4	0.416501549044593\\
7.8	0.349396635622841\\
};
\addlegendentry{M2};

\addplot [color=blue,solid,mark=square,mark options={solid}]
  table[row sep=crcr]{%
0.2	0.373563604209984\\
0.6	0.322934054320483\\
1	0.30924369897901\\
1.4	0.332545022201717\\
1.8	0.381515729507604\\
2.2	0.408637400876412\\
2.6	0.414837422208425\\
3	0.456967076862377\\
3.4	0.463904091517869\\
3.8	0.48686247658574\\
4.2	0.500344559662765\\
4.6	0.46941799636809\\
5	0.517392593226537\\
5.4	0.492264530592811\\
5.8	0.512370330073841\\
6.2	0.522837794160144\\
6.6	0.458377112136548\\
7	0.469381968020622\\
7.4	0.464062329770399\\
7.8	0.477651505061038\\
};
\addlegendentry{M3};

\end{axis}
\end{tikzpicture}%